\documentclass[aps,preprint,nofootinbib,preprintnumbers,eqsecnum]{revtex4-1}

\usepackage[usenames, dvipsnames]{color}

\newcommand{\nb}[1]{\color{blue}}

\newcommand{\hl}[1]{\color{magenta}}

\usepackage{geometry,tikz}

\usepackage[
	  pagebackref=false,
	  colorlinks=true,
      linkcolor=blue,
      urlcolor=blue,
      filecolor=black,
      citecolor=red,
      pdfstartview=FitV,
      pdftitle={},
        pdfauthor={},
        pdfsubject={},
        pdfkeywords={},
        pdfpagemode=None,
        bookmarksopen=true
      ]{hyperref}

\usepackage[normalem]{ulem}
\usepackage{amsmath}
\usepackage{enumerate}
\usepackage{amsfonts}
\usepackage{epsfig}
\usepackage{amssymb}

\usepackage{setspace}

\def\Tr{\mathop{\rm Tr}}

\newcommand\half{{\ensuremath{\frac{1}{2}}}}
\newcommand\p{\ensuremath{\partial}}

\newcommand\field[1]{{\ensuremath{\mathbb{{#1}}}}}

\newcommand\vev[1]{{\ensuremath{\left\langle{#1}\right\rangle}}}

\newcommand\ket[1]{\ensuremath{\lvert{#1}\rangle}}
\newcommand\bra[1]{\ensuremath{\langle{#1}\rvert}}

\newcommand{\imineq}[2]{\vcenter{\hbox{\includegraphics[height=#2ex]{#1}}}}

\newcommand{\FF}{\field{F}}
\newcommand{\OO}{\field{O}}

\newcommand{\HH}{\field{H}}

\newcommand{\MM}{\field{M}}

\newcommand{\RR}{\field{R}}

\newcommand{\TT}{\field{T}}
\newcommand{\XX}{\field{X}}

\newcommand{\be}{\begin{equation}}
\newcommand{\ee}{\end{equation}}
\newcommand{\bea}{\begin{eqnarray}}
\newcommand{\eea}{\end{eqnarray}}
\newcommand{\bega}{\begin{gather}}
\newcommand{\eega}{\end{gather}}

\newcommand{\bi}{\begin{itemize}}
\newcommand{\ei}{\end{itemize}}
\newcommand{\ben}{\begin{enumerate}}
\newcommand{\een}{\end{enumerate}}
\newcommand{\bca}{\begin{cases}}
\newcommand{\eca}{\end{cases}}
\newcommand{\bln}{\begin{align}}
\newcommand{\eln}{\end{align}}
\newcommand{\bst}{\begin{split}}
\newcommand{\est}{\end{split}}
\def\ie{\begin{equation}\begin{aligned}}
\def\fe{\end{aligned}\end{equation}}
\newcommand{\bma}{\le(\begin{matrix}}
\newcommand{\ema}{\end{matrix}\ri)}

\newcommand\al{{\alpha}}
\def\b{{\beta}}
\newcommand\ep{\epsilon}
\newcommand\sig{\sigma}

\newcommand\lam{\lambda}

\newcommand\om{\omega}
\newcommand\Om{\Omega}

\newcommand\ga{{\ensuremath{{\gamma}}}}
\newcommand\Ga{{\ensuremath{{\Gamma}}}}
\newcommand\de{{\ensuremath{{\delta}}}}
\newcommand\De{{\ensuremath{{\Delta}}}}

\newcommand\da{{\dagger}}

\newcommand\ov{\over}
\newcommand\ha{{\half}}

\def\le{\left}
\def\ri{\right}

\newcommand\sA{{\ensuremath{{\mathcal A}}}}
\newcommand\sB{{\ensuremath{{\mathcal B}}}}

\newcommand\sE{{\ensuremath{{\mathcal E}}}}
\newcommand\sF{{\ensuremath{{\mathcal F}}}}
\newcommand\sI{{\ensuremath{{\mathcal I}}}}
\newcommand\sG{{\ensuremath{{\mathcal G}}}}
\newcommand\sH{{\ensuremath{{\mathcal H}}}}

\newcommand\sM{{\ensuremath{{\mathcal M}}}}
\newcommand\sN{{\ensuremath{{\mathcal N}}}}
\newcommand\sO{{\ensuremath{{\mathcal O}}}}

\newcommand\sJ{{\mathcal J}}
\newcommand\sR{{\mathcal R}}
\newcommand\sS{{\mathcal S}}

\newcommand\sU{{\mathcal U}}

\newcommand\sX{{\mathcal X}}
\newcommand\sY{{\mathcal Y}}
\newcommand\sZ{{\mathcal Z}}

\newcommand{\bid}{\mathbf{1}}

\newcommand{\wt}{\widetilde}
\newcommand{\Nlim}{\widetilde{\lim_{N \to \infty}}}
\renewcommand{\ol}{\overline}
\newcommand\Fil[1]{{\ensuremath{\FF\{#1\}}}}

\newcommand{\Mp}{\MM_+}

\begin{document}

\title{``Filtering'' CFTs at large $N$: Euclidean Wormholes, Closed Universes, and Black Hole Interiors}

\preprint{MIT-CTP/5970}

\author{Hong Liu}
\affiliation{MIT Center for Theoretical Physics---a Leinweber Institute,\\ 
Massachusetts Institute of Technology, \\
77 Massachusetts Ave.,  Cambridge, MA 02139\\}


\begin{abstract}
 
 Despite its many successes, the large-$N$ holographic dictionary remains incomplete. Certain features of gravitational path integrals---most notably the appearance of Euclidean wormholes and the associated failure of factorization---lack a clear interpretation within the conventional large-$N$ framework. A related challenge is the possibility of erratic 
$N$-dependence in CFT observables, behavior with no evident semiclassical gravitational counterpart.

In this paper, we argue that these puzzles point to a missing ingredient in the holographic dictionary: a large-$N$
filter. This filter acts as a projection that removes the erratic $N$-dependence of CFT quantities when mapping them to their semiclassical gravitational duals, providing an intrinsic boundary definition of the gravitational ``averages.''  It also yields a boundary explanation of the origin of wormhole contributions and a boundary prediction of their amplitudes, thereby supplying a natural resolution of the factorization puzzle.
In addition, we derive an infinite tower of inequalities constraining wormhole amplitudes, and we argue that internal wormholes do not induce random couplings in the low-energy effective theory.
 
Beyond resolving factorization, the large-$N$ filter offers a generalized framework from which richer Lorentzian spacetime structures can emerge. Illustrative examples include the appearance of closed universes and black hole interiors.
We argue that, as a consequence of erratic large-$N$ behavior, both closed universes and black hole interiors are quantum volatile, and that an AdS spacetime entangled with a baby universe is likewise quantum volatile.
This quantum volatility may allow an observer in the AdS spacetime to infer the existence of the baby universe, whereas for an infalling observer crossing a black hole horizon, the ability to make measurements may become fundamentally limited---although they may not live long enough to recognize this limitation.

\end{abstract}

\today

\maketitle

\tableofcontents

\linespread{0.6}

\section{Introduction: a filtered large-$N$ dictionary}

The map between large-$N$ conformal field theories and semiclassical gravity has been one of the defining achievements of holographic duality~\cite{Mal97,GubKle98,Wit98}. It provides a precise realization of the idea that smooth spacetime geometry can emerge from the collective dynamics of quantum fields. Through this correspondence, a wide range of gravitational phenomena---from black hole dynamics to the statistical origin of gravitational entropy---have acquired a microscopic interpretation.

Yet despite its many successes, the large-$N$ holographic dictionary remains incomplete. Certain features of gravitational path integrals---most notably the appearance of Euclidean wormholes and the associated failure of factorization---do not admit a straightforward explanation within the conventional large-$N$ framework~\cite{WitYau99,MalMao04}. Another issue is the possibility of erratic $N$-dependence in CFT quantities, behavior that does not seem to have a direct counterpart on the gravitational side~\cite{SchWit22,Liu25}.

In this paper, we argue that these puzzles point to a missing element in the dictionary: a large-$N$ filter, a device that ``projects'' out the erratic $N$-dependence of CFT quantities in their map to the gravitational dual in the semiclassical limit. We show that such a filter provides a natural resolution of the factorization puzzle by giving an intrinsic boundary definition of the gravitational ``averages,'' and explaining the appearance of wormholes and their amplitudes. In addition, it offers a framework for understanding the emergence of closed universes and black hole interiors.

Below we begin by reviewing the two key problems that motivate it: the factorization problem and possible erratic-$N$ behavior in large-$N$ CFTs.

\subsection{The factorization and erratic-$N$ problems} 

The AdS/CFT duality~\cite{Mal97,GubKle98,Wit98} states that a quantum gravity theory in $(d+1)$-dimensional anti-de Sitter (AdS) spacetime is dual to a $d$-dimensional conformal field theory (CFT) defined on its boundary. The $\mathrm{CFT}_d$ parameter $N$, which measures the number of degrees of freedom, is related to the bulk Newton constant by $N^2 \propto G_N^{-1}$.\footnote{Different realizations of the duality may feature distinct relations between their degrees of freedom and $G_N$. For notational convenience, we define $N$ such that $N^2 \propto G_N^{-1}$ holds universally.}




A central entry in the AdS/CFT dictionary is the relation
\be \label{cenT}
Z_{\mathrm{CFT}}[M] = Z_{\mathrm{gravity}}[M] ,
\ee
where $Z_{\mathrm{CFT}}[M]$ denotes the partition function of the boundary CFT on a Euclidean manifold $M$, and $Z_{\mathrm{gravity}}[M]$ is that of the corresponding bulk quantum gravity theory with boundary conditions specified by $M$. In~\eqref{cenT}, we leave implicit the dependence on $N$ and $G_N$ on both sides, and the equality is \emph{assumed} to hold for each value of these parameters. The manifold $M$ should be understood broadly---it may be disconnected or possess nontrivial topology. We also allow for the inclusion of heavy operator insertions---operators with dimensions of order $O(N^2)$---as part of the definition of $M$.\footnote{In general, $Z_{\mathrm{CFT}}[M]$ can be complex, satisfying $(Z_{\mathrm{CFT}}[M])^* = Z_{\mathrm{CFT}}[\ol{M}]$, where $\ol{M}$ denotes the manifold $M$ with reversed orientation and any (heavy) operator insertions replaced by their conjugate.}
Light operator insertions---operators with dimensions of order $O(N^0)$---will be made explicit.



Since the bulk partition function is not yet defined in a precise nonperturbative sense, what is most commonly employed in the duality
is the large $N$ limit of~\eqref{cenT}, 
\be \label{larN}
\lim_{N \to \infty} Z_{\mathrm{CFT}}[M] = \lim_{G_N \to 0} Z_{\mathrm{gravity}}[M]  \ .
\ee
Here, the limits on both sides should not be interpreted as strict mathematical limits. Rather, standard practice has been to interpret equation~\eqref{larN} as an equality between asymptotic expansions in $1/N^2$ and $G_N$, respectively.

More explicitly, the right-hand side of~\eqref{larN} is conventionally defined in terms of Euclidean gravitational path integrals, 
which in turn is defined as a formal device to extract the asymptotic expansion in $G_N$,
\be
\label{EGP0}
\lim_{G_N \to 0} Z_{\mathrm{gravity}}[M] = Z_{\rm GPI} [M], 
\ee
where 
\begin{align} \label{EGP1}
 Z_{\rm GPI} [M] & \equiv  \int_{\text{$M$ as boundary}} D \Phi \, e^{ - \sI_E [\Phi]} \\
 & = \sum_{i} e^{- \sI_E[\sM_i]} Z_1[\sM_i]
\left(1 + G_N Z_2[\sM_i] +  \cdots \right)  \ .
 \label{EGP}
\end{align}
In the above equation\footnote{Possible $\log G_N$ terms have been suppressed for notational simplicity. The same applies to equation~\eqref{cftE} below.}, the sum over $i$ runs over all Euclidean saddle points $\sM_i$ that solve the equations of motion derived from the Euclidean action $\sI_E$, subject to the boundary condition $\p\sM_i = M$, including all possible bulk topologies.
The action $\sI_E$ is taken to be proportional to $1/G_N$, and the factor $Z_1[\sM_i]$  ($G_N$-independent) is the one-loop partition function of small fluctuations---both matter and metric---around the background $\sM_i$, while $Z_2, \cdots$ arise from higher-loop corrections. In equation~\eqref{EGP}, we should include all subdominant saddles. 

The structure~\eqref{EGP}, which we will refer to as exhibiting ``smooth behavior'' as  $G_N \to 0$, follows from the fact that {\it $G_N$ appears only as an overall coupling in the prefactor of the Euclidean gravity action $\sI_E$.} The term ``smooth behavior'' is not used in the mathematical sense of differentiability with respect to $G_N$, but rather refers more generally to situations in which $Z_{\mathrm{gravity}}[M]$ exhibits the specific dependence on $G_N$ of the form~\eqref{EGP}. This notion does not require a well-defined $G_N \to 0$ limit, since the Euclidean action $\sI_E$ may be negative, or complex when evaluated on a complex saddle. 
Mathematically, the series~\eqref{EGP}, defined as a formal sum of asymptotic expansions multiplied by exponential weights, is called a transseries. 


In string and $M$-theory, one can also encounter corrections of the form $e^{-O(1/\sqrt{G_N})}$, arising from D-branes or D-instantons in string theory and from membranes in $M$-theory. The inclusion of such terms would not change our discussion below; we mention them here for completeness but do not include them in what follows.

The CFT side of~\eqref{larN} has been conventionally assumed to have the structure of a transseries, i.e.,
\be \label{cftE00}
\lim_{N \to \infty} Z_{\rm CFT}[M] = \sum_{i} e^{-N^2 W_0^{(i)}} Z_1^{(i)} \le( 1 + {1 \ov N^2} Z_2^{(i)} + \cdots \ri)
\ee
where $i$ labels the different large-$N$ saddle points. Equation~\eqref{larN} then implies that the double expansion---the sum over saddles and the sum over power-law corrections---in~\eqref{EGP} and~\eqref{cftE00} must match term by term, including exponentially suppressed subdominant terms. We will also refer to~\eqref{cftE00} as smooth dependence.


There are, however, two potential challenges to equation~\eqref{larN}, interpreted as an equality between asymptotic expansions in $1/N^2$ and $G_N$.

\subsubsection{The factorization problem}\label{sec:fact}

Consider a manifold $M = M_1 \cup M_2$ consisting of disjoint components $M_1$ and $M_2$. By definition, one expects from~\eqref{larN}
\be \label{fact}
\lim_{N \to \infty} Z_{\mathrm{CFT}}[M_1 \cup M_2]
= \left(\lim_{N \to \infty} Z_{\mathrm{CFT}}[M_1]\right)
\left(\lim_{N \to \infty} Z_{\mathrm{CFT}}[M_2]\right)
= \lim_{G_N \to 0} Z_{\mathrm{gravity}}[M_1 \cup M_2] \ .
\ee
However, as first discussed in~\cite{WitYau99,MalMao04}, the right-hand side of~\eqref{fact} generally fails to factorize due to contributions from Euclidean wormhole solutions connecting $M_1$ and $M_2$. We stress that it is immaterial whether such a Euclidean wormhole solution is the dominant saddle or not: its mere existence, even as a subdominant contribution, spoils the factorization structure. This issue---known as the {factorization problem}---has attracted renewed attention following the realization that such wormholes are crucial for reproducing the Page curve~\cite{PenShe19,AlmHar19}. 

The problem is better understood in Jackiw-Teitelboim (JT) gravity~\cite{Jackiw,Tei83,AlmPol14}, where the Euclidean path integral can, in principle, be used to provide a definition of the theory. In this setting, summing over wormholes leads to an ensemble average over random boundary Hamiltonians~\cite{SaaShe19}, so the boundary theory does not factorize. There are also indications that something similar may occur in pure AdS$_3$ gravity, which has been argued to be described by random CFTs~\cite{CotJen20} (see also~\cite{ColEbe23,ColEbe23a,ColEbe24,BeldeB23,JafRoz24,JafRoz25,Har25a,Har25b,HunJia25,BorDiU25,Cha25}).

However, for duality examples in string/M theory---such as type IIB superstring theory on AdS$_5 \times S_5$ or AdS$_3 \times S_3 \times K3$---which possess specific boundary CFT duals, and for which the Euclidean gravitational path integral is notoriously ill-defined, the situation remains far from clear.
 




There are various logical possibilities for addressing the factorization problem, each with its own challenges. One possibility is to exclude wormhole configurations. This would require a substantial modification of the usual semiclassical prescription, even at arbitrarily small $G_N$, despite the absence of an independent reason for such a modification.\footnote{The familiar limitations of effective field theory in Lorentzian settings involving horizons do not directly address the factorization problem considered here, which arises already for Euclidean path integrals.} Moreover, it is unclear how such an exclusion could be reconciled with the successful Page-curve calculations based on replica wormholes~\cite{PenShe19,AlmHar19}.

Another possibility is that additional, ``hidden'' contributions combine with the wormhole contributions to reproduce the exactly factorized answer~\eqref{fact}, thereby preserving the validity of~\eqref{larN} and~\eqref{EGP0}--\eqref{EGP}. This mechanism has been exhibited in a fixed-coupling, zero-dimensional reduction of SYK~\cite{SaaShe21}, where additional saddle points---dubbed ``half-wormholes''---were identified whose contributions depend sensitively on the microscopic couplings and are non-self-averaging.
However, precisely because these configurations depend on the microscopic data of the theory, it is unclear how they generalize to conventional gravity theories or higher dimensions. Related approaches have been discussed in~\cite{MarMax20,BloMer19,SaaShe21b,GesMar24}.

A third possibility is that the left-hand side of~\eqref{larN} should be modified to include an average\footnote{See e.g.~\cite{KraMal16,CarMal17,ColMal19,Sta20,CheGor20,AfkCoh20,MalWit20,PolRoz20,LiuVar20b,ColMal21,ChaCol22,SchWit22,ChaHar22,BeldeB20,BeldeB21,BeldeB21b,AnoBel21,Sas22,BalLaw22,CotJen22,AntSas23,deBLis23,HaeRee23,DiUPer23,DiUPer23a,Cha23,deBLis24,ChaHar24,Cha24,Liu25,KudWit25,WanWan25,Yan25} for discussions of averages of CFTs and wormholes. \label{ft:ave}}, denoted by an overline, 
\be\label{aveG}
\lim_{N \to \infty} \ol{Z_{\rm CFT} [M]} = \lim_{G_N \to 0} Z_{\mathrm{gravity}}[M]  ,
\ee
so that the left-hand side of~\eqref{fact} no longer factorizes
\be \label{nonfa}
\lim_{N \to \infty} \ol{Z_{\rm CFT} [M_1 \cup M_2]} \neq  \left(\lim_{N \to \infty} \ol{Z_{\mathrm{CFT}}[M_1]} \right)
\left(\lim_{N \to \infty} \ol{ Z_{\mathrm{CFT}}[M_2]}\right) \ .
\ee
At a heuristic level, this seems physically reasonable: since semiclassical gravitational computations cannot resolve the detailed microscopic structure of the boundary theory, some degree of coarse-graining may naturally arise in mapping the large-$N$ CFT to its gravitational description. 

To date, there is no intrinsic boundary definition for the left-hand side of~\eqref{aveG}. The object $\overline{Z_{\rm CFT}[M]}$ is defined by the gravitational path integral and is presumed to represent some kind of boundary average---perhaps over Hamiltonians, states, observables, or $N$. Yet there are basic difficulties in formulating such an average within the field theory itself:




\ben
\item {\it No natural ensemble exists}

There is no canonical ensemble associated with a theory such as the $\mathcal N =4$ super-Yang-Mills (SYM), the boundary dual of type IIB superstring theory on AdS$_5 \times S^5$. The theory is rigidly defined, with no intrinsic source of randomness or family of nearby variants from which an ensemble could be constructed (except perhaps $N$~\cite{SchWit22}). 

We stress that, given the fundamental nature of the relation~\eqref{larN}, the averaging procedure used to address the factorization problem should be universal across all holographic systems. The relation cannot be fundamental if different averaging procedures must be invoked for different theories, or for different observables within the same theory.

\item {\it Large-$N$ structure may not be preserved by averaging.}

The large-$N$ expansion of the average would have to reproduce a structure precisely of the form~\eqref{EGP}, including the organization of all subdominant saddles. In general, however, even if each individual theory in an ensemble possesses its own large-$N$ expansion, the averaging procedure can spoil this structure. 


\item {\it Precise bulk-boundary matching relies on non-averaged theories.}

There exist detailed agreements between boundary BPS observables and bulk computations---such as Wilson loops, sphere partition functions, and supersymmetric indices---performed for specific (non-averaged) boundary theories (see~\cite{Pes16,Zaf19} for reviews). Any averaging risks destroying these matches. If the averaging were to apply only to a restricted class of observables, the criterion defining this class would itself need to be universal and independent of the internal structure of any particular theory.

\item {\it Potential difficulties with unitarity.}

Any introduction of an average must respect the fundamental holographic principle that bulk unitarity is guaranteed by the unitarity of the boundary theory. In particular, one must understand whether relations such as~\eqref{cenT}, valid at finite $N$, should be modified---and if so, in what way. Moreover, if the bulk theory corresponds to an ensemble average over boundary theories, then the boundary description represents a statistical mixture rather than a single unitary quantum system: unitarity holds for each member of the ensemble but not necessarily for the ensemble average itself.

\een
{The lack of an intrinsic boundary definition of the ``average''~\eqref{aveG} leads to ambiguities and difficulties, owing to the fact that the rules for gravitational path integrals are not uniquely determined. For example, one prescription implies that the Hilbert space of a closed universe is one-dimensional~\cite{UsaWang24,UsaZha24,HarUsa25,AbdAnt25}, while another results in a nontrivial Hilbert space~\cite{MarMax20,AbdAnt25}.}



\subsubsection{The erratic-$N$ problem}





Implicit in the relation~\eqref{larN} is the {\it assumption} that $Z_{\mathrm{CFT}}[M]$ exhibits ``smooth behavior''~\eqref{cftE00} in the $N \to \infty$ limit. However, in contrast to the gravitational path integral~\eqref{EGP}, in the CFT partition function---which can be written schematically as
\be\label{cftP}
\lim_{N \to \infty} Z_{\mathrm{CFT}}[M] = \lim_{N \to \infty} \int D \Phi \, e^{- {N \ov \lam} \Tr (\cdots) },
\ee
$N$ enters not only through the coupling but also through {\it the number of field-theoretic degrees of freedom}.
Here $\Phi$ collectively denotes the boundary fundamental fields, whose ranks depend on $N$, and the boundary action depends  on $N$ through the trace. There is no general principle guaranteeing that~\eqref{cftP} {\it always} leads, for arbitrary holographic CFTs on any $M$, to smooth $N \to \infty$ behavior of the form~\eqref{cftE00}, i.e., to a transseries expansion.

Consequently, $Z_{\mathrm{CFT}}[M]$ may exhibit ``non-smooth'' or ``erratic'' dependence on $N$. We will use these terms to refer to any behavior that cannot be expressed in terms of $1/N^2$ asymptotic expansion.\footnote{
We note that in~\eqref{EGP} and in~\eqref{cftE00} there can in principle be complex saddle points. For a real $Z_{\rm CFT}$, such saddles must appear in complex-conjugate pairs, leading to oscillatory behavior of the form $Z_{\rm CFT} \sim \cos(N^2 \#)  \sim \cos(\#/G_N)$. We will still refer to this oscillatory behavior in the partition function, when it arises from complex saddles, as smooth. However, oscillatory behavior of the form $\cos(N^2 \#)$ may also arise from mechanisms unrelated to complex saddles; in such cases, it should be regarded as non-smooth or erratic.}

At a microscopic level, erratic behavior in $N$ is generally expected for states of the boundary CFT with energies of $O(N^2)$. The usual assumption is that such erratic $N$-dependence becomes self-averaged in the large-$N$ limit upon summing over states in the partition function. Gregory Moore pointed out long ago the possibility of such erratic behavior in microcanonical ensembles of various supersymmetric indices (see, e.g.,~\cite{Moo98a,Moo98b,MilMoo99,DijMal00}). Recent discussions of fortuity (see, e.g.,~\cite{ChaLin24}) may likewise be viewed as further examples of this phenomenon in microcanonical ensembles. Here we question whether non-smooth behavior in $N$ can survive the summation over states and persist in $Z_{\mathrm{CFT}}[M]$. More recently, this possibility has also been emphasized in~\cite{SchWit22,Liu25,KudWit25} in different contexts.

When $Z_{\mathrm{CFT}}[M]$ exhibits erratic dependence on $N$, the two sides of~\eqref{larN} exhibit different structures, if the right hand side of~\eqref{larN} is given by~\eqref{EGP0}--\eqref{EGP}. Consequently, equations~\eqref{larN}--\eqref{EGP} are not self-consistent. 


This issue---which we will refer to as the erratic-$N$ problem---has received relatively little attention, possibly because no explicit example of erratic-$N$ dependence has been clearly identified in the boundary partition functions.


In this paper, we argue that erratic-$N$ dependence typically appears when $Z_{\mathrm{CFT}}[M]$ depends on OPE coefficients of heavy operators with dimensions of order $O(N^2)$. Examples include partition functions involving heavy-operator insertions and, in two-dimensional CFTs, those defined on manifolds of genus $g \ge 2$.

 
 

\subsection{Wormholes and correlations of erratic $N$-dependence}


One purpose of this paper is to propose that the factorization and erratic-$N$ problems are, in fact, the same and admit a unified resolution within a single framework.

We {\it postulate} that, as $N\to\infty$, the boundary partition function $Z_{\rm CFT}[M]$ admits the asymptotic decomposition
\begin{equation}\label{deCom}
Z_{\rm CFT}[M]
\underset{N\to\infty}{\simeq}
Z_{\rm CFT}^{(\rm smooth)}[M]
+
Z_{\rm CFT}^{(\rm erratic)}[M] \, .
\end{equation}
Here, $N\to\infty$ denotes the large-$N$ asymptotic regime and should not be interpreted as asserting the existence of a mathematical limit. The right-hand side approximates the CFT partition function at finite but large $N$. The smooth part $Z_{\rm CFT}^{(\rm smooth)}[M]$ is assumed to have the structure of a transseries in $1/N^2$, i.e.,~\eqref{cftE00}, which we reproduce here for convenience:
\be
\label{cftE}
Z_{\rm CFT}^{(\rm smooth)} [M] = \sum_{i} e^{-N^2 W_0^{(i)} } Z_1^{(i)} \le(1 + {1 \ov N^2} Z_2^{(i)} + \cdots \ri) \ .
\ee
The term $Z_{\rm CFT}^{(\rm erratic)}[M]$ denotes the remaining contribution that is not captured by this transseries structure and will be referred to as the erratic part. This contribution may be exponentially suppressed at large $N$ relative to the smooth part.

We stress that the decomposition~\eqref{deCom} is to be made on the basis of the mathematical structures of the two terms, with no averaging involved. See Appendix~\ref{sec:filter} for an illustrative example.

To address the erratic-$N$ problem, we propose that, in mapping to semiclassical gravity, the smooth part of $Z_{\rm CFT}[M]$ in the large $N$ regime be identified with the gravity partition function,
\be
Z^{\rm (smooth)}_{\mathrm{CFT}}[M] = \lim_{G_N \to 0} Z_{\mathrm{gravity}}[M] \ .
\ee
In other words, in matching to semiclassical gravity, we require a filter
\be \label{CC0}
\Fil{Z_{\mathrm{CFT}}[M]} \equiv Z^{\rm (smooth)}_{\mathrm{CFT}}[M]
\ee
that projects onto the smooth part and filters out the erratic contribution, thereby replacing~\eqref{larN} with
\be \label{mlarN}
\Fil{Z_{\mathrm{CFT}}[M]} = \lim_{G_N \to 0} Z_{\mathrm{gravity}}[M] \ .
\ee
This provides an intrinsic boundary definition of the left-hand side of~\eqref{aveG} in terms of $Z^{\rm (smooth)}_{\mathrm{CFT}}[M]$.

Denote 
\be \label{errd}
\de_e Z = Z_{\rm CFT}^{(\rm erratic)}[M] \ .
\ee
 By definition, under the filtering
\be
\Fil{\de_e Z} = 0 ,
\ee
but the information contained in $\de_e Z$ is not entirely lost. For example, if $\de_e Z$ contains an erratic phase, then such a phase may cancel in $\de_e Z^*  \de_e Z$, which can then have a smooth part, leading to a nonzero $\Fil{\de_e Z^* \de_e Z}$. In this case, we say that $\de_e Z$ and $\de_e Z^*$ have a nonzero ``correlation'' under filtering.
Similar statements apply to higher powers of $\de_e Z$, or to products of $\de_e Z$ with the erratic part of the CFT partition function on other manifolds. By studying such ``correlations'' of the erratic components of CFT partition functions on different manifolds under $\Fil{\cdot}$, one can extract the physics encoded in them.

By definition $\FF$ is a projection, i.e., 
\be 
\FF^2 = \FF , 
\ee
and satisfies 
 \be\label{CC2} 
\Fil{a Z_1 + b Z_2}
= a \Fil{Z_1}+ b \Fil{Z_2}  
\ee
where $a$ and $b$ are quantities that depend smoothly on $N$ (i.e. can be written as a transseries).  We will further assume the decomposition~\eqref{deCom} is such that 
the filter $\Fil{\cdot}$ 
is positive semi-definite
\be\label{posi}
\Fil{Z^* Z} \geq 0 \ .
 \ee
The above condition is motivated by (partial) preservation of unitarity in the large $N$ limit. Note that~\eqref{posi} further implies that  
\be \label{CC3}
(\Fil{Z})^* =\Fil{Z^*} \ .
\ee


We now explain how~\eqref{mlarN}  addresses the factorization problem. Consider a manifold $M = M_1 \cup M_2$ consisting of disjoint components $M_1$ and $M_2$. 
We have factorization 
\be
\text{finite $N$}: \quad Z_{\rm CFT} [M_1 \cup M_2] = Z_1 Z_2, \quad Z_1 \equiv Z_{\mathrm{CFT}}[M_1] ,  
\ee
and equation~\eqref{mlarN}  takes the form
\be \label{mul1} 
\Fil{Z_1 Z_2} = \lim_{G_N \to 0} Z_{\mathrm{gravity}}[M_1 \cup M_2] , 
\ee
where the right-hand side involves a sum over all gravitational saddles with asymptotic boundary $M_1 \cup M_2$, including possible wormhole configurations that connect $M_1$ and $M_2$. 

From~\eqref{deCom} and~\eqref{CC2}, we can write the left hand side of~\eqref{mul1} as 
\be \label{heno}
\Fil{Z_1 Z_2}  = \Fil{Z_1} \; \Fil{Z_2} + \Fil{\de_e Z_1 \de_e Z_2}, 
\ee
where we have used the notation~\eqref{errd}. The first term in~\eqref{heno}, which is equal to
$Z_{\mathrm{gravity}}[M_1] Z_{\mathrm{gravity}}[M_2]$, corresponds to the contribution from disconnected bulk manifolds.
Comparing~\eqref{mul1} and~\eqref{heno}, we then find that 
\be \label{imRe}
\Fil{\de_e Z_1 \de_e Z_2} = Z^{(\rm wormhole)}_{\mathrm{gravity}}[M_1 , M_2] ,
\ee
where the right-hand side represents the sum over all wormhole saddles connecting $M_1$ and $M_2$, including the corresponding asymptotic expansions in $G_N$ about each saddle. We stress that on the right 
hand side of~\eqref{mul1}  the relative dominance of the wormhole 
configuration versus disconnected contributions is immaterial here; they capture 
different parts of the partition functions.

Equation~\eqref{imRe} can be straightforwardly generalized to cases with multiple disjoint boundaries and light operator insertions. Denote $\sX = (M, X)$, where $M$ is a compact boundary manifold (possibly with heavy operator insertions), and $X$ specifies the light operator insertions together with their locations. We then have
\bega \label{0nide00}
\Fil{\delta_e Z_{\mathrm{CFT}}[\mathcal{X}_1] \cdots \delta_e Z_{\mathrm{CFT}}[\mathcal{X}_n]}
= Z_{\mathrm{gravity}}^{(\mathrm{wormhole})}[\mathcal{X}_1, \ldots, \mathcal{X}_n] \ .
\end{gather}
Both sides of the above expression may include disconnected contributions (though, by definition, no ``one-point'' amplitudes appear on either side). 
The fully connected version of~\eqref{0nide00} is given by
\begin{equation} \label{0nide}
\Fil{\delta_e Z_{\mathrm{CFT}}[\mathcal{X}_1] \cdots \delta_e Z_{\mathrm{CFT}}[\mathcal{X}_n]}_c
= Z_{\mathrm{gravity}}^{(\mathrm{wormhole, c})}[\mathcal{X}_1, \ldots, \mathcal{X}_n].
\end{equation}
Here, the subscript $c$ and the superscript $c$ denote the fully connected contributions to the CFT and gravity quantities, respectively. Specifically, the right-hand side includes only those wormhole saddles that connect all the boundaries $M_1, \cdots, M_n$, excluding any disconnected components.




Equations~\eqref{imRe}--\eqref{0nide} elucidate the physical nature of wormhole configurations and conveys a central message of our proposal:

\smallskip
{\it wormholes encode correlations among the erratic large-$N$ behavior of the boundary CFT.}
\smallskip

\noindent In particular, these equations can be used to provide boundary ``predictions'' for when bulk wormholes appear and for their amplitudes.

Equations~\eqref{imRe}--\eqref{0nide} concern an {\it external wormhole}---a 
manifold connecting disjoint boundaries. We will also consider (in Sec.~\ref{sec:internal}) examples of {\it internal wormholes}---those connecting different parts 
of a Euclidean manifold with a single boundary---and argue 
that they encode correlations of erratic $N$-dependence as well. In particular, with the erratic microscopic behavior already filtered out in the mapping to gravity, the internal wormholes do not appear to induce random couplings in the low-energy effective theory of gravity, in contrast to the arguments of~\cite{Col88,GidStr88a,GidStr88}.
Furthermore, we will see that although wormholes reflect correlations in 
the erratic $N$-dependence of boundary quantities, these 
correlations do not always manifest as wormhole configurations.

Currently, we do not have explicit boundary calculations that allow us to check the split~\eqref{deCom} or the proposal~\eqref{mlarN} directly. Nevertheless, using the identification~\eqref{mlarN} together with~\eqref{imRe}--\eqref{0nide}, we can deduce information about possible erratic behavior of the boundary theory, which we will explore extensively in Sec.~\ref{sec:AP}. In particular, from the non-existence~\cite{WitYau99,CaiGal00} of wormholes with boundary of the form $S_1 \times S_{d-1}$ and from the expected behavior of the spectral form factor, we can make ``predictions'' for the analytic behavior of thermal partition functions. Such predictions can, in principle, be checked and thus used to confirm or falsify our proposal.

\subsection{An emergent Hilbert space and an infinite tower of inequalities for wormholes} \label{sec:emG}




We now discuss an immediate implication of the positivity~\eqref{posi} and the identification~\eqref{0nide}: an infinite tower of inequality constraints that wormhole amplitudes should satisfy. 

We can view $\delta_e Z[\mathcal X] \equiv Z_{\rm CFT}^{(\rm erratic)}[M;X]$ 
as a ``function'' on the configuration space $\mathcal X$ 
(i.e.\ the space of manifolds $M$ together with heavy and light operator insertions). 
Finite linear combinations and products of such functions, together with complex conjugation, 
generate a commutative $*$-algebra $\mathcal A$.

The functional $\mathbb F$ acts linearly on $\mathcal A$,
\begin{equation}
\omega_{\mathbb F} : \mathcal A \to \mathbb T, 
\qquad 
\omega_{\mathbb F}(a) \equiv \mathbb F(a),
\end{equation}
where $\mathbb T$ denotes the space of transseries. 
The positivity condition~\eqref{posi} implies that 
\begin{equation} \label{psone}
\omega_{\mathbb F}(a^* a) \in \Gamma,
\end{equation}
for all $a \in \mathcal A$, where $\Gamma \subset \mathbb T$ is the cone of positive real transseries. 
The cone $\Gamma$ carries a strict total order determined by asymptotic growth.
The modulus $|\om_\FF (a)|$ of $\om_\FF (a)$ is defined as 
\be \label{modu}
|\om_\FF (a)|^2 \equiv \om_\FF (a) (\om_\FF (a))^* = \Fil{a} (\Fil{a})^* \in \Ga \ .
\ee

This setup directly generalizes that for the usual GNS construction, 
except that the ``state'' now takes values in $\mathbb T$ rather than in $\mathbb C$.

As in the standard GNS construction, we define the null space $\sJ = \{ a \in \sA : \om_{\FF}(a^\ast a) = 0 \}$, and associate a vector $\ket{a}$ with an element $a \in \mathcal A / \mathcal J$, with the inner product 
\be
\vev{a|b} \equiv \omega_{\mathbb F}(a^* b) = \Fil{a^*b}
\ee
now taking values in $\mathbb T$. Equation~\eqref{psone} implies that
the associated norm takes value in $\Ga$,
\begin{equation}
\| \ket{a} \| \equiv \sqrt{\omega_{\mathbb F}(a^* a)} \in \Ga \ .
\end{equation}
Since $\Gamma$ is strictly ordered, this defines a non-Archimedean norm. 
Completing $\mathcal A / \mathcal J$ with respect to the topology 
induced by this norm yields a space 
$\mathcal H_{\omega_{\mathbb F}}^{(\rm GNS)}$, 
which may be viewed as a transseries-valued generalization 
of the usual GNS Hilbert space.

We denote by $|\mathcal X\rangle$ the vector in 
$\mathcal H_{\omega_{\mathbb F}}^{(\rm GNS)}$ corresponding to 
$\delta_e Z[\mathcal X]$, and similarly
\begin{equation}\label{stees}
|\mathcal X_1,\dots,\mathcal X_n\rangle
\end{equation}
for the equivalence class of the product
$\delta_e Z[\mathcal X_1]\cdots \delta_e Z[\mathcal X_n]$.
The bra $\bra{\sX}$ denotes the equivalence class of $(\de_e Z [\sX])^* = \de_e Z [\ol{\sX}]$. 
By definition, the overlaps among~\eqref{stees} take values in $\TT$.



Later (in Sec.~\ref{sec:allClose}) we will interpret $\sH_{\omega_{\FF}}^{(\rm GNS)}$ as a ``semiclassical Hilbert space of all isolated closed universes'' described by the boundary CFT. Here we consider the implications of this Hilbert space structure for the wormhole amplitudes
through the identification~\eqref{0nide}.

For states $\ket{a}, \ket{a'} \in \sH_{\omega_{\mathbb F}}^{(\rm GNS)}$, with the definition of modulus in~\eqref{modu} and the 
positivity of the state~\eqref{psone}, 
the standard argument yields the Cauchy--Schwarz inequality
\begin{equation}\label{csi}
|\vev{a'|a}|^2 \leq \vev{a|a}\,\vev{a'|a'},
\end{equation}
where the inequality is understood in the sense that
\begin{equation}
\vev{a|a}\,\vev{a'|a'}
-
|\vev{a'|a}|^2
\in \Gamma \ .
\end{equation}

Taking $\ket{a} = \ket{\sX}$ and $\ket{a'} = \ket{\sX'}$, we obtain 
\be 
|\Fil{(\de_e Z[\sX'])^* \de_e Z[\sX]}|^2 \leq \Fil{(\de_e Z[\sX])^* \de_e Z[\sX]} \; \Fil{(\de_e Z[\sX'])^* \de_e Z[\sX']} , 
\ee
which from~\eqref{imRe} gives\footnote{We have learned that equation~\eqref{ineq0} was independently proposed in~\cite{DiUIli23} from considerations of positivity of gravitational path integrals, where it has also been verified to hold in various examples.}
\be  \label{ineq0} 
|Z^{(\rm wormhole)}_{\mathrm{gravity}}[\ol{ \sX'} , \sX] |^2 \leq Z^{(\rm wormhole)}_{\mathrm{gravity}}[\ol{ \sX }, \sX] \, 
Z^{(\rm wormhole)}_{\mathrm{gravity}}[\ol{\sX'} , \sX'] \ .
\ee

Similarly, by taking $\ket{a}, \ket{a'}$ to be $\ket{\sX_1 , \cdots , \sX_n}$ for different choices of $\sX_i$ and $n$, 
 we  get an infinite tower of inequalities from~\eqref{0nide00}, e.g., 
 \bega \label{ineq1}
 |Z^{(\rm wormhole)}_{\mathrm{gravity}}[\ol{\sX_3} , \sX_1, \sX_2] |^2 \leq Z^{(\rm wormhole)}_{\mathrm{gravity}}[\ol{\sX_1} , \ol{\sX_2} , \sX_1 , \sX_2] \, 
Z^{(\rm wormhole)}_{\mathrm{gravity}}[\ol{\sX_3} , \sX_3], \\
\label{ineq2}
|Z^{(\rm wormhole)}_{\mathrm{gravity}}[\ol{\sX_3} , \ol{\sX_4} , \sX_1, \sX_2] |^2 \leq Z^{(\rm wormhole)}_{\mathrm{gravity}}[\ol{\sX_1} , \ol{\sX_2} , \sX_1 , \sX_2] \, 
Z^{(\rm wormhole)}_{\mathrm{gravity}}[\ol{\sX_3} , \ol{\sX_4} , \sX_3 , \sX_4] , \\
\cdots \cdots \cdots \nonumber
\end{gather}
Note that for more than three boundaries, $Z^{(\rm wormhole)}_{\mathrm{gravity}}$ includes contributions from geometries that are both connected and disconnected (recall that in any disconnected contribution, each factor necessarily involves at least two boundaries).
For example, 
\bega
Z^{(\rm wormhole)}_{\mathrm{gravity}}[\ol{\sX_1} , \ol{\sX_2} , \sX_1 , \sX_2]    =   Z^{(\rm wormhole)}_{\mathrm{gravity}}[ \ol{\sX_1}, \sX_1]  Z^{(\rm wormhole)}_{\mathrm{gravity}}[ \ol{\sX_2}, \sX_2]  \cr
  +
    Z^{(\rm wormhole)}_{\mathrm{gravity}}[ \sX_1, \sX_2]  Z^{(\rm wormhole)}_{\mathrm{gravity}}[ \ol{\sX_1}, \ol{\sX_2} ]   \cr
+  Z^{(\rm wormhole)}_{\mathrm{gravity}}[ \ol{\sX_1}, \sX_2]  Z^{(\rm wormhole)}_{\mathrm{gravity}}[ \ol{\sX_2}, \sX_1]  
+
  Z^{(\rm wormhole,c)}_{\mathrm{gravity}}[\ol{\sX_1} , \ol{\sX_2} , \sX_1 , \sX_2]   \ .
  \label{ineq3}
\end{gather}

\subsection{Further comments on the factorization problem}


We may interpret the passage from~\eqref{cenT} to~\eqref{mlarN} as implementing a filtering procedure on both sides of~\eqref{cenT}. 
On the gravity side, the filtering is implemented by the gravitational path integral itself, understood semiclassically.

There are many indications that, in a complete theory of quantum gravity, $G_N$ plays a role beyond that of a conventional coupling constant: it also controls the effective number of degrees of freedom, as reflected, for example, in black hole entropy and in the stringy exclusion principle. It thus appears reasonable to say that when we approximate the full theory by a semiclassical evaluation of the gravitational path integral, we are effectively projecting onto a restricted sector of the full quantum gravitational answer (even at arbitrarily small $G_N$).

This perspective unifies and sharpens the two logical possibilities discussed earlier regarding the factorization problem:

\bi 

\item {\bf Factorization at arbitrarily small \(G_N\)} 

Previously, it was unclear where the semiclassical gravitational path integral approximation might fail, and whether restoring factorization would require excluding wormhole contributions (which are essential, for example, in Page curve calculations). In our proposal, 
the semiclassically defined Euclidean gravitational path integral is treated as valid within its domain of applicability, and wormhole saddles should be included. 

At every finite \(G_N\), including arbitrarily small \(G_N\), exact factorization should be restored once the semiclassical result is supplemented by contributions that are erratic in \(G_N\)---namely, the gravitational counterparts of $Z_{\rm CFT}^{\rm (erratic)}[\sX]$. 
In other words, instead of~\eqref{EGP0}, we have 
\begin{equation}\label{deComG}
Z_{\rm gravity}[M]
\underset{N\to\infty}{\simeq}
Z_{\rm GPI}[M]
+
Z_{\rm grav}^{(\rm erratic)}[M] \, .
\end{equation}

\item {\bf The role of averages} 

The filtering proposal does not require introducing an ensemble of theories or performing averages over states, observables, or $N$. 
It is, in principle, defined intrinsically for each individual CFT.\footnote{As mentioned earlier, equation~\eqref{deCom} should be understood as an approximation for a CFT with a given fixed, but large $N$.}  It thus avoids the conceptual and technical difficulties associated with averages that were enumerated after~\eqref{nonfa}.

Nevertheless, the proposal is compatible with the heuristic expectation that some degree of coarse-graining is involved in relating a large-$N$ CFT to semiclassical gravitational saddle-point computations. In particular, it is consistent with the extensive literature on CFT averages and wormholes (see references in footnote~\ref{ft:ave}). Rather than explicitly implementing the decomposition~\eqref{deCom}, it may be technically convenient in specific problems to use suitable averaging procedures to isolate $Z_{\rm CFT}^{\rm (smooth)}[\sX]$, since $Z_{\rm CFT}^{\rm (erratic)}[\sX]$ may cancel or be suppressed under such averages. Moreover, there may exist multiple distinct averaging schemes that extract the same smooth component in a given context.

This is analogous to the use of ensemble averages in statistical mechanics: an ensemble can serve as a mathematical device for extracting macroscopic observables that characterize individual systems.

\ei




\subsection{Filtered Lorentzian emergence: baby universes and black hole interiors} \label{sec:Loren}

We now consider extending~\eqref{mlarN} to Lorentzian signature.  

Let us first review the standard story, i.e., the Lorentzian counterpart of~\eqref{larN}. In this setting, we associate a Lorentzian gravitational solution $\sM$---together with a corresponding ``vacuum'' state $\ket{0}_{\sM}$ for matter fields on $\sM$---with the 
``large-$N$ limit'' of a boundary theory (semiclassical) state $\ket{\Psi}$, or more precisely, a sequence of states ${\ket{\Psi^{(N)}}}$, one for each value of $N$. In the Lorentzian context, the set of observables is much richer, and we can schematically express the analogue of~\eqref{larN} as
\be \label{larN1}
\lim_{N \to \infty}  \text{observables in} \, \ket{\Psi^{(N)} }= \text{semiclassical gravity quantities} ,
\ee
where the right-hand side should be understood broadly as quantities extracted from the bulk spacetime through suitable procedures, and once again admits an asymptotic expansion in $G_N$, i.e., displays a ``smooth'' $G_N$-dependence.

We may also pose the reverse question: given a sequence of states $\ket{\Psi^{(N)}}$, how can we determine whether it describes a semiclassical gravitational configuration in the large-$N$ limit---namely, whether bulk geometry emerges? This question is not yet fully understood. A useful working hypothesis is the following (see Sec.~VI~A of~\cite{Liu25b} for a more detailed discussion).
Consider correlation functions of single-trace operators,
\be \label{singQ}
\vev{\Psi^{(N)}| \sO_1 \sO_2 \cdots \sO_n |\Psi^{(N)}},
\ee
where each $\sO_i$ denotes a single-trace operator, with its one-point function in $\ket{\Psi^{(N)}}$ subtracted.
We say that $\ket{\Psi^{(N)}}$ admits a semiclassical gravity description if, in the large-$N$ limit, the correlator~\eqref{singQ} factorizes into products of two-point functions (i.e., exhibits Gaussian factorization), and if these two-point functions admit a well-defined large-$N$ limit---for example, converge pointwise.

In parallel with the earlier Euclidean discussion, equation~\eqref{larN1} cannot hold if the CFT observables have erratic dependence on $N$.  For a Lorentzian observable $G (N) $ (for example a two-point function of single-trace operators) in a state $\ket{\Psi^{(N)}}$, we again make the split 
\be \label{deiL}
G (N) 
\underset{N\to\infty}{\simeq}
G^{\rm smooth} + G^{\rm erratic} ,
\ee
where $G^{\rm smooth}$ denotes the part with a ``smooth'' $N$-dependence. The filter $\Fil{\cdot}$ is similarly defined as the projection to the smooth part 
\be 
\Fil{G} \equiv  G^{\rm smooth},
\ee
and we propose to replace~\eqref{larN1}  by 
\be\label{mlarN1} 
\Fil{G} =
\text{semiclassical gravity result}   \ .
\ee

Equation~\eqref{mlarN1} defines a weaker notion of the large-$N$ limit than~\eqref{larN1}. It leads to a generalized framework---``filtered Lorentzian emergence''---within which more general Lorentzian spacetime structures may arise:

\emph{We say that a sequence of CFT states $\ket{\Psi^{(N)}}$ admits a semiclassical gravity description if the filtered large-$N$ limit of their single-trace correlation functions exists. In particular, these correlation functions need not converge pointwise in the large-$N$ limit.}

Similarly, we say that a boundary operator admits a semiclassical description if its correlation functions, evaluated in such a (generalized) semiclassical state, possess a filtered large-$N$ limit. The resulting operator algebra $\sA_{\Psi}$ is, in general, larger than the algebra generated purely by single-trace operators.

As illustrative examples of ``filtered Lorentzian emergence,'' and building on the earlier discussions of~\cite{Liu25,KudWit25}, we will argue that {\it erratic large-$N$ behavior in the boundary CFT is responsible for the emergence of closed universes and black hole interiors.}

\begin{figure}
\begin{center}
\includegraphics[width=14cm]{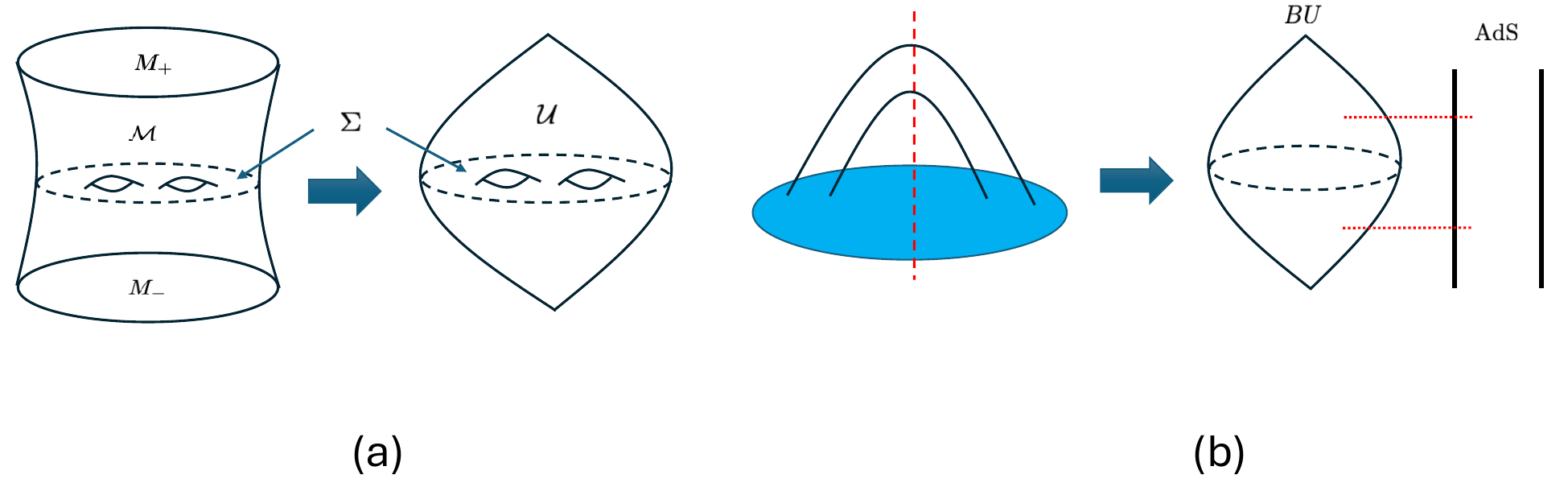}
\caption{\small (a) Schematic illustration of the analytic continuation from a reflection-symmetric external wormhole $\sM$, with boundaries $M_+ = M_- = M$, to a closed universe $\sU$. The reflection-symmetric slice $\Sigma$ of $\sM$ continues to the time-reflection-symmetric slice of $\sU$.
(b) Schematic illustration of the analytic continuation from a reflection-symmetric internal wormhole (with respect to the black dashed line), producing a baby closed universe (BU) that is entangled (shown as red dotted lines) with an asymptotic AdS spacetime. The reflection-symmetric cross section of the handle continues to the time-reflection-symmetric slice of the baby universe.
}
\label{fig:anaE}
\end{center}
\end{figure}

A particularly concrete setting in which this emergence can be seen is the boundary description of closed universes obtained via Lorentzian analytic continuation of reflection-symmetric Euclidean wormholes.
There are two ways to obtain a closed universe: 

\ben 

\item {\it Isolated closed universes from Lorentzian analytic continuation of an external Euclidean wormhole}

Analytically continuing an external wormhole produces a Lorentzian geometry with no boundary, corresponding to either a single closed universe or multiple closed universes entangled with one another~\cite{MalMao04}; see Fig.~\ref{fig:anaE}(a) for an illustration.
We refer to such closed universes as ``isolated.''

From the analytic continuation, one can show that the semiclassical Hilbert space of the resulting closed universes can be constructed from correlations of the erratic components of Euclidean partition functions. For example, in the situation depicted in Fig.~\ref{fig:anaE}(a), one may define an (overcomplete) basis of states $\ket{X}$ in the Hilbert space of the closed universe $\sU$ by inserting light operators $X$ on the lower boundary $M_-$ of the Euclidean wormhole geometry. The inner products between these states are given by
\be \label{ad0}
\vev{X'|X} =
\Fil{(\de_e Z_{\rm CFT}[M;X'])^{*} \,\de_e Z_{\rm CFT}[M;X]} ,
\ee
where $Z_{\rm CFT}[M;X]$ denotes the Euclidean partition function on $M$ with a configuration $X$ of light operator insertions.
Equation~\eqref{ad0} shows that the semiclassical Hilbert space of the closed universe---including its full operator algebra---arises from the erratic $N$-dependence of the boundary partition function. 

Furthermore, it implies that the Hilbert space is {\it not} one-dimensional. 
{The one-dimensional-Hilbert-space conclusion of~\cite{UsaWang24,UsaZha24,HarUsa25,AbdAnt25} was reached by allowing nontrivial averages of the form $\overline{\vev{X|X'} \vev{X'|X}} \neq \overline{\vev{X|X'}} \; \overline{\vev{X'|X}}$.}
In contrast, in~\eqref{ad0}, the quantity $\vev{X'|X}$ is already a filtered object, and the product $\vev{X'|X}\vev{X|X'}$ is simply the usual product of complex numbers---there is no further nontrivial averaging to be performed.\footnote{The difference with replica wormholes for an evaporating black hole is discussed at the end of Sec.~\ref{sec:CUext}.}

The Hilbert space $\sH_{\omega_{\FF}}^{(\rm GNS)}$ discussed in Sec.~\ref{sec:emG} can in fact be interpreted as a semiclassical Hilbert space of {\it all} isolated closed universes, with inner products among different states capturing transition and branching amplitudes between different universes.

In the absence of boundaries, isolated closed universes do not appear to admit a description as the large-$N$ limit of any sequence of finite-$N$ boundary CFT states.



\item {\it Baby closed universes entangled with AdS}

Analytically continuing an internal wormhole leads to a Lorentzian geometry with boundary, resulting in closed universe(s) entangled with asymptotic AdS spacetime(s), as illustrated in Fig.~\ref{fig:anaE}(b).

An explicit example is the Antonini-Sasieta-Swingle (AS$^2$) cosmology~\cite{AntSas23}, which is the candidate gravity description of the partially entangled thermal state (PETS)~\cite{GoeLam18} below the Hawking-Page temperature. In this setup, a baby closed universe is entangled with two copies of asymptotic AdS spacetime (see Fig.~\ref{fig:baby}(a)). Using this example as an illustration, and expanding on the discussion in~\cite{Liu25,KudWit25}, we will argue that the baby universe---including its semiclassical Hilbert space and the operator algebra in the closed universe---owes its very existence to the erratic $N$-dependence of the corresponding state. 
Furthermore, the full spacetime, including both the baby universe and the AdS regions, is quantum-volatile~\cite{EngLiu23}: the geometry exhibits $O(G_N^0)$ fluctuations controlled by the erratic $N$-dependent sector. 


Such $O(G_N^0)$ fluctuations may be interpreted as an intrinsic limitation on the ability of an observer in the AdS region to make arbitrarily precise measurements, even within the regime of quantum field theory in curved spacetime.
If an observer in the AdS region were able to detect this limitation, they might, in principle, infer the existence of the baby universe!

\een

Using as an illustration the PETS~\cite{GoeLam18} above the Hawking-Page temperature, which describes a long two-sided black hole, we will also argue that the interior of a generic black hole likewise emerges from erratic $N$-dependence, in the sense that the very definition of local bulk operators in the interior involves quantities with intrinsically erratic $N$-dependence. 
Furthermore, the interior geometry can be quantum volatile.

\bigskip

\noindent{\bf Plan of the paper} 

\medskip

The remainder of the paper explores the implications of the proposal~\eqref{mlarN} and its Lorentzian counterpart~\eqref{mlarN1}, and is organized as follows.


In Sec.~\ref{sec:AP}, we use~\eqref{mlarN} and~\eqref{imRe}--\eqref{0nide} to deduce general properties of the erratic components of boundary partition functions. Our discussion uses the theorems of~\cite{WitYau99,CaiGal00} and draws heavily on the elegant analyses of~\cite{Sas22,ChaCol22}.  We further show that known wormhole solutions are consistent with the hypothesis that partition functions involving OPE coefficients of heavy operators exhibit erratic $N$-dependence, and we use the bulk wormhole action to infer the implied correlations among these OPE coefficients. In Sec.~\ref{sec:internal}, we examine the internal wormhole solutions of~\cite{ChaCol22}  and argue that they do not induce random couplings in the bulk low-energy theory.

In Sec.~\ref{sec:PhysI}, we discuss several implications of the proposal~\eqref{mlarN}. In Sec.~\ref{sec:deriv}, we present a general formal analysis of the new contributions to holographic R\'enyi and entanglement entropies that arise from the filtering procedure, showing in particular that the replica wormholes of~\cite{PenShe19,AlmHar19}---which compute the entropies of radiation subsystems in models of evaporating black holes---are naturally incorporated. Interpreting these contributions as encoding correlations among the erratic components of boundary partition functions offers a new boundary perspective on the physical origin of the turnaround in the Page curve. In Sec.~\ref{sec:glob}, we consider the boundary interpretation of global symmetry violation in quantum gravity mediated by wormholes.

In Sec.~\ref{sec:MM}, we study closed universes obtained from analytic continuation of external wormholes and their connection to erratic $N$-dependence. We explain the emergence of the closed-universe Hilbert space from the boundary and interpret $\sH_{\omega_{\FF}}^{(\rm GNS)}$, introduced in~Sec.~\ref{sec:emG}, as a ``semiclassical Hilbert space of all closed universes'' described by the boundary CFT.

In Sec.~\ref{sec:ASS}, we analyze closed baby universes arising from the analytic continuation of internal wormholes. Using the AS$^2$ cosmology~\cite{AntSas23} as an illustrative example, we argue that both the baby universe and the AdS spacetimes entangled with it are quantum volatile.

In Sec.~\ref{sec:BHI}, we argue that the interior of a generic black hole likewise emerges from erratic $N$-dependence and can be quantum volatile.

Finally, in Sec.~\ref{sec:conc}, we conclude with a brief summary and outlook. We speculate on possible interpretations of singularities in closed universes and black hole interiors, and discuss the nature of the semiclassical approximation on the gravity side.

In Appendix~\ref{sec:filter}, we give some illustrative examples  of 
the split~\eqref{deCom}.



\bigskip

\noindent 
{\bf Notations and conventions} 

\medskip

\noindent $\bullet\,$ $\sB (\sH)$ denotes the set of bounded operators on a Hilbert space $\sH$.

\noindent $\bullet\,$ Throughout the paper we consider a gravitational system in AdS$_{d+1}$ described by a $d$-dimensional boundary CFT$_d$. We assume that the CFT has a parameter $N$, related to the bulk Newton constant by $G_N \propto 1/N^2$.

\noindent $\bullet\,$ $\de_e Z_{\rm CFT}[M]$ denotes the erratic part of $Z_{\rm CFT}[M]$, and $\de_e Z_{\rm CFT}[M,X]$ denotes the corresponding quantity with additional light operator insertions $X$.

\noindent $\bullet\,$ $Z_{\mathrm{gravity}}[M]$ denotes the gravitational partition function with boundary specified by $M$.

\noindent $\bullet\,$ $\sX = (M, X)$ denotes a compact manifold $M$ with light operator insertions specified by $X$.

\noindent $\bullet\,$  $Z_{\mathrm{gravity}}^{(\sM)}$ denotes the contribution to the gravitational partition function from the saddle $\sM$, including the classical action of $\sM$ and perturbative corrections from matter fluctuations around it. 

\noindent $\bullet\,$ $\p \sM$ denotes the boundary of a manifold $\sM$. 

\noindent $\bullet\,$  $\sI_E [M_1, M_2, \cdots, M_n]$ denotes the Euclidean action of the {\it leading} wormhole saddle connecting $M_1, M_2, \cdots, M_n$.

\noindent $\bullet\,$ General energy eigenstates of the CFT are written as $|n\rangle$ or $|m\rangle$. Heavy states with energies $O(N^2)$ are denoted by $|i\rangle, |j\rangle$, and light states of energy $O(N^0)$ by $|a\rangle, |b\rangle$.

\section{Erratic $N$-dependence from wormholes} \label{sec:AP}

In this section, we discuss examples where wormholes are known to exist, and interpret them in terms of the presence or absence of erratic-$N$ behavior in the corresponding partition functions.

\subsection{Thermal partition function and the spectral form factor} \label{sec:vac}

Consider the boundary CFT$_d$ defined on a $d$-dimensional compact Euclidean manifold $M$, with partition function $Z_{\rm CFT}[M]$. According to the proposal~\eqref{mlarN} and~\eqref{imRe}--\eqref{0nide}, if a Euclidean manifold $M$ appears as a boundary component in a wormhole configuration, then its partition function must necessarily contain a non-vanishing erratic component,
\be
\label{equnon0}
\de_e Z_{\rm CFT}[M] \neq 0 \ .
\ee

The wormholes considered in the literature generally fall into two categories:

\ben
\item Genuine saddles of the gravitational path integral, which we will refer to as on-shell wormholes.

\item ``Constrained'' wormholes~\cite{CotJen20b,CotJen21,CotJen22}, which are saddles up to a finite number of parameters that must be integrated over. Since the gravitational path integral is not rigorously defined in higher-dimensional gravity, constrained saddles are perhaps the most one could hope for.\footnote{Note that wormholes in JT gravity~\cite{SaaShe19} can also be interpreted as being constrained.}
\een

Whether a genuine on-shell saddle exists depends on the specific theory and matter configurations. In~\cite{WitYau99,CaiGal00}, powerful general theorems restrict their existence for manifolds of nonnegative curvature.\footnote{We say that  $M$ has positive (nonnegative, zero) scalar curvature if the conformal class of metrics on $M$ contains a representative metric of positive~(nonnegative, zero) scalar curvature.} Suppose $\sM$ is a $(d+1)$-dimensional complete Riemannian manifold with metric $g_{AB}$ and $d$-dimensional conformal boundary $N = \p \sM$, and suppose further that the Ricci tensor $\sR_{AB}$ of $\sM$ satisfies two conditions:

(a) $\sR_{AB} n^A n^B \geq - d$  for any unit vector $n^A$;

(b)  $\sR_{AB} \to - d \, g_{AB}$ sufficiently fast as one approaches infinity.

\noindent If $N$ has a component $M$ of {\it nonnegative scalar curvature}, then it has been proven that~\cite{WitYau99,CaiGal00}:\footnote{A third conclusion states that any loop in $\sM$ can be deformed to a loop in $M$, which is not directly relevant for us.}

\bi
\item  $M$ must constitute the full boundary, i.e. $N = M$. Thus there is no external wormhole solution connecting $M$ to any other disjoint boundary component.

\item  There is no internal wormhole, i.e., no nontrivial topological connection between different regions of $\sM$ with a single boundary $M$.
\ei

For a bulk gravity theory, condition (a) is satisfied by solutions of the equations of motion with ``non-exotic'' matter configurations, while condition (b) requires that the matter fields fall off sufficiently fast near the boundary (i.e., that no sources are turned on in the boundary theory). Denote by $\Mp$ the set of boundary manifolds with nonnegative scalar curvature. The above statements then imply that there is no on-shell wormhole associated with $M \in \Mp$~(in the absence of  boundary sources that can back react to the geometry at the leading order).

Despite this restriction on on-shell saddles, off-shell wormholes appear to exist. In~\cite{CotJen20,CotJen20b,CotJen21,CotJen22}, constrained wormholes were constructed which connect two copies of $M = S_1 \times \TT_{d-1}$ or two copies of $M = S_1 \times S_3$ with a different $S_1$ size for each copy. Here $\TT_{d}$ denotes a $d$-dimensional torus. An important difference between an on-shell wormhole and the two-boundary off-shell wormholes found in~\cite{CotJen20,CotJen20b,CotJen21,CotJen22} is their scaling with $G_N$, 
\begin{gather}
Z_{\rm gravity}^{\rm (wormhole, on-shell)} [M_1, M_2] \sim e^{O(G_N^{-1})}, \\
Z_{\rm gravity}^{\rm (wormhole, off-shell)} [M_1, M_2] \sim O(G_N^0) \ .
\end{gather} 

For boundary gauge theories such as $\sN = 4$ SYM on $M = S_1 \times S_{d-1}$ (where $S_1$ has circumference $\b$, the inverse temperature), the thermal partition function takes the form $Z_\b = \sum_m e^{-\b E_m}$, summing over all energy eigenstates. Above the Hawking-Page temperature, this sum is dominated by black hole microstates with $E_m \sim O(N^2)$, whose density of states is expected to be erratic in $N$. 
The existence of off-shell wormholes connecting two copies of $S_1 \times S_{d-1}$ implies that 
\be \label{offw}
\Fil{\de_e Z_{\b_1} \de_e Z_{\b_2}} = Z_{\rm gravity}^{\rm (wormhole, off-shell)} [\b_1, \b_2] \sim O(G_N^0) \neq 0  ,
\ee
which can be understood as a consequence of the erratic density of states in the black hole sector. Notably, the magnitude of this ``erratic contribution'' is of order $O(N^0)$.  
 
\begin{figure}[h]
        \centering
		\includegraphics[width=8cm]{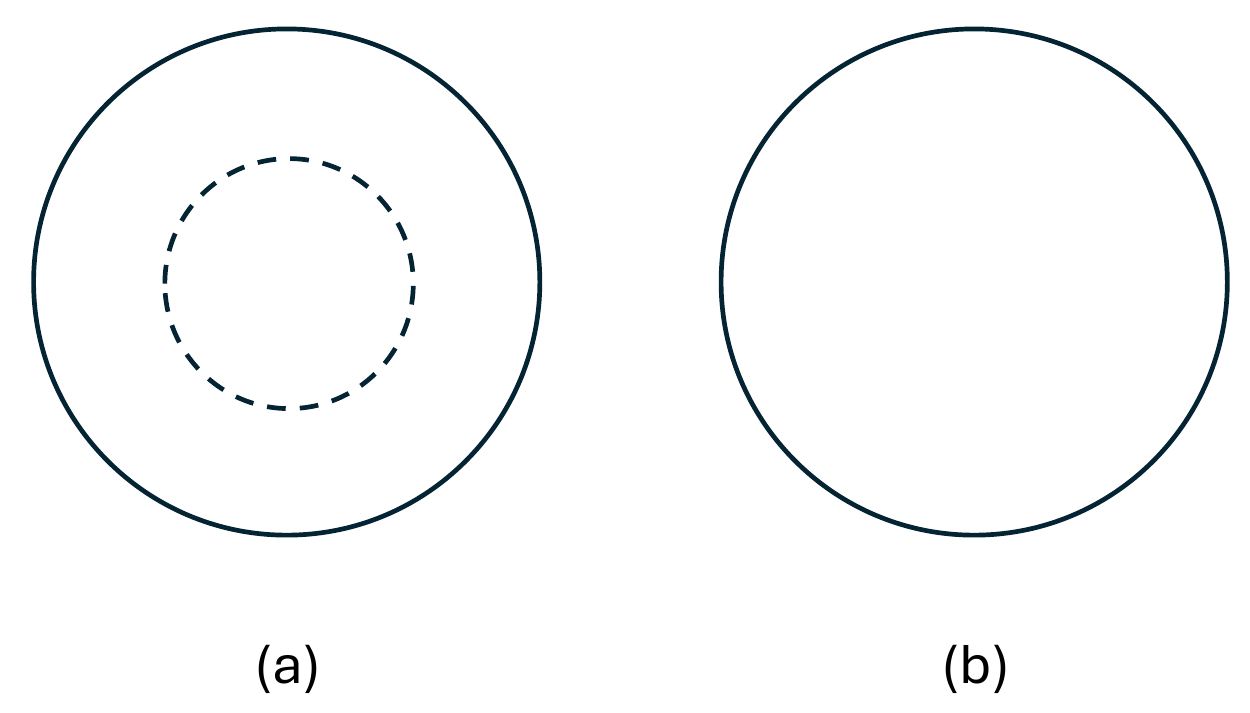}
                \caption[  ]
        {\small  Bulk geometries describing the thermal partition function $Z_\b$. Each point in the plots represents an $S_{d-1}$. The angular direction corresponds to the Euclidean time, with the outer circle representing the boundary of topology $S_1 \times S_{d-1}$.
(a) The Euclidean thermal AdS saddle that dominates the computation of $Z_\b$ below the Hawking-Page temperature. The inner 
circle~(dashed line) indicates the center where $S_{d-1}$ shrinks to zero size and the space terminates smoothly.
(b) The Euclidean black hole geometry that dominates the computation of $Z_\b$ above the Hawking-Page temperature, where $S_1$ shrinks to zero size smoothly at the center of the disk, while $S_{d-1}$ remains of finite size everywhere.
      } 
\label{fig:thermal}
\end{figure}

For later comparison, we recall in Fig.~\ref{fig:thermal} the bulk geometries describing $Z_\b$ below and above the Hawking-Page temperature.

To probe the erratic nature of these highly excited energy levels more sensitively, we can introduce a Lorentzian time evolution by considering the complexified partition function:
\be \label{complZ}
Z_{\b + i t} = \sum_m e^{- (\b + i t) E_m}, \quad \b \in \RR_+, \; t \in \RR \ .
\ee
Here, the rapidly oscillating Lorentzian phases $e^{-i E_m t}$ provide a sharper probe of the spectrum. Products of~\eqref{complZ} defines  the spectral form factor, 
\be\label{sFF}
K (\b, t) \equiv \overline{Z_{\beta + it} Z_{\beta - it}} \ , 
\ee
where the overline denotes an average over either ensembles or a Lorentzian time window. In chaotic systems, this quantity famously exhibits the characteristic dip-ramp-plateau structure.

Here we are interested in possible erratic $N$-dependence of~\eqref{complZ} without performing any average in $t$ or any ensemble average. It has been proposed in~\cite{SaaShe18} that in a holographic boundary system, the ramp behavior of~\eqref{sFF} (for $\b =0$) can be described on the gravity side by 
the double-cone geometry---obtained by periodically identifying the Lorentzian times of an eternal black hole~(see also~\cite{CotJen22,CheIvo23}). From~\eqref{imRe}, we can make the identification,
\be\label{fspe}
\Fil{\de_e Z_{it} \de_e Z_{- it}} = Z_{\text{gravity}}^{(\text{double cone})}   ,
\ee
for which neither ensemble nor explicit time average is needed.\footnote{Note that in the double-cone geometry, there is no explicit time average.}
The double-cone geometry can be obtained from the off-shell wormholes~\cite{CotJen20,CotJen20b,CotJen21,CotJen22} entering~\eqref{offw} by analytic continuation of $\b_1, \b_2$ to $it$ and $-it$ respectively, and thus~\eqref{fspe} is compatible with~\eqref{offw}. 
Since the double-cone geometry accounts only for the ramp behavior, we expect~\eqref{fspe} to hold for time scales $t$ that are not too large, namely before the onset of the plateau.

For the rest of the section, we will explore various on-shell wormholes.\footnote{Equations~\eqref{offw} and~\eqref{fspe} have since been explored in detail in~\cite{Liu26}.}

\subsection{Euclidean wormholes and erratic $N$-dependence}\label{sec:EuW}

On-shell wormhole solutions connecting disjoint boundaries can arise if one considers boundary manifolds with negative curvature or introduces sources that backreact on the bulk geometry, thereby evading the conditions of~\cite{WitYau99,CaiGal00}. Examples of such constructions include~\cite{MalMao04,BetKir19,MarSan21,Sas22,ChaCol22,IliKol21,AstGau23}.

A wormhole solution can exhibit perturbative instabilities associated with negative modes, or nonperturbative instabilities due to bulk brane nucleation~\cite{SeiWit99}. The nonperturbative instabilities reflect that the corresponding boundary theory is not well-defined. Therefore, if we consider well-defined boundary theory partition functions, they should be automatically absent.
The interpretation of perturbative instabilities is more delicate. Should we discard such wormhole solutions? From the perspective of~\eqref{imRe}, the wormhole configuration reflects correlations among the erratic parts of partition functions, so the unstable modes may simply reflect features of these correlations rather than signal a fundamental inconsistency. As already mentioned in the Introduction, it is also not important whether the wormhole configuration is dominant or subdominant compared to disconnected contributions; they merely represent different types of contributions.

Below, we analyze various wormhole solutions of~\cite{Sas22} and argue that they can be attributed to the erratic behavior of OPE coefficients involving heavy operators. For $d=2$, more universal structure in the erratic behavior of these OPE coefficients can be extracted, and this will be discussed separately in the next subsection.



\begin{figure}[h]
        \centering
		\includegraphics[width=8cm]{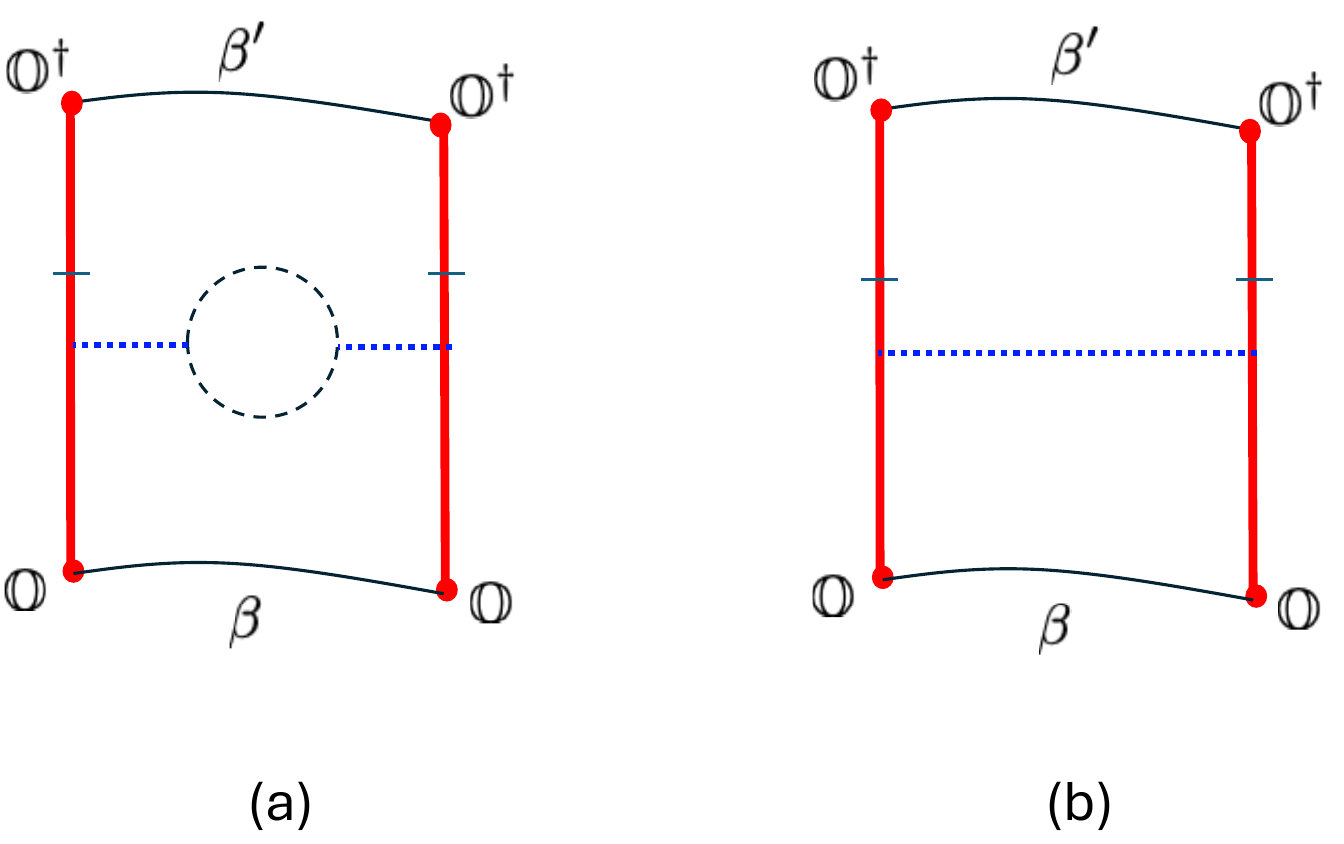}
                \caption[  ]
        {\small Cartoons of the external wormhole configurations describing $\Fil{\vev{\OO}_\b \vev{\OO^\da}_{\b'}}_c$ below and above the Hawking-Page temperature. The conventions for the plots are the same as in Fig.~\ref{fig:thermal}. The geometries connect two disjoint boundaries, each of topology $S_1 \times S_{d-1}$ (upper and lower boundaries).
        Each boundary $S_1$ is shown as a line with identified endpoints, which we take to be the locations of the $\OO$ or $\OO^\da$ insertions. The thick red lines, which are identified, denote the mass shell resulted from the insertion of $\OO$. When $\b = \b'$ (and the insertion points of $\OO$ and $\OO^\da$ coincide), the geometries are reflection-symmetric about the central horizontal dotted line.
(a) The geometry below the Hawking-Page temperature. The $S_{d-1}$ smoothly shrinks to zero size at the dashed circle. 
(b) The geometry above the Hawking-Page temperature. In this case, there is nowhere in the geometry where $S_{d-1}$ shrinks to zero size, and all horizontal slices have topology $S_1 \times S_{d-1}$.
            } 
\label{fig:wormM}
\end{figure}

\begin{figure}[h]
        \centering
		\includegraphics[width=8cm]{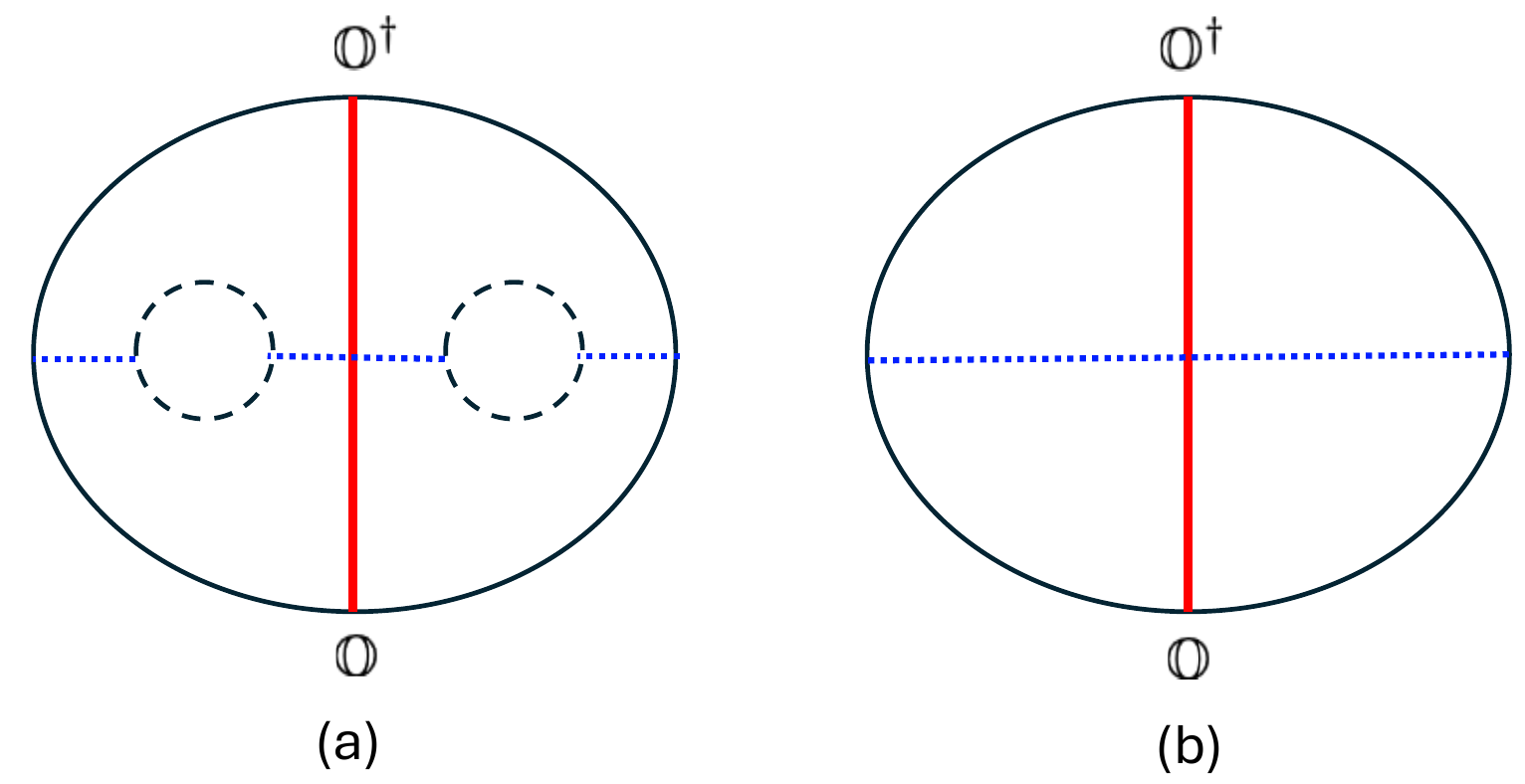}
                \caption[  ]
        {\small  Cartoons of the gravity configurations describing $\Fil{G_\OO(\tau; \b)}_c$ below and above the Hawking-Page temperature proposed by~\cite{Sas22}.
      The conventions for the plots are the same as in Fig.~\ref{fig:thermal}.
      The geometries are reflection-symmetric about the central horizontal dotted line.
(a) The geometry below the Hawking-Page temperature, which can be viewed as adding a ``handle,'' i.e., an internal wormhole, to the thermal AdS geometry of Fig.~\ref{fig:thermal}(a).
(b) The geometry above the Hawking-Page temperature, which does not contain wormholes of either type.} 
\label{fig:PETSL}
\end{figure}

\begin{figure}[h]
        \centering
		\includegraphics[width=8cm]{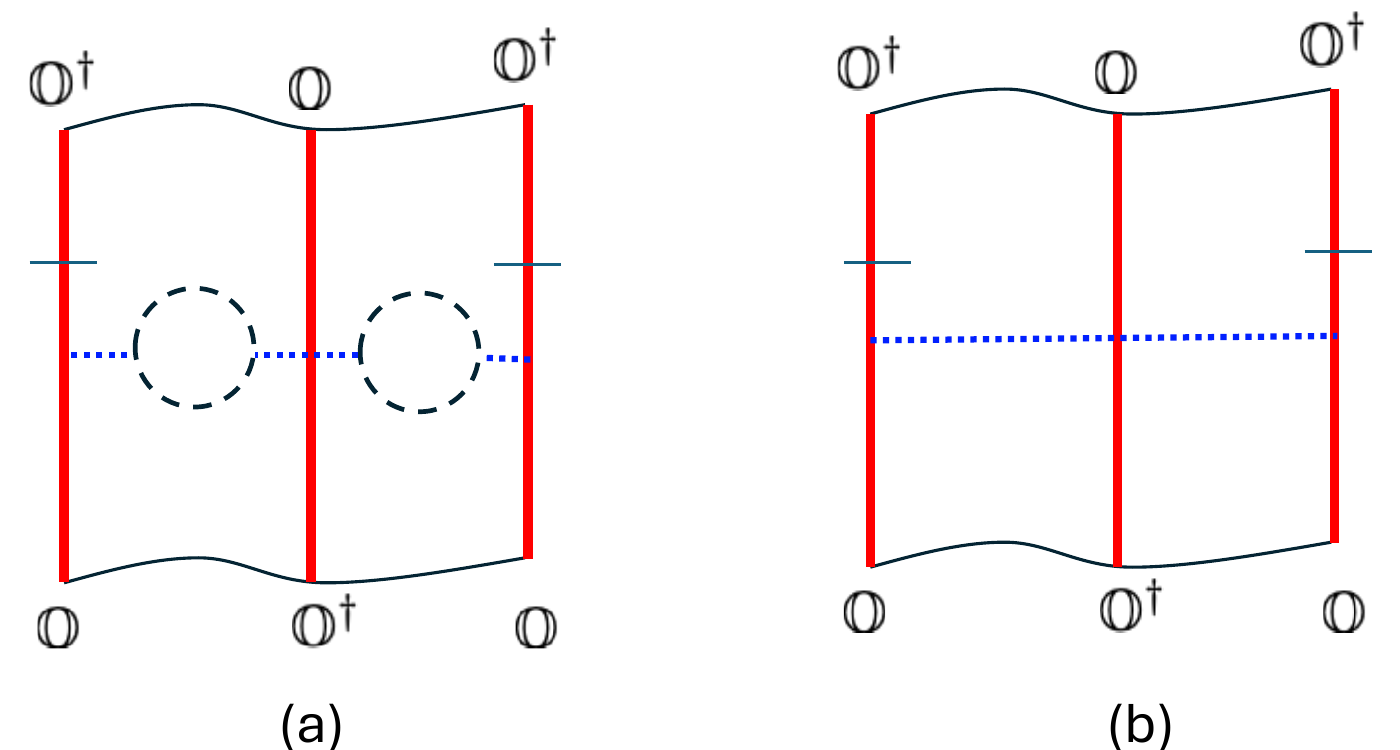}
                \caption[  ]
        {\small  Cartoons of the external wormhole configurations describing $\Fil{\de_e G_\OO (\tau; \b)  \de_e G_\OO (\tau'; \b') }$ below and above the Hawking-Page temperature. The conventions for the plots are the same as in Fig.~\ref{fig:thermal} and Fig.~\ref{fig:wormM}. 
        The geometries connect two disjoint boundaries, each of topology $S_1 \times S_{d-1}$ (upper and lower boundaries) and with insertions of $\OO$ and $\OO^\da$.  The left and right  thick red lines are identified. 
        When $\tau =\tau'$ and $\b = \b'$, the geometries are reflection-symmetric about the central horizontal dotted line.
        (a) The geometry below the Hawking-Page temperature. (b) The geometry above the Hawking-Page temperature.} 
\label{fig:PETSH}
\end{figure}

As a simple example, consider correlation functions of a generic heavy operator $\OO$ of dimension $O(N^2)$ in the thermal ensemble---for example, one- and two-point functions\footnote{We always treat the argument $t$ of an operator $\OO (t)$ as Lorentzian time.},
\begin{gather} \label{Zein}
\vev{\OO}_\b = \frac{1}{Z_\b} \Tr \left(e^{-\b H} \OO \right) = \frac{1}{Z_\b} \sum_n e^{-\b E_n} \OO_{nn} , \\
\label{Ecor1}
G_\OO (\tau; \b) = \langle \OO^\da (-i \tau) \OO(0) \rangle_\b = \frac{1}{Z_\b} \sum_{n,m} e^{-(\b -\tau) E_n - \tau E_m } |\OO_{mn}|^2 \ .
\end{gather}
For operators $\OO$ whose gravitational duals can be modeled by a mass shell uniform along the boundary spatial directions, 
wormhole configurations describing ``fluctuations'' of $\vev{\OO}_\b$ and $G_\OO(\tau; \b)$ were previously found in~\cite{Sas22}. 
See Fig.~\ref{fig:wormM}--\ref{fig:PETSH}.
From~\eqref{imRe}, we interpret these wormholes as reflecting the presence of erratic-$N$ dependence in these observables.

The erratic behavior in~\eqref{Zein}--\eqref{Ecor1} arises from two possible sources: the density of states (as discussed in Sec.~\ref{sec:vac}) and the matrix elements $\OO_{mn}$\footnote{$\OO_{mn}$ may be heuristically understood as analogous to an OPE coefficient between $\OO$ and the operators corresponding to the states $\ket{m}$ and $\ket{n}$, even though $\OO$ is in general not a local primary operator. With an abuse of terminology, we will still often refer to such matrix elements as OPE coefficients.}. Given their distinct physical origins, it is reasonable to expect that the ``correlations'' between the density of states and the matrix elements under the filter operation $\Fil{\cdot}$ are negligible. Furthermore, as shown in Sec.~\ref{sec:vac}, the erratic part of the density of states yields off-shell wormholes with an amplitude of $O(G_N^0)$. Since we are dealing here with on-shell wormholes characterized by an exponential dependence of $e^{O(1/G_N)}$, we expect their leading behavior to stem entirely from the erratic nature of $\OO_{mn}$ alone. 
Consequently, in what follows we set the density of states to its smooth component and focus solely on the erratic behavior of $\OO_{mn}$.


The amplitudes of wormholes in Fig.~\ref{fig:wormM}--\ref{fig:PETSH} can then be respectively identified with 
\bega 
\label{t001}
\Fil{\vev{\OO}_\b \vev{\OO^\da}_{\b'}}_c= \frac{1}{Z_\b Z_{\b'}} \sum_{n,m} e^{-\b E_n- \b' E_m} \Fil{\OO_{nn} \OO^*_{mm}}_c, \\
\label{t000}
\Fil{G_\OO (\tau; \b) }_c 
= \frac{1}{Z_\b} \sum_{n,m} e^{-(\b -\tau) E_n - \tau E_m } \Fil{|\OO_{mn}|^2 }_c, \\
\Fil{\de_e G_\OO (\tau; \b) \de_e G_\OO (\tau'; \b')} = 
 \frac{1}{Z_\b Z_{\b'}} \sum_{n,m,p,q} e^{-(\b -\tau) E_n - \tau E_m -(\b' -\tau') E_q - \tau' E_p }  \Fil{|\OO_{mn} |^2 | \OO_{pq} |^2 }_c\ .
 \label{t002}
\end{gather} 
In~\eqref{t002}, $\Fil{|\OO_{mn} |^2 | \OO_{pq} |^2 }_c$ denotes the connected part of the correlation between $|\OO_{mn} |^2$ and $| \OO_{pq} |^2$,  which encodes erratic $N$-dependence of the magnitude of $\OO_{mn}$. On general grounds we expect $\Fil{\OO_{mn}} =0$ for $m \neq n$ (as in the Eigenstate Thermalization Hypothesis~\cite{Sre94}).  

In~\cite{Sas22}, it was suggested that a nontrivial gravity solution with a single insertion of $\OO$ in the thermal ensemble does not exist.
This would imply
\be
\label{0ooco}
\Fil{\vev{\OO}_\b} = 0  \quad \to \quad  \Fil{\OO_{mm}} = 0 \ .
\ee
Here, the zero on the right-hand side should not be taken strictly, but rather understood to mean a value too small to source a nontrivial gravity solution. This would also imply that on the left hand side of~\eqref{t000}, we can replace $\Fil{G_\OO (\tau; \b) }_c$ by $\Fil{G_\OO (\tau; \b) }$ as  $\Fil{|\OO_{mn}|^2 } = \Fil{|\OO_{mn}|^2 }_c$. 

\begin{figure}[h]
        \centering
		\includegraphics[width=8cm]{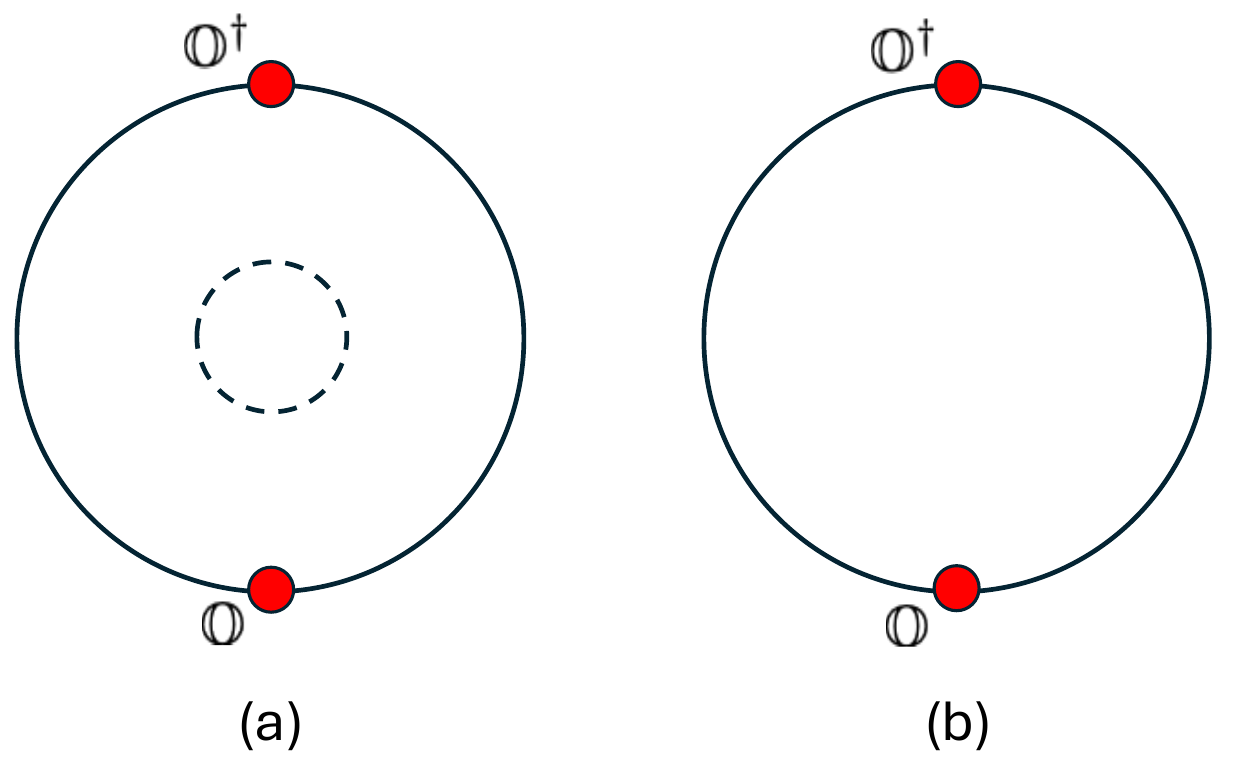}
                \caption[  ]
        {\small Schematic illustrations of the additional gravity solutions contributing to $\Fil{G_\OO(\tau; \b)}$ from the disconnected term in~\eqref{hdm}. The configurations are shown for temperatures (a) below and (b) above the Hawking-Page temperature~\cite{VanLop26}. In such solutions, the operator insertions $\OO$ and $\OO^\da$ undergo ``self-contraction'' (denoted by solid red dots) rather than mutually contracting.
                 } 
\label{fig:newsol}
\end{figure}

A more recent paper~\cite{VanLop26} proposed that gravity solutions dual to $\Fil{\vev{\OO}_\b}$ could, in fact, generically exist, implying a nonzero $\Fil{\OO_{mm}}$. This does not affect our discussion of wormholes below, since equations~\eqref{t001}--\eqref{t002} involve only connected correlations under the filter. However, it leads to nontrivial consequences for $\Fil{G_\OO (\tau; \b)}$, which can now receive a contribution from the disconnected part of $\Fil{\OO_{mn} \OO_{mn}^*}$, 
\be \label{dis}
\Fil{\OO_{mn} \OO_{mn}^*}_{\rm dis} = \de_{mn} \Fil{\OO_{mm}} \Fil{\OO_{mm}^*} \ .
\ee
That is,
\be \label{hdm}
\Fil{G_\OO (\tau; \b)} = \frac{1}{Z_\b} \sum_m e^{-\b E_m } |\Fil{\OO_{mm}}|^2 + \Fil{G_\OO (\tau; \b)}_c \ .
\ee
Now in addition to those in Fig.~\ref{fig:PETSL}, there can be additional gravity solutions corresponding to the disconnected term in~\eqref{hdm} (i.e., arising from the self-contraction of $\OO$)~\cite{VanLop26}. See Fig.~\ref{fig:newsol}. We will discuss further implications of such solutions in Sec.~\ref{sec:ASS} and~\ref{sec:BHI}.

As with the standard thermal partition function, the gravity configurations describing~\eqref{t001}--\eqref{t002} exhibit a Hawking-Page-like transition~\cite{Sas22}. There exists a critical temperature (which we will also refer to as the Hawking-Page temperature) that separates two distinct regimes. Below this temperature, the sums in~\eqref{t001}--\eqref{t002} are dominated by states $\ket{a}$ with energies of order $O(N^0)$, and thus the wormhole geometries reflect the erratic behavior of $\OO_{ab}$. Above this temperature, the sums are instead dominated by states $\ket{i}$ with energies of order $O(N^2)$, meaning the wormhole geometries reflect the erratic behavior of $\OO_{ij}$.

 The following features on the filtering of powers of the OPE coefficients $\OO_{mn}$ can be deduced from the gravity analysis of~\cite{Sas22}: 

\ben

\item Below the Hawking-Page temperature, the bulk geometries describing~\eqref{t001}--\eqref{t002} are shown respectively 
by Fig.~\ref{fig:wormM}(a), Fig.~\ref{fig:PETSL}(a), and Fig.~\ref{fig:PETSH}(a). 

Note that the gravity configuration depicted in Fig.~\ref{fig:PETSL}(a) for $\Fil{G_\OO(\tau; \b)}$ contains an internal wormhole, in contrast to the external wormhole configurations shown in Fig.~\ref{fig:wormM}(a) and Fig.~\ref{fig:PETSH}(a).
All these wormhole configurations can be regarded as different manifestations of the same correlations, $\Fil{\OO_{ab}\OO_{ab}^*}_c$, as can be seen from the fact that their corresponding Euclidean gravity actions can be reproduced by inserting
\begin{gather}
\label{ooco}
\Fil{\OO_{ab}\OO_{cd}^*}_c = \delta_{ac}\delta_{bd} e^{-N^2 f_\OO^{(0)} + \cdots},
\end{gather}
into~\eqref{t001}--\eqref{t002}~\cite{Sas22}.
The quantity $f_\OO^{(0)}$ is an $O(N^0)$ constant (depending on the dimension of $\OO$), and the ellipsis indicates possible additional $O(N^0)$ terms.

\item Above the Hawking-Page temperature, the Euclidean gravity actions of Fig.~\ref{fig:wormM}(b), Fig.~\ref{fig:PETSL}(b), and Fig.~\ref{fig:PETSH}(b) can be reproduced by inserting into~\eqref{t001}--\eqref{t002}~\cite{Sas22} 
\begin{gather} \label{oij1}
\Fil{\OO_{ij} \OO_{kl}^*}_c = \delta_{ik}  \delta_{jl} e^{-N^2 f_\OO (\epsilon_i, \epsilon_j) + \cdots} , \quad
\epsilon_i = \frac{E_i}{N^2} \ .
\end{gather}
Here $i, j, \cdots$ denotes states with energies of order $O(N^2)$ which dominate the sums above the Hawking-Page temperature,
$f_\OO (\epsilon_i, \epsilon_j)$ is an appropriate $O(N^0)$  function\footnote{Note $f_\OO^{(0)} = f_\OO (0, 0)$.}, and $\cdots$ denotes  subleading corrections.

\item  The wormholes of Fig.~\ref{fig:PETSH} can be interpreted as coming from 
\be \label{ooco1}
\Fil{\OO_{mn} \OO_{mn}^*  \OO_{pq} \OO_{pq}^* }_c 
= \de_{mp} \de_{nq} \le(\Fil{\OO_{mn} \OO_{mn}^*}_c\ri)^2 ,
\ee
i.e., capturing the Gaussian part of the connected correlations.

\item Although wormholes reflect correlations of erratic-$N$ dependence, the converse does not have to be true: 
above the Hawking-Page temperature, while $\Fil{G_\OO (\tau; \b) }$ contains the correlation $\Fil{\OO_{ij} \OO_{ij}^*}_c$,  
its gravity description Fig.~\ref{fig:PETSL}(b) does not involve a wormhole.\footnote{The nonvanishing of $\Fil{\de_e G_\OO(\tau; \b) \de_e G_\OO(\tau; \b)}$ indicates that $G_\OO(\tau; \b)$ possesses an erratic component, implying that the erratic dependence of $\OO_{ij}\OO_{ij}^*$ is not fully self-averaged by the sums over $i$ and $j$.}
$\OO_{ij}$ is an example of an ``OPE coefficient'' $C_{ijk}$ where all indices refer to heavy states with energies of order $O(N^2)$. 
In Sec.~\ref{sec:BHI} we will argue that there are nevertheless important Lorentzian implications.



\een 

The existence of gravity solutions in Fig.~\ref{fig:newsol} arising from self-contractions of $\OO$ and $\OO^\da$ imply ``one-point function'' for $\OO_{mn}$ of the form 
\bega\label{onP1}
\Fil{\OO_{ab}} = \de_{ab} e^{-N^2 g_0} h (E_a) + \cdots,  \\
\Fil{\OO_{ij}} = \de_{ij} e^{- N^2 g (\ep_i) + \cdots} , \quad
\epsilon_i = \frac{E_i}{N^2}  \ .
\label{onP2}
\end{gather} 
In~\eqref{onP1}, $g_0$ is an $O(1)$ constant with no energy-dependence, and in~\eqref{onP2}, $\cdots$ denote terms of order $O(N^0)$. 

Other wormhole solutions discussed in~\cite{MalMao04,BetKir19,MarSan21,IliKol21,AstGau23}, which involve turning on more complicated boundary sources that backreact on the bulk geometry, may be interpreted in a similar manner.
A source that backreacts on the geometry introduces energy scaling as $O(N^2)$ (or $O(1/G_N)$), potentially inducing erratic $N$-dependence among the microstates contributing to the partition function. It is clearly of interest to find calculable examples where the proposal~\eqref{imRe} can be directly tested. On the boundary side, there exist calculations of superconformal and topologically twisted indices for two-dimensional orbifold CFTs and for ABJM theory on a wider class of boundary manifolds~\cite{Pes16,Zaf19} (see also~\cite{BobHon22,BobCho24} for more recent developments), which potentially provide candidates for such explicit comparisons.

\subsection{Erratic $N$-dependence in CFT$_2$ and Liouville CFT} \label{sec:2dEW}




We now specialize to $d = 2$, where the general structure of boundary CFTs is better understood and many more wormhole solutions are known~\cite{MalMao04,ChaCol22}. We reinterpret these results through the lens of erratic $N$-dependence.


As discussed in Sec.~\ref{sec:vac}, there is no on-shell wormhole solution when any of the disjoint boundary components is a sphere or a torus with only light operator insertions. 
Wormhole solutions connecting multiple boundary manifolds ${M_i}$ can exist only when the Euler number of each $M_i$ is negative, 
\be \label{negEu}
\chi = 2 - 2g - \ga < 0 , 
\ee
where $g$ is the genus of a manifold $M$ and $\ga$ comes from heavy operator insertions.

For a CFT$_2$, heavy operators---those whose dimensions scale with the central charge $c$---can be separated into two types: operators with dimensions $\Delta \geq {c \over 12}$, which are above the black hole threshold, and defect operators with $\Delta < {c \over 12}$. Defect operators are particularly convenient, as their insertions can be described on the gravity side by defects with conical deficit angles $\de = 2 \pi \le(1-\sqrt{1 - {12 \De \ov c}} \ri)$. It is convenient to parameterize $\De$ as $\De = 2 h$ with $h = {c \ov 6} \eta (1-\eta)$, in terms of which $\de = 4 \pi \eta$, and since $\de < 2 \pi$, 
\be \label{boeta}
\eta < \ha \ .
\ee
Note that each bulk defect contributes $2\eta$ to $\ga$ in~\eqref{negEu}. 

We now argue that equation~\eqref{negEu} precisely characterizes situations in which OPE coefficients among heavy operators are involved and can, in principle, introduce erratic $N$-dependence in the corresponding partition functions. Here are some explicit examples from~\cite{ChaCol22}:

\ben 
\item{$g=0$ and $n=3$} 

Consider insertions of three defect operators $\OO_{1,2,3}$  on the sphere, with the boundary partition function given by  
\be 
G_{123} (x_i) = \vev{\OO_1 (x_1) \OO_2 (x_2) \OO_3 (x_3)} \propto C_{123} ,
\ee
where $C_{123}$ is the OPE coefficient among the defect operators. In this case, we have 
\be 
\ga = 2 (\eta_1 + \eta_2 + \eta_3), 
\ee
and equation~\eqref{negEu} can be satisfied for $\eta_1 + \eta_2 + \eta_3 > 1$. Note that, given~\eqref{boeta}, equation~\eqref{negEu} cannot be satisfied with only two defect insertions. Indeed, we do not expect erratic dependence in the vacuum two-point functions which are determined by conformal symmetry.

\begin{figure}[h]
        \centering
		\includegraphics[width=12cm]{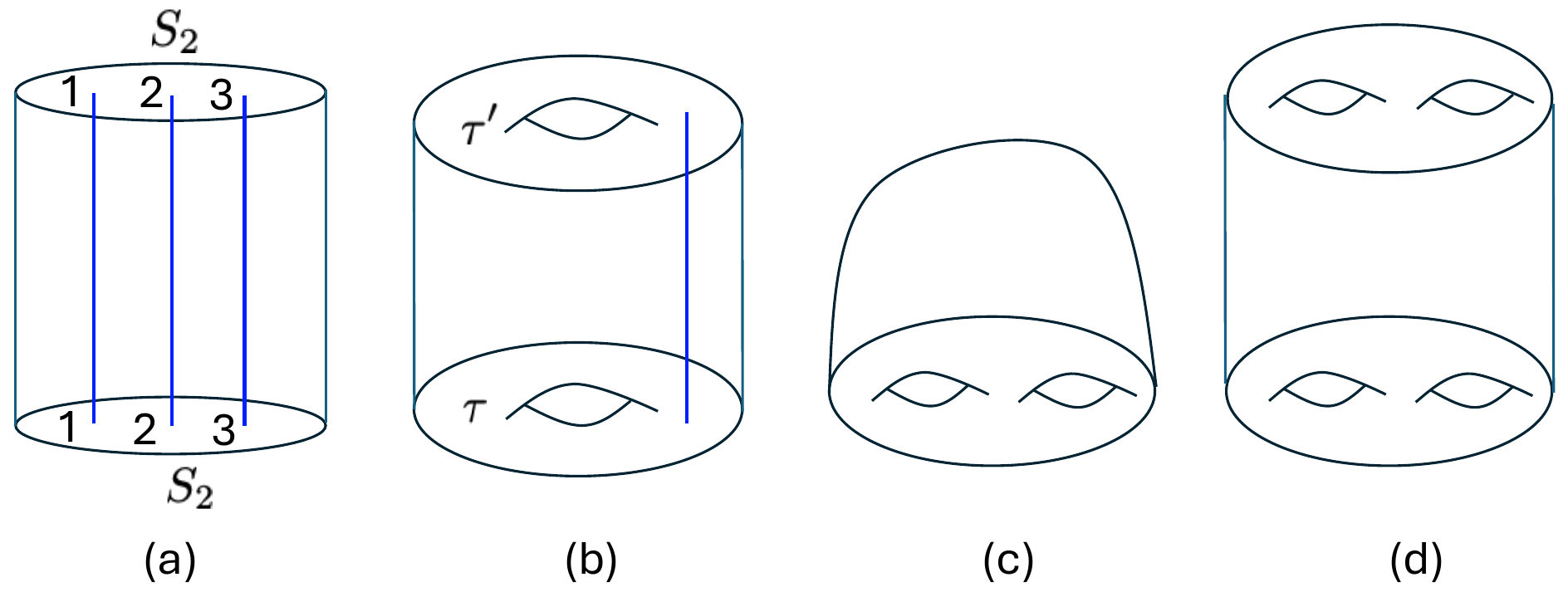}
                \caption[  ]
        {\small  (a): Wormhole configuration connecting two $S_2$ describing $\Fil{G_{123} G_{123}^*}$, with blue solid lines denoting defect propagators. (b) Wormhole configuration describing $\Fil{\vev{\OO}_\tau \vev{\OO^\da}_{\tau'}}$. (c) Single-boundary saddle ending on a genus-2 boundary. (d) Wormhole connecting two disjoint copies of a genus-2 boundary.
    } 
\label{fig:2dw}
\end{figure}

There exists a wormhole connecting two boundaries, with three defect lines, as illustrated in Fig.~\ref{fig:2dw}(a),
implying that 
\be 
\Fil{C_{123}} =0 , \quad \Fil{G_{123} G_{123}^*} \propto \Fil{C_{123} C_{123}^*} \neq 0   \ .
\ee

\item {$g=1$ and $n=1$} 

Now consider one-point function of a defect operator on torus, 
\be
\vev{\OO}_\tau = \Tr \le(q^{L_0 - {c \ov 24}} \bar q^{\bar L_0 - {c \ov 24}} \OO \ri) 
= \sum_j \OO_{jj}  q^{h_j - {c \ov 24}} \bar q^{\bar h_j - {c \ov 24}}, \quad
q= e^{2 \pi i \tau}
 \ ,
\ee
which is the $d=2$ counterpart of~\eqref{Zein} (for a defect operator), with $\tau$ the modular parameter of the torus. 
With $g = 1$,~\eqref{negEu} is satisfied for arbitrary defect operator insertions, and a wormhole saddle exists connecting two boundaries, with a defect line~(see Fig.~\ref{fig:2dw}(b)). This is consistent with
\be
\Fil{\OO_{jj}} = 0 , \qquad \Fil{\OO_{jj}\OO_{kk}^*} \neq 0 , .
\ee

\item{$g=2$ and $n=0$} 

As illustrated in Fig.~\ref{fig:2dw}(c)-(d), this case admits both a single-boundary saddle and two-boundary wormhole solutions.
The boundary genus-2 partition function can be written as 
\begin{equation}
    \label{eq:Zgenus2}
    Z_{\rm CFT}^{(g=2)}(\Omega,\bar\Omega) = \sum_{i,j,k} |C_{ijk}|^2\; \sF^{(2)}_{ijk}(\Omega)\; (\sF^{(2)}_{ijk}(\Omega))^*,
\end{equation}
where $C_{ijk}$ are the OPE coefficients of primaries, $\Omega$ is the genus-2 period matrix, and 
$\sF^{(2)}_{ijk}$  is a genus-2 Virasoro conformal block. The single-boundary saddle Fig.~\ref{fig:2dw}(c) and the two-boundary wormhole configuration Fig.~\ref{fig:2dw}(d) can be interpreted as reflecting correlations
$\Fil{ |C_{ijk}|^2}$ and $\Fil{ |C_{ijk}|^2 |C_{lmn}|^2}$, respectively. Note that Fig.~\ref{fig:2dw}(c) does not involve a wormhole, which is the two-dimensional analogue of Fig.~\ref{fig:PETSL}(b).


\een

Parallel discussions apply to other wormhole solutions whose boundaries satisfy~\eqref{negEu}, linking the existence of wormholes to the appearance of OPE coefficients of heavy states in the corresponding partition functions.


In~\cite{ChaCol22}, the wormhole contributions have been attributed to ensemble averages of CFT$_2$ with OPE coefficients $C_{ijk}$ 
of heavy operators obeying a Gaussian distribution with~\cite{ColMal19} 
\be 
\ol{C_{ijk}} = 0 , \quad \ol{|C_{ijk}|^2} = C_0 (h_i, h_j , h_k) C_0 (\bar h_i, \bar h_j , \bar h_k)  ,
\ee 
where the dimension of $i$-th primary is labeled by $(h_i, \bar h_i)$ and $C_0$ is  a universal function depending only on the central charge and the dimensions, and can be expressed  in terms of the structure constants of the Liouville CFT. In the large $c$ limit, it has the structure  
\be 
C_0 (h_i, h_j , h_k)  = e^{c f (\hat h_i, \hat h_j , \hat h_k) + O(c^0)}, \quad \hat h_i \equiv h_i/c ,  
\ee
for some function $f$~\cite{ChaCol22}. 
In our interpretation, the wormholes reflect correlations among erratic $N$-dependence of OPE coefficients. Assuming the erratic parts of various partition functions can all be attributed to those in $C_{ijk}$, we can conclude from the discussion of~\cite{ChaCol22} that 
\be \label{Liou}
\Fil{C_{ijk}} = 0 \quad \text{and}  \quad \Fil{C_{ijk} C^*_{ijk}} = C_0 (h_i, h_j , h_k) C_0 (\bar h_i, \bar h_j , \bar h_k)  \ .
\ee
Equation~\eqref{Liou} is remarkable because, in large-$c$ CFT$_2$, the Liouville structure constants and density of states provide a universal description of the two-point correlations arising from the erratic $N$-dependence of $C_{ijk}$, independent of microscopic details.

\subsection{More examples of internal wormholes and their interpretations} \label{sec:internal}

It was proposed in~\cite{Col88,GidStr88a,GidStr88} that microscopic (internal) wormholes could give rise to low-energy effective theories with random couplings. The wormholes discussed here, by contrast, are macroscopic saddles of the Euclidean path integral. Nevertheless, they may shed light on the important question of random couplings in the low-energy bulk theory~\cite{ArkOrg07,ChaCol22}.

We have already encountered one example of an internal wormhole, shown in Fig.~\ref{fig:PETSL}(a), whose interpretation is no different from that of external wormholes---both represent manifestations of the erratic behavior of boundary quantities, and there is no indication of random couplings in the low-energy theory. Internal wormholes can be constructed from symmetric external ones using a simple procedure, and many such examples were given in~\cite{ChaCol22}. In what follows, we examine some of these examples to illustrate their boundary interpretations.

Consider an external wormhole geometry $\mathcal{M}$ that connects two asymptotic boundaries $M_1$ and $M_2$.  
Near each boundary $M_i$, choose a half-ball-shaped region $D_i \subset \mathcal{M}$ such that its boundary $\partial D_i = B_i$ intersects $M_i$ along a closed codimension-one surface $A_i = B_i \cap M_i$.  
Now remove the regions $D_1$ and $D_2$ from $\mathcal{M}$, and identify their remaining boundary components $B_1$ and $B_2$ by a suitable diffeomorphism.  The resulting manifold
\be
\mathcal{M}' = \frac{\mathcal{M} \setminus (D_1 \cup D_2)}{B_1 \sim B_2}
\ee
has a single connected boundary $M'$, obtained by gluing the original boundary components $M_1$ and $M_2$ together along the surfaces $A_1$ and $A_2$.  
The identification of the bulk boundaries $B_1$ and $B_2$ effectively creates a ``tunnel'' that connects the regions which were previously located near $M_1$ and $M_2$.  
Through this gluing, new nontrivial loops are generated in the topology of the manifold, indicating that $\mathcal{M}'$ contains an internal wormhole connecting the two formerly separate boundary regions.

Now suppose that $\mathcal{M}$ satisfies the Einstein equations.  
For the resulting geometry $\mathcal{M}'$ to also satisfy the equations of motion without introducing a matter shell at the gluing surface, the induced metric and the extrinsic curvature must both be continuous across the junction.  
A simple way to ensure these matching conditions is to take $\mathcal{M}$ to be reflection-symmetric between its two boundaries, so that $M_1 = M_2 = M$, and to choose the gluing surfaces $B_1$ and $B_2$ to be identical and extremal.  
The identical choice guarantees continuity of the induced metric, while the extremality condition ensures continuity of the extrinsic curvature, thereby yielding a smooth Einstein geometry after the gluing.

Now consider some explicit examples in $d=2$ discussed in~\cite{ChaCol22}:

\ben 

\begin{figure}[h]
        \centering
			\includegraphics[width=12cm]{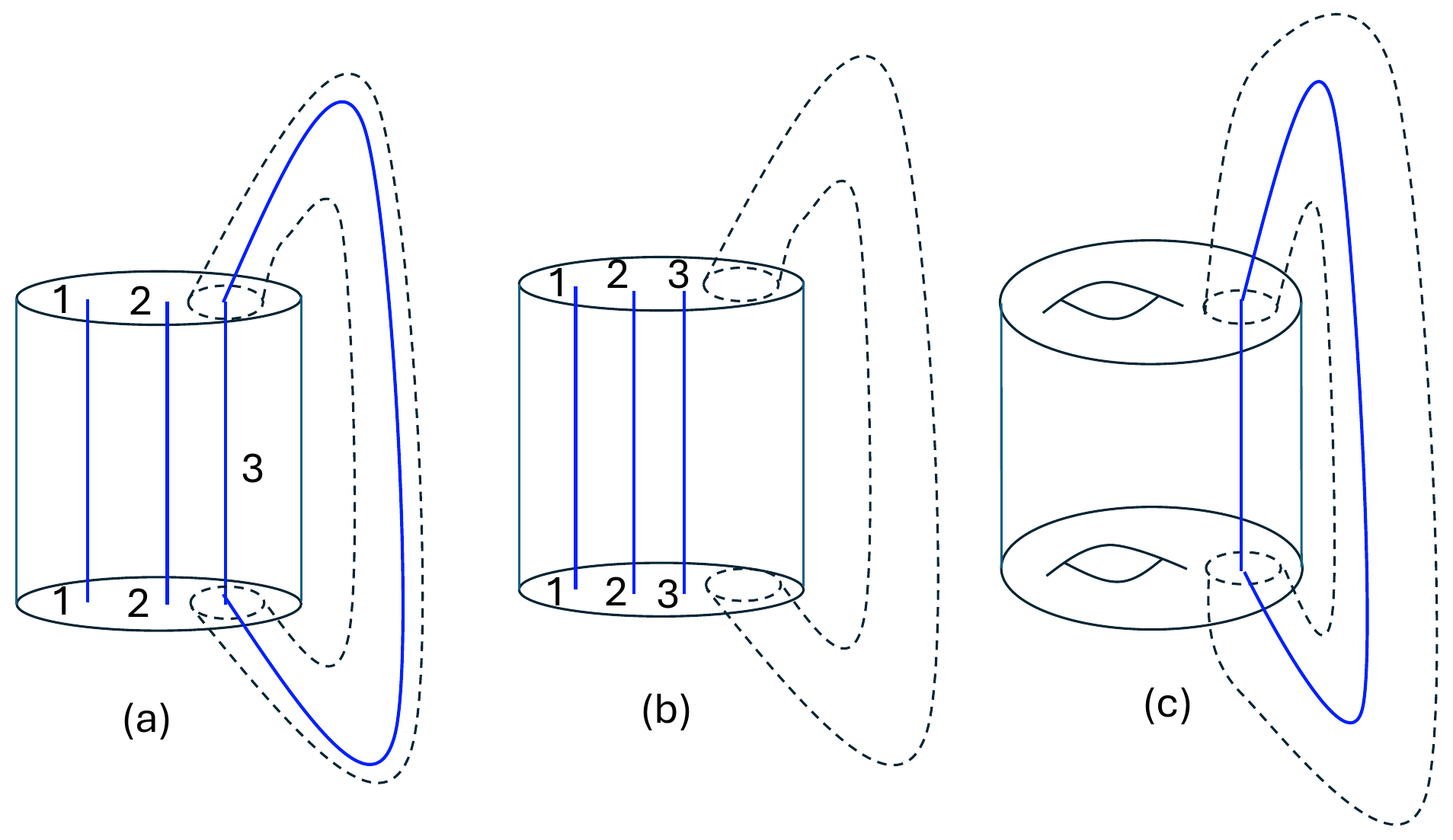}
                \caption[  ]
        {\small Examples of manifolds with an internal wormhole, obtained by removing the interior of an extremal surface $B_i$ near each boundary $M_i$ ($i=1,2$) of an external wormhole and identifying $B_1$ and $B_2$.
(a) Obtained from Fig.~\ref{fig:2dw}(a) by taking $B_i$ to be around $\OO_3$.
(b) Obtained from Fig.~\ref{fig:2dw}(a) by taking $B_i$ not to be around any defect operator.
(c) Obtained from Fig.~\ref{fig:2dw}(b) by taking $B_i$ to be around the defect operator.
    } 
\label{fig:intW}
\end{figure}

\item  Consider Fig.~\ref{fig:2dw}(a), and take $B_i$ to be an extremal surface surrounding one of the defect operator insertions, say $\mathcal{O}_3$, near the boundary $M_i$, which in this case is an $S^2$.  A cartoon of the resulting sewn geometry $\mathcal{M}'$ is shown in Fig.~\ref{fig:intW}(a). The boundary $M' = \partial \mathcal{M}'$ remains topologically $S^2$, but now contains four defect operator insertions, corresponding to the correlator
\be
G_{1122}  = \vev{\OO_1 \OO_1 \OO_2 \OO_2}   \ .
\ee
The saddle point described by $\mathcal{M}'$ corresponds to the contribution to $G_{1122}$ arising from the exchange of the operator $\OO_3$ in the intermediate channel.  
In terms of the OPE, this process can be represented schematically as
\be 
\OO_1 \OO_2 \sim C_{123} \OO_3  (x)  + \cdots \quad \Longrightarrow  \quad \vev{\OO_1 \OO_1 \OO_2 \OO_2} = |C_{123}|^2
| \sF_{1221} (h_3, x)|^2
+ \cdots
\ee
where $x$ denotes the insertion point of $\OO_3$, and $\mathcal{F}_{1221}(h_3, x)$ is the conformal block associated with the corresponding exchange channel. More explicitly, the gravity partition function evaluated around $\sM'$ should then give 
\be 
Z_{\rm gravity}^{(\sM')} =   \Fil{|C_{123}|^2} |\sF_{1221} (h_3, x)|^2
\ee
where we have assumed that $ |\sF_{1221} (h_3, x)|^2$ has a well-defined limit in the large-$c$ limit. 
Note that there is no need to perform any averaging in the bulk description. Rather, the averaging occurs when filtering the microscopic CFT data to obtain the corresponding gravitational description.

\item Now take $B_i$ to be an extremal surface near the boundary $M_i$ in Fig.~\ref{fig:2dw}(a) that avoids any defect operator insertions. A cartoon of the resulting sewn geometry $\mathcal{M}'$ is shown in Fig.~\ref{fig:intW}(b). The boundary of $\sM'$ is still  $S^2$, with now six defect operator insertions, corresponding to the correlator
\be
G_{123123}  = \vev{\OO_1 \OO_2 \OO_3 \OO_1 \OO_2 \OO_3}   \ .
\ee
Now we have 
\be 
Z_{\rm gravity}^{(\sM')} =   \Fil{|C_{123}|^2} \le|\imineq{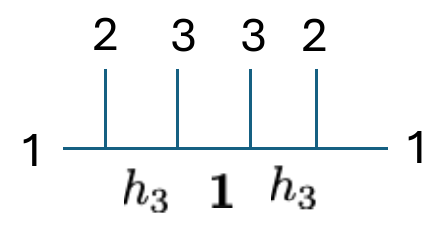}{7} \ri|^2
\ee
where, on the right-hand side, we have represented the corresponding conformal block by a diagram illustrating the relevant OPE channel.


\item Take $B_i$ to be an extremal surface near the boundary $M_i$ in Fig.~\ref{fig:2dw}(b) surround the defect operator insertion. 
 A cartoon of the resulting sewn geometry $\mathcal{M}'$ is shown in Fig.~\ref{fig:intW}(c), whose boundary is now a genus-2 surface, and whose bulk contains a defect line loop corresponding to $\OO$. Now we have 
 \be 
Z_{\rm gravity}^{(\sM')} =  \sum_{i,j} \Fil{|C_{ij\OO}|^2} \le|\imineq{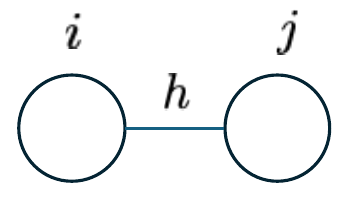}{7}\ri|^2
\ee
where $h$ denotes the conformal dimension of $\OO$. 
\een

Many further examples can be constructed using this sewing procedure.  
The general lesson remains the same: the resulting internal wormhole geometry encodes correlations among the erratic parts of the OPE coefficients of heavy operators.  

In~\cite{ChaCol22}, bulk couplings among defect lines were taken to be proportional to the OPE coefficients $C_{ijk}$, and internal wormholes---such as the one shown in Fig.~\ref{fig:intW}(a)---arising from ensemble averages of the $C_{ijk}$'s were interpreted as generating bulk random couplings to be averaged over.
Our interpretation differs: the OPE coefficients $C_{ijk}$ exhibit erratic $N$-dependence and therefore cannot be directly identified with any smooth gravitational quantity.  
Consequently, there is no notion of random coupling in the gravity description. This is also consistent with the conclusion of~\cite{ArkOrg07}.

\section{Some physical implications} \label{sec:PhysI}

In this section, we explore further implications of the identification~\eqref{mlarN}. It provides a systematic way to incorporate possible non-perturbative corrections to the standard holographic R\'enyi and entanglement entropies~\cite{RyuTak06,HubRan07,EngWal14,LewMal13,FauLew13} via replica wormholes~\cite{PenShe19,AlmHar19}. We then offer some remarks on the violation of global symmetries in quantum gravity arising from wormhole effects.

\subsection{Non-perturbative corrections to holographic R\'enyi and entanglement entropies} \label{sec:deriv}




The replica wormhole analyses of~\cite{PenShe19, AlmHar19}, originally formulated for specific setups such as evaporating black holes coupled to baths, strongly suggest that additional wormhole contributions may need to be included more generally in holographic R\'enyi and entanglement entropies. In JT gravity, this is natural, since the boundary theory corresponds to an ensemble average, and such contributions have recently been emphasized and discussed in detail in~\cite{AntIli25,AntIli25b}. Replica wormhole contributions have also been identified in various higher-dimensional contexts; see, for instance,~\cite{AntSas23,BalLaw22}. 
In~\cite{LiuVar20b}, a projection was introduced to compute R\'enyi  and entanglement entropies of equilibrated pure states in the thermodynamic limit, incorporating in particular the relevant replica wormhole contributions.\footnote{The projection introduced there can in fact be viewed as a baby version of~\eqref{deCom} and~\eqref{deiL} in that specific context.}

In this subsection, we present a general discussion using the prescription~\eqref{mlarN}.


Consider the boundary theory on a manifold $M$ in a state $|\Psi\rangle$, described on the gravity side by the pair $(\sM, |\psi\rangle)$, where $\sM$ denotes the bulk geometry (which may include nontrivial matter configurations) and $|\psi\rangle \in \sH_{\Phi_c}^{\rm (Fock)}$ is a state in the Fock space obtained by quantizing small excitations around $\sM$. By definition, $\partial \sM = M$, and we assume that $\sM$ admits a time-reflection-symmetric Cauchy slice $\sY$ with $\partial \sY = Y \subset M$. Both $M$ and $\sM$ are Lorentzian. For notational simplicity, we will use the same symbols to denote their Euclidean analytic continuations; the intended meaning should be clear from context. We assume the boundary theory has a short-distance cutoff $\epsilon$, so that the entanglement entropy (EE) and R\'enyi entanglement entropies (REEs) are well defined.



Now consider a boundary subregion $A$ lying on the time-reflection symmetric slice $Y$, with reduced density operator 
\be
\rho_A = {1 \ov \sZ} \hat \rho_A
\ee
where $\sZ$ is the partition function of the boundary system (in state $\ket{\Psi}$) and $\hat \rho_A$ is the un-normalized 
reduced density operator for subsystem $A$ defined using Euclidean path integral.  
The REE $S^{(n)}_A$ and EE are then obtained by 
\bega \label{coment}
e^{- (n-1) S^{(n)}_A} \equiv \Tr \rho_A^n = {1 \ov \sZ^n} \sZ_A^{(n)} ,\quad
\sZ_A^{(n)} \equiv \Tr \hat \rho_A^n , \qquad n =2 ,3 \cdots    \\
\label{comen1}
S_A = \lim_{n \to 1}  S^{(n)}_{A}  = - \le[\p_n (\log \sZ_A^{(n)} - n \log \sZ)  \ri]_{n=1}  \ .
\end{gather} 
In~\eqref{comen1}, we have assumed that $\sZ_A^{(n)}$ as well as $S^{(n)}_{A} $ can be continued to general real $n$. 
All quantities above should be understood as defined for each $N$, which we have suppressed for notational convenience. 

To extract them from gravity, we consider\footnote{Note that we can exchange the filtering operation $\Fil{\cdot}$ and derivative with respect to $n$ due to the linearity property~\eqref{CC2} of $\FF$.}  
\bega \label{uen1}
\Fil{ S^{(n)}_A }  = -{1 \ov n-1} \le(\Fil{\log  \sZ_A^{(n)} }- n \, \Fil{\log  \sZ} \ri)
, \\
\Fil{ S_A } 
= - \p_n \le(\Fil{\log \sZ_A^{(n)}} \ri)_{n=1} + \Fil{\log  \sZ}  \ .
\label{uen2} 
\end{gather} 
If the large $N$ limit is pointwise, we can simply exchange the large-$N$ filter $\Fil{\cdot}$ and $\log$ operations. 
But if $ \sZ_A^{(n)}$ or $\sZ$ contains erratic terms in $N$, the two procedures in general do not commute, 
and $\Fil{\log  \sZ_A^{(n)}} \neq \log (\Fil{ \sZ_A^{(n)}})$. 

We can simplify the expressions in~\eqref{uen1}--\eqref{uen2} by using another replica trick, writing the logarithm of a quantity $a$ as 
\be 
\log a = {\p a^k \ov \p k} \biggr|_{k=0}  ,
\ee
which leads to 
\bega\label{rep1} 
\Fil{S^{(n)}_A} = -{1 \ov n-1} {\p \ov \p k} \le(\Fil{(\sZ_A^{(n)})^k}   - n \Fil{\sZ^k} \ri)_{k=0}  ,\\
\Fil{S_A}   = -  {\p \ov \p n} \le[{\p \ov \p k}   \le(\Fil{(\sZ_A^{(n)})^k}   - n \Fil{\sZ^k} \ri)_{k=0} \ri]_{n=1}  \ .
\label{rep2} 
\end{gather}

Applying~\eqref{mlarN}, we have 
\bega \label{znpath}
\Fil{\sZ_A^{(n)}} = \lim_{G_N \to 0}  Z_{\rm gravity} [M_n] , \\
\label{znpath1}
\Fil{(\sZ_A^{(n)})^k}= \lim_{G_N \to 0}  Z_{\rm gravity} [M_n^{\sqcup k}] , 
  \\
\label{znpath2}
\Fil{\sZ^k} =  \lim_{G_N \to 0}  Z_{\rm gravity} [M^{\sqcup k}] , 
\end{gather} 
where $M_n$ is the boundary replica manifold obtained by patching together $n$ copies of $M$ along $A$, and $M_n^{\sqcup k}$ denotes $k$ disjoint copies of $M_n$.

The discussion of~\cite{LewMal13,FauLew13} amounts to the so-called annealed approximation 
\be \label{anneal} 
\Fil{(\sZ_A^{(n)})^k} \approx (\Fil{\sZ_A^{(n)}})^k , \quad \Fil{ \log \sZ_A^{(n)}} \approx  \log (\Fil{\sZ_A^{(n)}} )  , \quad  \Fil{\log \sZ} \approx \log (\Fil{\sZ} ), 
\ee
i.e., including only the fully disconnected contributions. 
We will denote the expressions for $\Fil{S^{(n)}_A}$ and $\Fil{S_A}$ under this approximation as 
\bega \label{ann1}
\Fil{S^{(n)}_A} \approx \le(S^{(n)}_A \ri)_a \equiv -{1 \ov n-1} \le(\log \Fil{  \sZ_A^{(n)} }- n \, \log \Fil{  \sZ} \ri)
, \\
\Fil{S_A} \approx \le(S_A \ri)_a \equiv  - \p_n \le(\log  \Fil{\sZ_A^{(n)}} \ri)_{n=1} + \log \Fil{  \sZ}  \ .
\label{ann2}
\end{gather} 
Corrections in~\eqref{uen1}--\eqref{uen2} beyond those in~\eqref{ann1}--\eqref{ann2} capture effects arising from the erratic $N$-dependence of $\sZ_A^{(n)}$ and $\sZ$, appearing as contributions from (replica) wormhole configurations connecting the various $M_n$'s and $M$'s.

Clearly, any additional contributions depend on the specific system under consideration.
Let us now consider a simplifying situation in which the saddles contributing to $\Fil{\sZ^k}$ and $\Fil{(\sZ_A^{(n)})^k}$ are replica-symmetric in $k$. In this case, the saddles must be either fully disconnected or fully connected, i.e.,
\be 
\Fil{\sZ^k}  = (\Fil{\sZ})^k + z_k , \quad \Fil{(\sZ_A^{(n)})^k}  = (\Fil{\sZ_A^{(n)}})^k + z_{A, k}^{(n)} \ .
\ee
From~\eqref{0nide}, we have
\be 
z_k = \Fil{(\de_e \sZ)^k} , \quad z_{A, k}^{(n)} = \Fil{(\de_e \sZ_A^{(n)})^k}, 
\ee 
i.e., they capture the fully connected correlations of the moments of the erratic $N$-dependence in $\sZ$ and $\sZ_A^{(n)}$, respectively.

From~\eqref{rep1}, we then find 
\be
\Fil{S^{(n)}_A} = \le(S^{(n)}_A \ri)_a  + \le(S^{(n)}_A \ri)_r, \quad
\le(S^{(n)}_A \ri)_r \equiv  -{1 \ov n-1} {\p \ov \p k} \le(z_{A, k}^{(n)}   - n z_k \ri)_{k=0} , 
\ee
where $\le(S^{(n)}_A \ri)_r$ comes from the contribution of full connected replica wormholes. 
Similarly, we find 
\be
\Fil{S_A} =  \le(S_A \ri)_a  + \le(S_A \ri)_r, \quad
\le(S_A \ri)_r \equiv \lim_{n \to 1} \le(S^{(n)}_A \ri)_r  \equiv  -\p_n \le[ {\p \ov \p k} \le(z_{A, k}^{(n)}   - n z_k \ri)_{k=0} \ri]_{n=1} \ .
\ee
It would be worthwhile to investigate whether $\le(S_A \ri)_r$ admits a geometric description, as $\le(S_A \ri)_a$ does~\cite{LewMal13,FauLew13}.  Of course, there can be other non-replica symmetric configurations, which will give further corrections. 

The above discussion concerned a purely holographic system. Let us now consider a more general situation in which a holographic system $B$ is entangled with a non-holographic system $R$, such as in an evaporating black hole, where $B$ corresponds to the black hole subsystem~(which we assume has a holographic description) and $R$ corresponds to the radiation. The state of the combined system can be written as
\be 
\ket{\Psi} =\sum_i  \ket{\psi_i}_{B} \otimes \ket{\chi_i}_R ,
\ee
where $|i\rangle_B$ are boundary CFT microstates describing the black hole. The reduced density operator $\rho_R$ for $R$ can be written as 
\be 
\rho_R = \sum_{i,j} \rho_{ij} \ket{\chi_i} \bra{\chi_j}, \quad \rho_{ij} = \vev{\psi_j|\psi_i}_B  \ .
\ee
The gravity prescription for calculating the second R\'enyi entropy for the radiation subsystem is then given by 
\be \label{rreny}
\Fil{S^{(2)}_R}  = \Fil{\vev{\psi_j|\psi_i}_B \vev{\psi_i|\psi_j}_B}  \ .
\ee
The replica wormhole arises from connected correlations of $\de_e \langle \psi_j | \psi_i \rangle_B$, the erratic-$N$ component of $\langle \psi_j | \psi_i \rangle_B$. In other words, the turning-around of the Page curve---produced by such replica-wormhole contributions---can be attributed to the erratic $N$-dependence of boundary partition functions. It would be instructive to understand this connection more explicitly and directly.

\subsection{Remarks on violation of global symmetries in quantum gravity} \label{sec:glob}

It is widely believed that quantum gravity admits no exact global symmetries. In AdS/CFT, where a theory of quantum gravity is described by a boundary CFT, global symmetries of the boundary theory are mapped to asymptotic symmetries in the bulk---heuristically, the ``global'' part of bulk gauge symmetries---leaving no room for additional fundamental bulk global symmetries. Any global symmetry on the gravity side is therefore expected to be accidental, i.e., emergent in the semiclassical limit and broken at a more fundamental level. From the boundary perspective, such a symmetry emerges in the $N \to \infty$ limit.

While a global symmetry in the low-energy effective theory of gravity can be broken in many different ways---for example, by higher-dimensional operators that break the symmetry at some order $O(G_N^n)$ for an integer $n$---such power-law breaking is generally non-universal and system-dependent. In contrast, wormhole contributions~\cite{HsiIli20, BeldeB20b, CheLin20, BahChen22}, although exponentially small in $G_N$, provide a universal mechanism for violating global symmetries, independent of the detailed structure of the underlying theory.  More explicitly, consider a boundary operator $\sO$ that carries a charge under some global symmetry in the bulk. While there is no bulk solution corresponding to a single insertion of $\sO$ on the boundary, there can exist a wormhole connecting two disjoint boundaries with $\sO$ inserted on one boundary and $\sO^\dagger$ inserted on the other.

Under the prescription~\eqref{mlarN}, we conclude that 
\be
\Fil{\vev{\sO}} = 0, \quad \Fil{\de_e \vev{\sO} \de_e \vev{\sO^\da}} \neq 0 , 
\ee
signaling that $\vev{\sO} = \de_e \vev{\sO} \neq 0$, containing a nonzero erratic part.  The large-$N$ filter perspective on wormholes clarifies how a bulk global symmetry can appear in the semiclassical limit while being absent in the finite-$N$ boundary theory. In this view, the filtering procedure produces the emergent symmetry, and the wormhole contributions show that the would-be symmetry violations are erratic in $N$ (although their overall magnitude can still be exponentially small in $N^2$, depending on the wormhole amplitude).


\section{ 
Closed universes from external wormholes} \label{sec:MM}

We now explore how the erratic $N$-dependence of the boundary system manifests geometrically in Lorentzian signature. In this section, we consider the Maldacena-Maoz (MM) cosmology~\cite{MalMao04}, generalizing the discussion of~\cite{Liu25}.


\subsection{Examples}

The Maldacena-Maoz (MM) cosmology~\cite{MalMao04}---an isolated closed universe (or collection of isolated closed universes) with a negative cosmological constant---is obtained by analytically continuing a reflection-symmetric external wormhole solution~$\sM$ to Lorentzian signature.  A simple class of examples is obtained by starting with the hyperbolic slicing of Euclidean AdS,
\be
ds^2_{\rm EAdS} = d\tau^2 + \cosh^2 \tau \, ds^2_{\HH_d} ,
\ee
and quotienting the hyperbolic space $\HH_d$ by a free-acting discrete group $\Ga$ such that the resulting quotient manifold $M = \HH_d / \Ga$ is {\it compact}. For $d=2$, $M$ is a constant curvature Riemann surface of genus $g \geq 2$. This gives
\be
ds^2 = d\tau^2 + \cosh^2 \tau \, ds^2_{M} , 
\label{euw1}
\ee
which is an external wormhole connecting two copies of $M = \HH_d / \Ga$. Analytically continuing~\eqref{euw1} to Lorentzian signature by taking $\tau =it$ leads to a big-bang/big-crunch closed universe with spatial section given by $M$, 
\be\label{euw}
ds^2 = -dt^2 + \cos^2 t \, ds^2_{M} \ .
\ee

The closed universes obtained by analytically continuing the reflection-symmetric cases of the external wormholes discussed in Sec.~\ref{sec:2dEW} are examples of the form~\eqref{euw}.  
For instance, the configuration shown in Fig.~\ref{fig:2dw}(a) leads to $M$ being a sphere with three defects, the configuration in Fig.~\ref{fig:2dw}(b) gives $M$ as a torus with a defect, and the configuration in Fig.~\ref{fig:2dw}(c) yields $M$ as a genus-2 surface.

\begin{figure}
\begin{center}
\includegraphics[width=9cm]{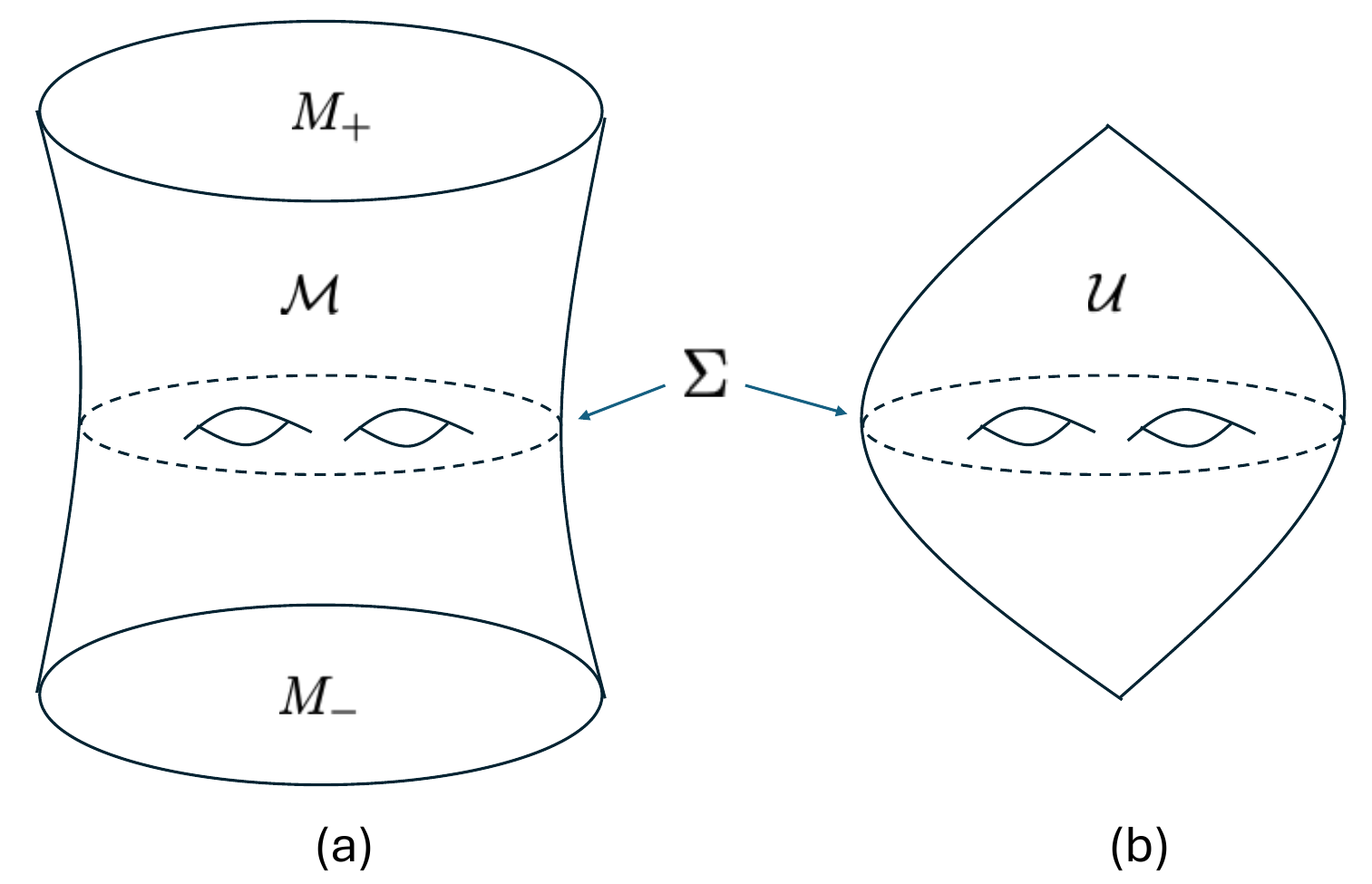}
\caption{\small Cartoon of analytic continuation from a reflection-symmetric external wormhole $\sM$ (plot (a)) to a closed universe $\sU$ (plot (b)). The reflection-symmetric slice $\Sigma$ is deliberately drawn to have a nontrivial topology, highlighting that it need not match the topology of the boundaries $M_\pm$.
}
\label{fig:genA}
\end{center}
\end{figure}

More generally, consider a reflection-symmetric external wormhole $\mathcal{M}$ with a disjoint boundary $M_- \cup M_+$, where $M_- = M$ and $M_+ = \overline{M}$.  Here $M$ is a compact manifold that may contain heavy operator insertions, and $\overline{M}$ denotes the same manifold\footnote{An orientation reversal may also be required.} with the corresponding operator insertions taken with the appropriate conjugation. Denote the reflection-symmetric slice of $\mathcal{M}$ by $\Sigma$.  
The reflection symmetry of $\mathcal{M}$ ensures that it admits a well-defined analytic continuation to a Lorentzian closed universe $\mathcal{U}$ possessing a time-reflection symmetry, with $\Sigma$ corresponding to the time-reflection-symmetric slice.  See Fig.~\ref{fig:genA}.
Note that, in general, $\Sigma$ is not only distinct from $M$ but may also possess a different topology.

\begin{figure}[h]
        \centering
		\includegraphics[width=8cm]{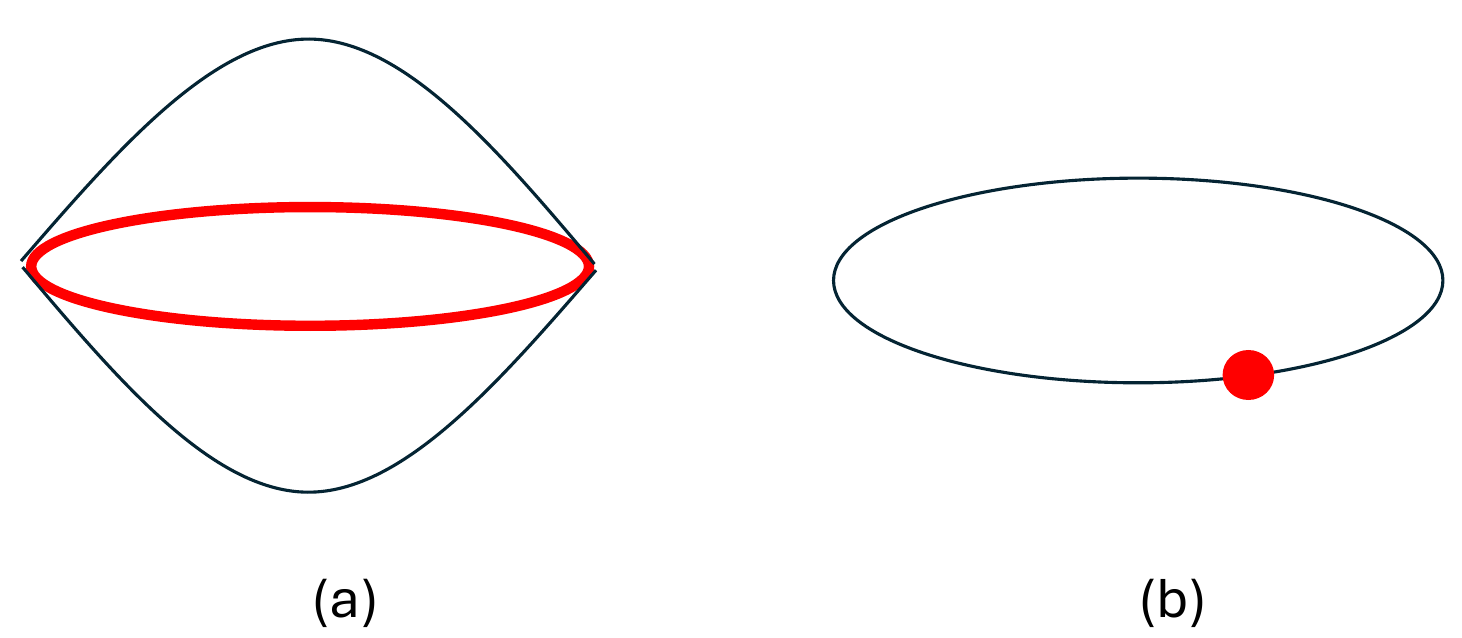}
                \caption[  ]
        {\small  Cartoons of the closed universes obtained by analytic continuation of those in Fig.~\ref{fig:wormM}. Only the 
        time-reflection-symmetric slice $\Sigma$, representing the spatial section of the universes, is shown.
(a) From Fig.~\ref{fig:wormM}(a): with the spatial section topologically $S_d$, obtained by gluing two $d$-dimensional disks $D_d$ together along their boundaries~(corresponding to the location of the mass shell, represented by the thick red line).
(b) From Fig.~\ref{fig:wormM}(b): the spatial section is topologically $S_1 \times S_{d-1}$. Only $S_1$ is shown, with each point in the plot representing an $S_{d-1}$. The red dot indicates the location of the mass shell (smeared uniformly over $S_{d-1}$) on $S_1$.} 
\label{fig:close1}
\end{figure}

More examples of closed universes can be obtained by analytically continuing the reflection-symmetric cases of the external wormholes discussed in Sec.~\ref{sec:EuW} and Sec.~\ref{sec:2dEW}. Consider  Fig.~\ref{fig:wormM}(a) in the reflection-symmetric case.  
The reflection-symmetric slice, indicated by the central horizontal dotted line, can be viewed as two $d$-dimensional disks glued together along their boundaries (corresponding to the location of the mass shell), resulting in a compact space topologically $S_d$ with a mass shell at the equator.  
The corresponding Lorentzian analytic continuation then yields a closed universe whose spatial section is topologically $S_d$.  
Similarly, the Lorentzian continuation of the reflection-symmetric configuration shown in Fig.~\ref{fig:wormM}(b) gives a closed universe with spatial topology $S_1 \times S_{d-1}$ with a mass shell~(uniformly distributed over $S_{d-1}$) at a point on $S_1$. See Fig.~\ref{fig:close1}.

\begin{figure}[h]
        \centering
\includegraphics[width=12cm]{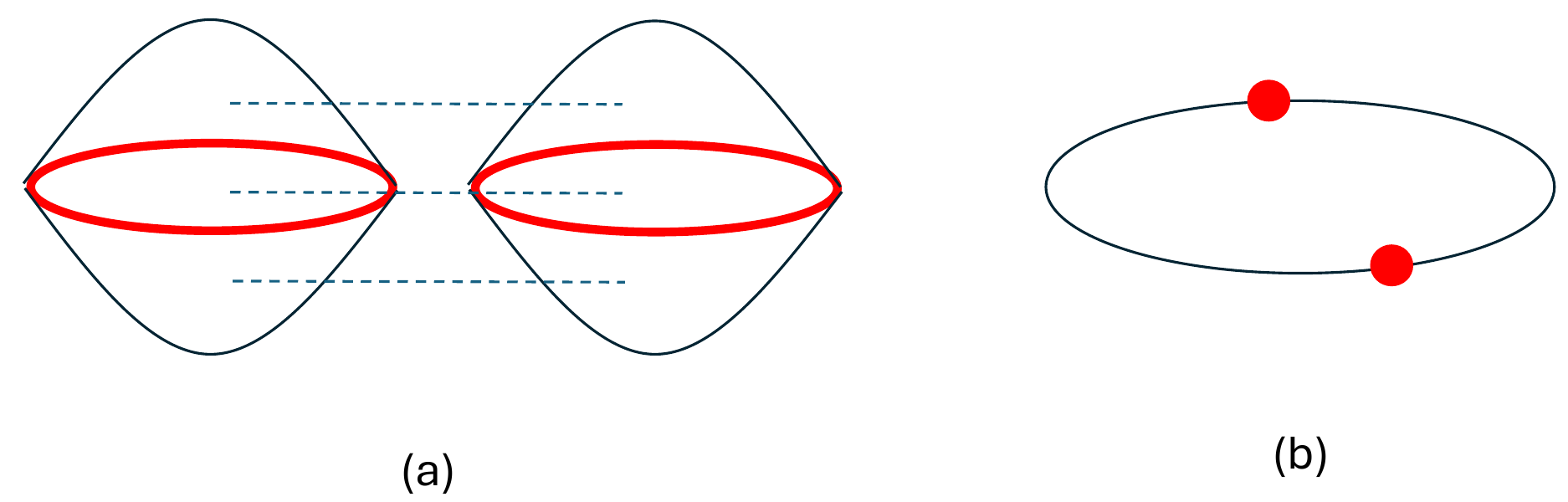}
                \caption[  ]
        {\small Cartoons of the closed universes obtained by analytic continuation of those in Fig.~\ref{fig:PETSH}. 
        The conventions are the same as in Fig.~\ref{fig:close1}.
(a) From Fig.~\ref{fig:PETSH}(a): two entangled closed universes, each with spatial section $S_d$ and a mass shell located at the equator. Dashed lines between them represent entanglement.
(b) From Fig.~\ref{fig:PETSH}(b): the spatial section is topologically $S_1 \times S_{d-1}$, with two mass shells (uniformly distributed over $S_{d-1}$) positioned at two distinct points along $S_1$.} 
\label{fig:close2}
\end{figure}

For the reflection-symmetric case of Fig.~\ref{fig:PETSH}(a), the reflection-symmetric slice has topology $S_d \times S_d$.  
Analytically continuing to Lorentzian signature then yields two entangled closed universes, each with spatial section $S_d$ and a mass shell located at the equator.  
Similarly, analytic continuation of Fig.~\ref{fig:PETSH}(b) results in a closed universe whose spatial section has topology $S_1 \times S_{d-1}$, with two mass shells (uniformly distributed over $S^{d-1}$) positioned at two distinct points along $S_1$.  
See Fig.~\ref{fig:close2}.

\subsection{Closed universe from erratic $N$-dependence} \label{sec:CUext}

Consider a closed universe $\sU$ obtained from analytically continuing a reflection-symmetric external wormhole $\sM$, as illustrated in Fig.~\ref{fig:genA}.  Denote the Fock space obtained by quantizing matter fields (including metric perturbations) in the closed universe as $ \sH_\sU^{(\mathrm{Fock})}$. Given the Euclidean description, an~(overcomplete) basis of states in  $ \sH_\sU^{(\mathrm{Fock})}$ can be constructed  via Euclidean path integral with boundary single-trace operators inserted on the Euclidean boundary $M_-$. 

\begin{figure}
\begin{center}
\includegraphics[width=10cm]{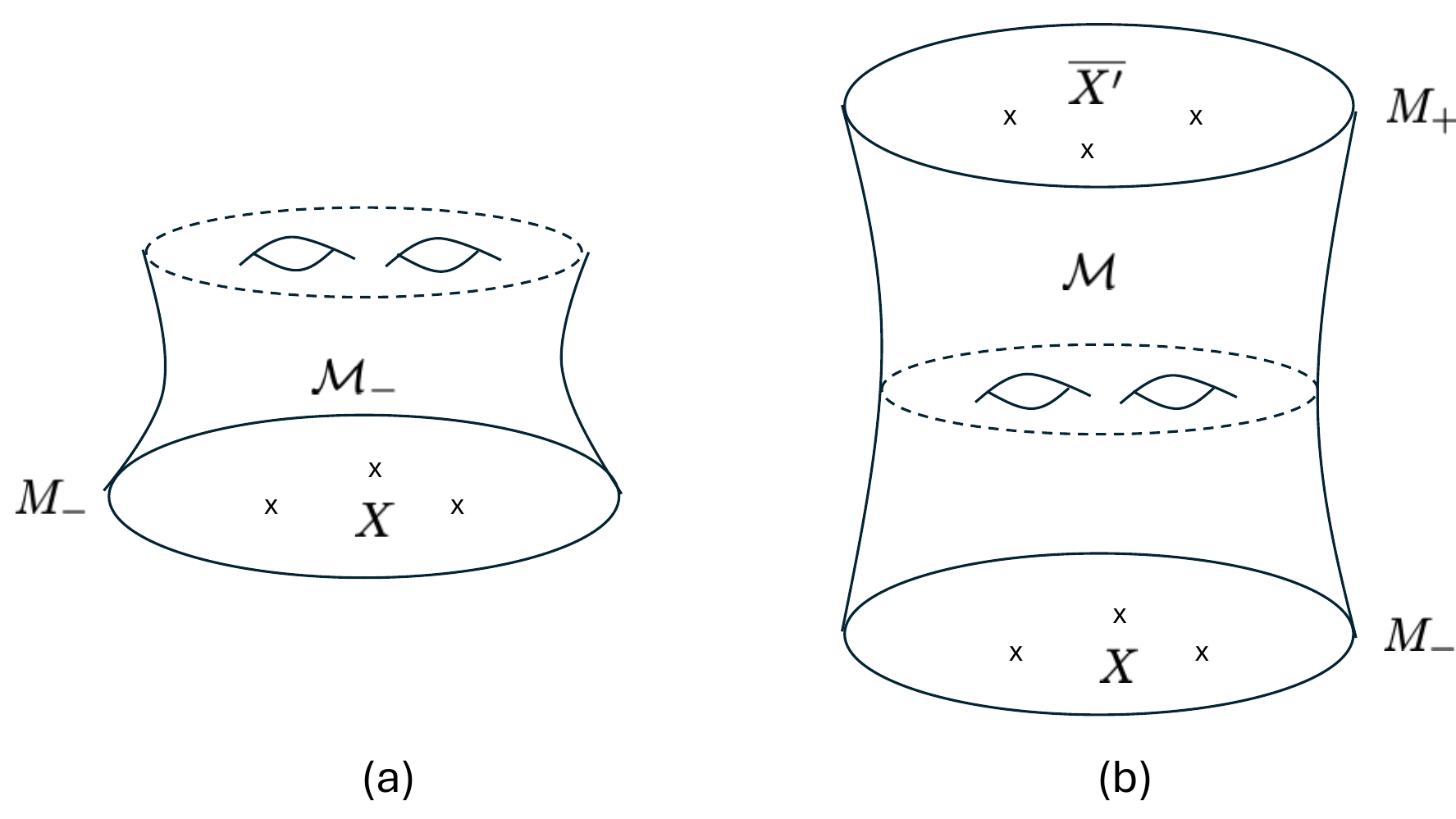}
\caption{\small   The conventions are the same as in Fig.~\ref{fig:genA}.
 (a) Euclidean path-integral description of the Fock space state~\eqref{Xsta}, obtained by inserting boundary operators at $M_-$ (represented by crosses), whose collection is denoted by $X$.
(b) Euclidean path-integral description of the overlap~\eqref{overl} between Fock space states.
}
\label{fig:AdSFRW}
\end{center}
\end{figure}

More explicitly, the bulk Euclidean action $\sI_E$ can be expanded around the wormhole $\sM$ as 
\be 
\sI_E = \sI_E [\sM] + \De \sI^{(\sM)} [\phi]  ,
\ee
where $\De \sI^{(\sM)}[\phi]$ includes the quadratic action for matter perturbations (which we collectively denote as $\phi$) around $\sM$, as well as their interactions. Denoting as $X$ the insertion data~(including both the operator types and positions), we define 
\be \label{Xsta}
\ket{X} \equiv \int_{(\sM_-, X)} D \phi  \, e^{- \De \sI^{(\sM)} [\phi] } \in \sH_\sU^{(\mathrm{Fock})},
\ee
where $(\sM_-, X)$ denotes integrating over the lower half of $\sM$ with
$X$ inserted on $M_-$.\footnote{$X$ specifies the boundary conditions for bulk fields $\phi$.} See Fig.~\ref{fig:AdSFRW}(a). 
The conjugate state $\bra{X}$ can be defined by integrating over the upper half $\sM_+$ of $\sM$, with $\overline{X}$ inserted on $\sM_+$, where $\overline{X}$ denotes the appropriately conjugated version of $X$.

The overlaps among $\ket{X}$ can be written in terms of Euclidean path integrals as 
\bega  \label{overl}
\vev{X'|X} = A (X', X) \equiv \int_{X}^{\ol{X'}}  D \phi  \, e^{-  \De \sI^{(\sM)} [\phi] }  , \qquad \ket{X}, \ket{X'} \in \sH_\sU^{(\mathrm{Fock})}, \\
 A (X', X) = A_0 (X', X)  + O(G_N),
\end{gather} 
where the path integral is over the full $\sM$ with $X, \ol{X'}$ inserted respectively on $M_-$ and $M_+$~(see Fig.~\ref{fig:AdSFRW}(b)). $A_0 (X',X)$ is the overlap in the (bulk) free theory limit (i.e. leading order in $G_N$ expansion). 

Now from the identification~\eqref{imRe}, the wormhole solution $\sM$ gives two-point correlations of the erratic part of $Z_{\rm CFT} [M;X]$ (the boundary CFT partition function on Euclidean manifold $M$ with insertions $X$), 
\bega \label{apar1}
 \Fil{(\de_e Z_{\rm CFT} [M;X'])^*\, \de_e Z_{\rm CFT} [M;X]}  = e^{- \sI_E [\sM]} A (X', X)   ,
\end{gather} 
where the conjugations on $\ol{M}$ and $\ol{X'}$ should be such that $Z_{\rm CFT} [M;X']^* = Z_{\rm CFT} [\ol{M}, \ol{X'}]$.
Recall that $X$ or $X'$ insertions do not affect $\sM$ and the leading term $\sI_E [\sM]$. 
Thus, up to an overall constant $ e^{- \sI_E [\sM]} $, correlations of the erratic part of $Z_{\rm CFT} [M;X]$ gives 
the Gram matrix of the basis $\{\ket{X}\}$ in $ \sH_\sU^{(\mathrm{Fock})}$. It is convenient to include the $X$-independent factor $e^{- \sI_E [\sM]} $ in the normalization of $\ket{X}$ and write\footnote{A form of this equation was given earlier in~\cite{Liu25}.}
\be \label{ueb1}
\vev{X'|X} =  \Fil{(\de_e Z_{\rm CFT} [M;X'])^*\, \de_e Z_{\rm CFT} [M;X]}  \ .
\ee
Equation~\eqref{ueb1} makes it clear that the very existence of the closed universe---together with all the states in and operators acting on $\mathcal{H}_{\mathcal{U}}^{(\mathrm{Fock})}$---can be regarded as a Lorentzian manifestation of the erratic $N$-dependence.

We define $\ket{0}$---the state with no boundary operator insertions---as the ``vacuum'' state, and interpret $\mathcal{I}_E(\mathcal{M})$ as the $\mathcal{O}(1/G_N)$ contribution that plays the role of a ``Casimir energy'' for the closed universe, while $-\log A(0,0)$ represents the contribution from matter fluctuations.\footnote{Since there is no time-translation symmetry in this setting, energy is not precisely defined; the term ``Casimir energy'' is used here in an analogue rather than a literal sense.}

It is instructive to contrast equation~\eqref{ueb1} with~\eqref{rreny} for replica wormholes. In~\eqref{ueb1}, $\langle X'|X\rangle$ is a semiclassical bulk quantity---i.e., a filtered object---and the product $\langle X'|X\rangle \langle X|X'\rangle$ is simply an ordinary product of complex numbers. No further averaging is required, nor should any additional wormhole configurations be included.
In contrast, $\langle \psi_j | \psi_i \rangle_B$ in~\eqref{rreny} is a boundary quantity defined at finite $N$. Filtering products of these matrix elements has nontrivial consequences: it generates replica wormholes. Consequently, the existence of replica wormholes does {\it not} imply that the Hilbert space of a closed universe is one-dimensional.\footnote{In~\cite{UsaWang24,UsaZha24,HarUsa25,AbdAnt25}, one sums over all wormhole configurations when computing $\langle X'|X\rangle \langle X|X'\rangle$ and higher moments. As a result, the wormhole configurations contributing to $\Tr Q^n$ and $(\Tr Q)^n$ coincide, where $Q = \langle X'|X\rangle$ is the Gram matrix. This implies that the two quantities must be proportional and $Q$ has rank~$1$.}

\subsection{An intrinsic boundary construction of the closed universe Hilbert space} \label{sec:CHilb}

In~\cite{Liu25}, we provided a boundary description of the closed universe~\eqref{euw}, obtained by quotienting the bulk AdS spacetime, in terms of the quotient of the boundary theory's algebra in the $N \to \infty$ limit.
Here we develop a more general holographic boundary framework, applicable to any closed universe $\sU$ obtained through analytic continuation of an external wormhole.

\subsubsection{Bulk Hilbert space from boundary} 

In the previous subsection, we used bulk path integrals to construct $\sH_\sU^{(\mathrm{Fock})}$ and to establish~\eqref{ueb1}.
Here, we show that it is possible to construct a Hilbert space intrinsically from the boundary theory, which we identify with $\sH_\sU^{(\mathrm{Fock})}$, thereby providing a holographic description of the closed universe.

Let $M$ be a compact Euclidean manifold and $X$ denote insertion data (light operator types and positions).
We focus on the erratic part $\de_e Z_{\rm CFT} [M;X]$ of the partition function. Under the large-$N$ filtering $\Fil{\cdot}$, we have 
\bega\label{0stM}
\Fil{\de_e Z_{\rm CFT} [M;X]}=0, \\
K(X',X) \equiv \Fil{(\de_e Z_{\rm CFT} [M;X'])^* \de_e Z_{\rm CFT} [M;X]}  = K^* (X, X') ,
\label{stM}
\end{gather}
where the last equality follows from~\eqref{CC3}. 
Denote by $\XX$ the space of possible insertions $X$ on $M$, equipped with the measure $dX$.\footnote{The space $\XX$ has a well-defined measure, involving a sum over light operator types (which form a discrete set in a CFT) and an integration over insertion positions with the measure inherited from $M$. Since the integration over $M$ and the summation over operator types is invariant under operator conjugation, the measure is invariant under conjugation on $\XX$, i.e. $dX = d\ol{X}$.} We should then have
\be
\int dX dY \, f^*(Y) \, K(Y,X) \, g(X) \geq 0 , 
\ee
by the semidefinite positivity condition~\eqref{posi} of $\Fil{\cdot}$,
where $f$ and $g$ are ``functions'' on $\XX$.
That is, $K$ is a Hermitian positive semidefinite kernel on $\XX$.

For a function $f(X)$ on $\XX$, we can associate another function 
\be 
\tilde f (X) = \int dY \, K(X, Y) f (Y) , 
\ee
and define an inner product 
\be \label{innP}
\vev{f|g} = \int dX \, \tilde f^* (X) g (X) =  \int dX dY \, f^* (Y) K (Y, X)  g (X)  \ .
\ee
We define basis vectors $\ket{X'}$ as $\de (X'-X)$ and denote $f (X)$ as $\ket{f}$. It then follows that
\be 
\vev{X'|X} = K (X', X), \quad \vev{X'|f} = \tilde f (X') ,  
 \quad
\ket{f} = \int dX \, f (X) \ket{X}  \ .
\ee
A Hilbert space $\sH_\XX$ is then obtained by factoring out the null states and completing the space with respect to the inner product~\eqref{innP}. Although $\sH_\XX$ is constructed from a commutative algebra of functions on $\XX$, it naturally carries a non-commutative structure encoded in the symplectic form
\begin{equation}
\sigma(f,g) \equiv 2 \,\mathrm{Im}\,\langle f|g\rangle \ .
\end{equation}

We can identify the Hilbert space $\sH_\XX$ with the bulk Fock space,
\be\label{idd}
\sH_\XX = \sH_\sU^{(\mathrm{Fock})} ,
\ee
which provides an intrinsic holographic description of the closed universe in the semiclassical limit.
With the identification~\eqref{idd}, the usual algebraic formulation of bulk reconstruction in AdS/CFT (see~\cite{Liu25b} for a review) can then be carried over without change, including the subregion-subalgebra duality which associates a bulk subregion with a boundary subalgebra. 


\subsubsection{Emergent time} 

Here we propose an emergent time for the closed universe. 

From the kernel~\eqref{stM}, we can define 
a operator
\begin{equation}
(\mathsf K f)(X) \equiv \int K(X, Y)f(Y)\,d Y = \tilde f (X) , 
\end{equation}
on the Hilbert space $\sH_\XX$. Note that 
\be
\vev{g|\mathsf K f} = 
\int dY dX dZ \, g^* (Y) K(Y,X) K(X,Z) f (Z) \ .
\ee 
which means that $\mathsf K$ is a positive Hermitian operator. 

We can then use $\mathsf K$ to generate a unitary flow
\be \label{ust}
U(s) = e^{- i \mathsf K s},
\ee
with $s$ interpreted as an emergent time for the closed universe.
We do not expect $s$ to coincide with a coordinate time such as $t$ in~\eqref{euw}, since the evolution in the closed universe depends explicitly on $t$, whereas $\mathsf{K}$ is independent of $s$.
Nevertheless, $U(s)$ implies the presence of a hidden one-parameter unitary group generated by a positive operator acting on $\sH_{\XX}$.
It is tempting to speculate that $s$ corresponds to a ``hidden'' time variable that evolves in the same direction as the evolution in $t$ and is related to it by an appropriate frame transformation.\footnote{An analogue that comes to mind is a time-dependent Hamiltonian in a rotating magnetic field that becomes time-independent in the rotating frame.} 

Since $\mathsf K$ is Hermitian, it admits a spectral decomposition of the form
\be
\mathsf K = \int_{\sig(\mathsf K)} \lam \, dE(\lam),
\ee
where $\sig(\mathsf K)$ denotes the spectrum of $\mathsf K$ and $E(\lam)$ is a projection-valued measure.
It is tempting to further speculate that the expansion and collapse of the universe---including the big bang and big crunch singularities---are encoded in the spectral structure of $\mathsf K$.

In $d = 2$, using~\eqref{Liou}, it may be possible to work out the kernel~\eqref{stM}, and thereby determine $\mathcal{H}_{\XX}$ as well as $\mathsf{K}$ explicitly.



\subsection{A semiclassical Hilbert space of all isolated closed universes} \label{sec:allClose}




There are also wormhole solutions connecting different compact manifolds (with possible operator insertions) and wormholes connecting multiple---more than two---compact boundaries. In this subsection, we briefly comment on their interpretations for closed universes.

In Secs.~\ref{sec:CUext}--\ref{sec:CHilb}, we discussed how a closed universe $\sU$ and its semiclassical Hilbert space $\sH_\sU^{\rm (Fock)}$ emerge from the erratic parts of $Z_{\rm CFT}[M]$ and $Z_{\rm CFT}[\sX]$, respectively, where $\sX = (M, X)$. It is then natural to interpret the wormhole amplitude $Z_{\rm gravity}^{(\mathrm{wormhole})}[\sX_1, \sX_2]$ as the transition amplitude for a closed universe $\sU_1$ (obtained from the external wormhole connecting $M_1$ and $\overline{M_1}$) in the state specified by $X_1$ to transition to a closed universe $\sU_2$~(obtained from the wormhole connecting $M_2$ and $\overline{M_2}$) in the state specified by $X_2$. Similarly, we may interpret $Z_{\rm gravity}^{(\mathrm{wormhole})}[\sX_1, \sX_2, \sX_3]$ as the branching amplitude for a closed universe $\sU_1$ to transition into two closed universes $\sU_2$ and $\sU_3$, with analogous interpretations for amplitudes involving more boundaries.

From~\eqref{0nide}, these transition amplitudes among closed universes can be described on the boundary in terms of ``correlations'' among the $\de_e Z_{\rm CFT}[\sX_i]$ under the filtering operation $\Fil{\cdot}$. In Sec.~\ref{sec:emG} we constructed a GNS Hilbert space $\sH_{\om_{\FF}}^{(\rm GNS)}$ based on the action of $\Fil{\cdot}$ on the commutative product algebra generated by the $\de_e Z_{\rm CFT}[\sX]$. The inner products in $\sH_{\om_{\FF}}^{(\rm GNS)}$ can then be interpreted as transition amplitudes among closed universes. Thus $\sH_{\om_{\FF}}^{(\rm GNS)}$ may be viewed as a ``semiclassical Hilbert space of all isolated closed universes'' described by the boundary CFT.

The index set ${\sX}$ ranges over all manifolds $M$ of arbitrary topology and metric---a vast space with no canonical finite measure. As a consequence, while one can meaningfully discuss norms and transition amplitudes within $\sH_{\om_{\FF}}^{(\rm GNS)}$, there is generally no canonical probability measure, and one cannot assert, for example, that the total transition probability from a normalized initial state $|\psi\rangle$ to all other states equals $1$. Nor is it possible to define a trace or to diagonalize the full structure. This is natural: the saddle-point approximation (even when subdominant saddles are included) underlying the definition of $Z_{\rm gravity}^{(\mathrm{wormhole})}[\sX_1, \cdots , \sX_n]$ is intrinsically semiclassical. The semiclassical Hilbert space $\sH_{\om_\FF}^{(\rm GNS)}$ should not be regarded as the physical Hilbert space at finite $N$; it is merely a mathematical device for organizing the semiclassical wormhole saddles. After all, for $d>2$ (and for $d=2$ with matter), it is widely believed that Einstein gravity coupled to matter is not UV complete, and hence the gravitational path integral cannot be consistently defined beyond the semiclassical approximation.

In the case of AdS$_3$, the structure of $\sH_{\om_\FF}^{(\rm GNS)}$ can be understood more concretely using~\eqref{Liou}. In particular, various wormhole amplitudes can be expressed in terms of Liouville CFT amplitudes~\cite{ChaCol22}.

\section{
Baby closed universes from internal wormholes} \label{sec:ASS}

In this section, we consider closed universes obtained via analytic continuation of internal wormholes. In this case, the resulting spacetime does not merely contain closed universes; rather, it consists of one or more closed ``baby'' universes entangled with other AdS universes that possess asymptotic boundaries. A prime example is the AS$^2$ cosmology~\cite{AntSas23}, which can be obtained by continuing the internal wormhole shown in Fig.~\ref{fig:PETSL}(b). Additional examples arise from analytic continuations of the geometries discussed in Sec.~\ref{sec:internal} (see also~\cite{BeldeB25,SasSwi25}).

With the closed universe now forming only a sector of the full boundary description, isolating its Hilbert-space structure becomes correspondingly more intricate. Nevertheless, building on the discussions in~\cite{Liu25} and~\cite{KudWit25}, we will argue that these closed-universe sectors emerge from the erratic $N$-dependence of matrix elements of heavy operators.

\subsection{Examples} \label{sec:exmInt}

We first review the AS$^2$ cosmology~\cite{AntSas23}. Take two copies of the boundary CFT, denoted CFT$_R$ and CFT$_L$, with (finite-$N$) Hilbert space $\sH_R \otimes \sH_L$. A partially entangled thermal state (PETS) is defined as~\cite{GoeLam18}:
\bea\label{0PETS} 
\ket{\Psi_\OO^{(\b_R, \b_L)}} 
& =& {1 \ov \sqrt{Z_\OO Z_\b}}  \sum_{m,n} e^{- \ha \b_L  E_m} e^{- \ha \b_R E_n} \OO_{nm} \ket{n}_R  \ket{\tilde m}_L  \\
& =&  {1 \ov \sqrt{Z_\OO}} \OO_R \le({i \ov 2} \b_R \ri) \ket{\Psi_\b} , \quad \b = \b_R + \b_L ,
\label{0PETS1} 
\eea
where $\OO$ is a heavy operator of dimension $\De_\OO \sim O(N^2)$, and $\OO_{nm} = \langle n|\OO|m\rangle$ are its matrix elements in the energy eigenbasis ${\ket{m}}$ with corresponding eigenvalues $E_m$.  $\ket{\Psi_\b}$ denotes the thermofield double (TFD) state with inverse temperature $\b$.
The normalization factor $Z_\OO$ in~\eqref{0PETS}--\eqref{0PETS1} is given by 
\be \label{parti}
Z_\OO =\vev{\OO^\da (- i \b_R/2) \OO (i \b_R/2)}_\b = G_{\OO} (\b_R; \b) ,
\ee
where in the last equality we have used the definition of~\eqref{Ecor1}.

A gravity description of~\eqref{parti} with a spatially uniform operator insertion $\OO$ was discussed in Sec.~\ref{sec:EuW}.
 Below the Hawking-Page temperature, it is described by an internal wormhole attached to Euclidean thermal AdS, as shown in Fig.~\ref{fig:PETSL}(a), whereas above the Hawking-Page temperature, it is described by a mass shell in a Euclidean black hole, as shown in Fig.~\ref{fig:PETSL}(b). Both geometries are reflection-symmetric with respect to the central horizontal slice.

\begin{figure}[h]
        \centering
			\includegraphics[width=14cm]{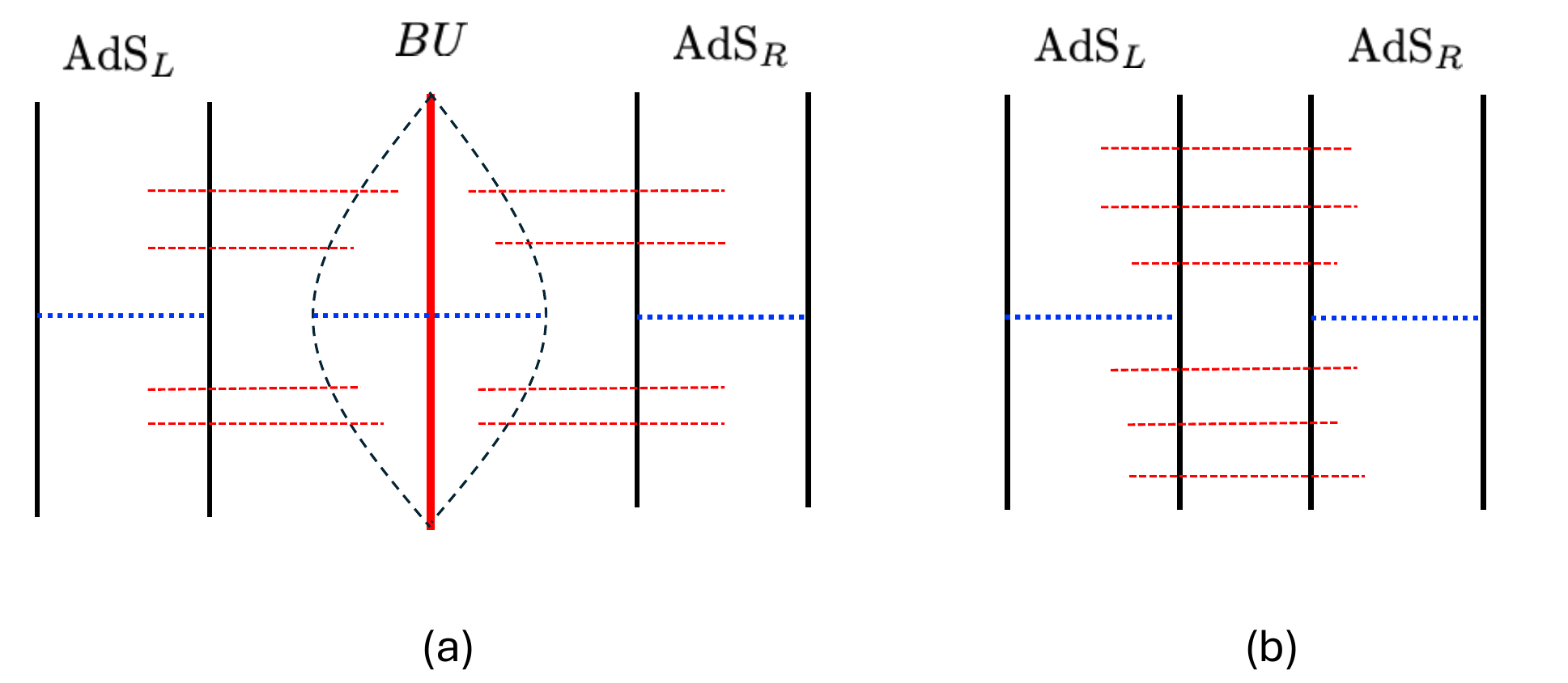}
                \caption[  ]
        {\small (a) The thermal AdS-baby-universe (TAdS-BU) spacetime obtained by analytically continuing the Euclidean gravity solution of Fig.~\ref{fig:PETSL}(a) to Lorentzian signature. The geometry consists of two copies of global AdS entangled with a baby universe (BU). The dotted blue lines indicate the reflection-symmetric slice, corresponding to that in Fig.~\ref{fig:PETSL}(a), and the dashed horizontal red lines represent entanglement between the baby universe and the AdS spaces. The thick red vertical line denotes the mass shell. The spatial section of the BU, which is topologically $S_d$, is obtained by gluing together two $d$-dimensional disks (whose centers are represented by dark dashed arcs) along their boundaries, corresponding to the location of the shell. (b) 
        Analytically continuing the Euclidean gravity solution of Fig.~\ref{fig:newsol}(a) to Lorentzian signature leads to two entangled copies of global AdS without a baby universe. 
        } 
\label{fig:baby}
\end{figure}

Below the Hawking-Page temperature, analytically continuing the Euclidean gravity solution of Fig.~\ref{fig:PETSL}(a) results in
the AS$^2$ cosmology illustrated in Fig.~\ref{fig:baby}(a), where a baby closed universe (of topology $S_{d}$)  from the internal wormhole is entangled with two copies of thermal AdS spacetime~\cite{AntSas23}.

In~\cite{VanLop26}, it was proposed that additional saddles contribute to the gravitational computation of~\eqref{parti}, and that the configuration shown in Fig.~\ref{fig:newsol}(a) in fact dominates over the saddle in Fig.~\ref{fig:PETSL}(a). Analytically continuing this new solution to Lorentzian signature yields two entangled copies of global AdS without a baby universe (see Fig.~\ref{fig:baby}(b)). If this proposal is correct, the dominant gravitational description of the PETS~\eqref{0PETS} does not contain a baby universe, relegating such topologies entirely to subdominant contributions. This picture is also intuitively appealing: the closed-universe sector would then be associated only with the exponentially suppressed tails of the state~\eqref{0PETS}, avoiding the potential tension of accommodating a closed universe within the very limited phase space available at low temperatures.

Nevertheless, understanding the holographic dual of the baby universe shown in Fig.~\ref{fig:baby}(a) remains a conceptually important question.

\begin{figure}[h]
        \centering
			\includegraphics[width=12cm]{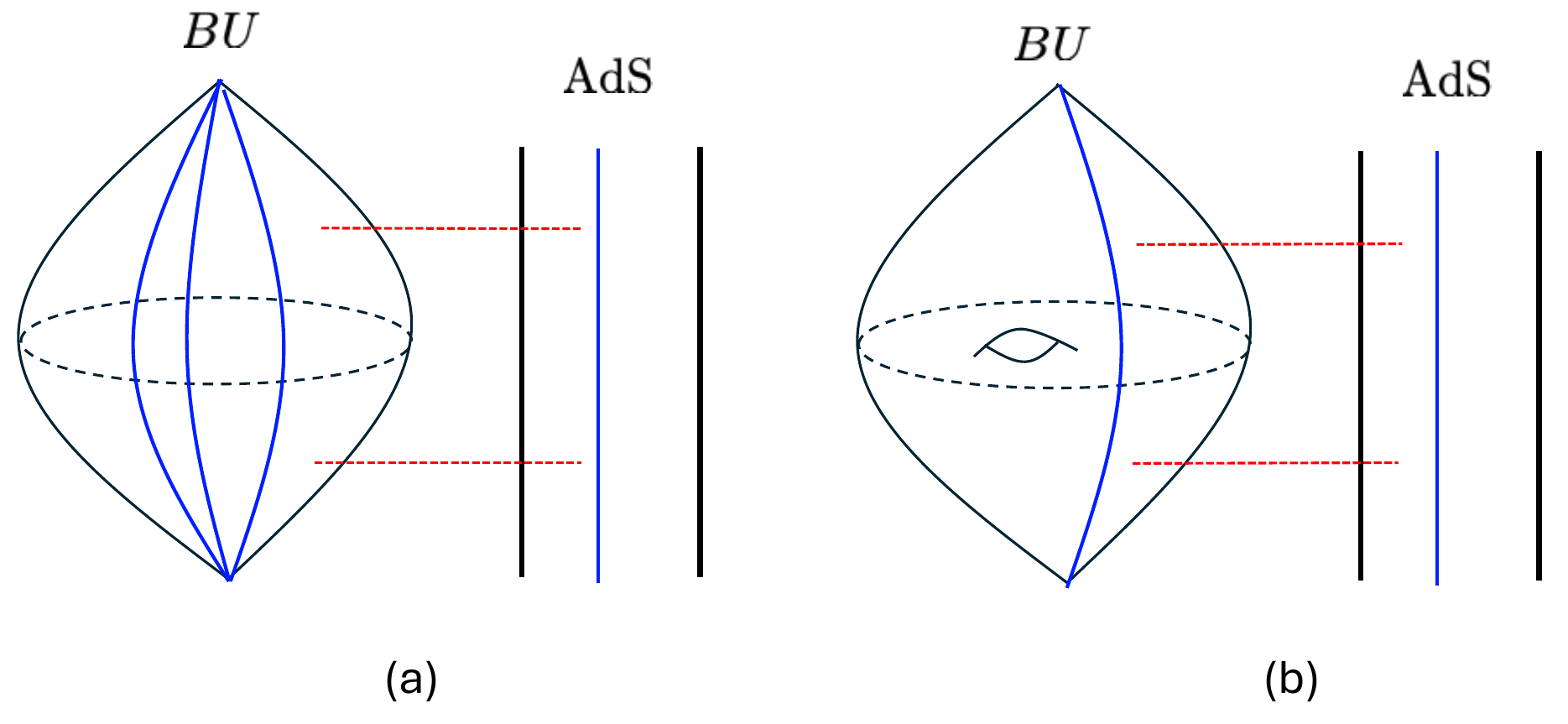}
                \caption[  ]
        {\small Examples of closed baby universes entangled with asymptotic AdS spacetimes, obtained by analytic continuation of the examples in Fig.~\ref{fig:intW}. Defects are represented by blue lines, and horizontal dashed red lines represent entanglement.
(a) From Fig.~\ref{fig:intW}(a): a baby universe with spatial section topologically $S_2$ (with three defects) entangled with an asymptotic AdS spacetime (with one defect).
(b) From Fig.~\ref{fig:intW}(c): a baby universe with spatial section topologically a torus (with one defect) entangled with an asymptotic AdS spacetime (with one defect).
        } 
\label{fig:babyN}
\end{figure}

More examples of baby universes can be found from analytic continuation of the configurations discussed in Sec.~\ref{sec:internal}, which are reflection-symmetric. 
Consider first the example shown in Fig.~\ref{fig:intW}(a).  
The identification generates a tunnel with topology $D_2 \times I$, connecting the previously disjoint boundaries.  
The time-reflection slice $\Sigma$ therefore has topology $D_2 \cup S_2$, with three defects on the $S_2$ component and one defect on the $D_2$ component.  
The $S_2$ originates from the central slice of the external wormhole prior to the identification, while the $D_2$ intersects the boundary of the full geometry along an $S_1$.  
Upon analytic continuation to Lorentzian signature, we obtain a baby universe with spatial section $S_2$ (containing three defects) entangled with an asymptotic AdS$_3$ spacetime that contains one defect; see Fig.~\ref{fig:babyN}(a).  
A similar Lorentzian geometry arises from the analytic continuation of Fig.~\ref{fig:intW}(b), except that the AdS$_3$ component in this case contains no defect. Analytic continuation of Fig.~\ref{fig:intW}(c) similarly yields a baby universe with spatial section a two-dimensional torus $T_2$ (containing one defect) entangled with an asymptotic AdS$_3$ spacetime that contains one defect; see Fig.~\ref{fig:babyN}(b).

\subsection{Baby universe from erratic $N$-dependence and quantum volatile AdS} \label{sec:AS22}

Using the AS$^{2}$ cosmology\footnote{For additional perspectives on the AS$^{2}$ cosmology, see~\cite{AntRat24,EngGes25,EngGes25b,Hig25,AntRat25,BeldeB25}.} as an illustrative example, we argue below that the closed baby universe arises from the erratic $N$-dependence of the matrix elements $\OO_{ab}$ of $\OO$ among the low-energy states $\ket{a}, \ket{b}$ appearing in~\eqref{0PETS}~\cite{Liu25,KudWit25}.

Below the Hawking-Page temperature, we can write~\eqref{0PETS} as
\be\label{trucP}
\ket{\Psi_\OO^{(\b_R, \b_L)}} 
= \frac{1}{\sqrt{Z_\OO Z_\b}} \sum_a e^{-\frac{1}{2} \b_L E_a} \ket{\hat a}_R \ket{a}_L = \sum_{a,b} f_{ab} \ket{a}_L \ket{b}_R
\ ,
\ee
where we have truncated the sum to energy eigenstates $\ket{a}$ with $E_a \sim O(N^0)$---such states dominate the sum---which span the GNS Hilbert space $\sH_{\Om}^{(\mathrm{GNS})}$ around the vacuum $\ket{\Om}$. We have also defined
\bega\label{defah}
\ket{\hat a} \equiv e^{-\frac{1}{2} \b_R H} \OO \ket{a}
= \sum_b e^{-\frac{1}{2} \b_R E_b} \OO_{ba} \ket{b} , \\
f_{ab} =  \frac{1}{\sqrt{Z_\OO Z_\b}} e^{-\frac{1}{2} \b_L E_a} e^{-\frac{1}{2} \b_R E_b} \OO_{ba} 
\ .
\label{fab}
\end{gather}

Fig.~\ref{fig:baby} indicates that the semiclassical bulk Hilbert space should factorize as
\be
\sH_{\rm bulk}^{(\mathrm{Fock})} = \sH_{\rm AdS_R}^{(\mathrm{Fock})} \otimes \sH_{\rm AdS_L}^{(\mathrm{Fock})} \otimes \sH_{BU}^{(\mathrm{Fock})}, 
\ee
with $\sH_{\rm AdS_R}^{(\mathrm{Fock})}$ and $\sH_{\rm AdS_L}^{(\mathrm{Fock})}$ identified with the boundary GNS Hilbert spaces
$\sH_{\Om,R}^{(\mathrm{GNS})}$ and $\sH_{\Om,L}^{(\mathrm{GNS})}$, respectively.
The central question is how the baby-universe Hilbert space $\sH_{BU}^{(\mathrm{Fock})}$ arises from the boundary perspective.

If the coefficients $f_{ab}$ converge to definite values in the large-$N$ limit, then the state~\eqref{trucP} lies entirely within $\sH_{\Om,R}^{(\mathrm{GNS})} \otimes \sH_{\Om,L}^{(\mathrm{GNS})}$ and describes an entangled bulk pure state in two AdS regions. Consequently, no additional baby-universe sector appears~\cite{AntRat24,EngGes25,Ges25}.
However, as we noted in Sec.~\ref{sec:EuW}, the matrix elements $\OO_{ab}$ are expected to display erratic dependence on $N$, and therefore the preceding conclusion does not apply: a baby universe may emerge~\cite{Liu25,KudWit25}.

The Lorentzian-filter generalization of the holographic dictionary~\eqref{mlarN1} makes the emergence of the baby universe precise.
As described in Sec.~\ref{sec:Loren} (see the discussion following~\eqref{mlarN1}), this extended dictionary broadens the set of observables and states---such as~\eqref{trucP}---that can be assigned a gravitational interpretation.
We now highlight several key features of this ``filtered emergence'' as it appears in AS$^{2}$ cosmology:

\ben

\item {\it Emergence of the baby universe}

Let $\sA$ denote the joint algebra of single-trace operators acting on the two CFTs.
One can show that the erratic behavior of $\OO_{ab}$ implies that the state $\ket{\Psi_\OO^{(\b_L,\b_R)}}$~\eqref{trucP} is mixed with respect to $\sA$; equivalently, $\sA$ possesses a nontrivial commutant within $\sB(\sH_{\rm bulk}^{(\mathrm{Fock})})$~\cite{Liu25,KudWit25}.
This nontrivial commutant encodes the presence of a baby-universe sector.

At a heuristic level, we can also see the emergence of the baby universe as follows~\cite{Liu25}. 
Despite its appearance, $\ket{\hat a}$ of~\eqref{trucP}--\eqref{defah} is not a state in $\sH_{\Om}^{(\mathrm{GNS})}$, owing to the erratic $N$-dependence of the OPE coefficients $\OO_{ab}$. Rather, $\OO \ket{a}$ should be viewed as belonging to a new sector of states that themselves lack a well-defined large-$N$ limit. It is these new sectors (on both the $R$ and $L$ sides) that can be thought of as generating the Hilbert space of the baby universe---in other words, the baby universe owes its existence to the erratic large-$N$ behavior of $\OO_{ab}$.

We emphasize that this discussion is unaffected by the possibility that the AS$^2$ cosmology is a subdominant configuration for the PETS; it merely implies that the features discussed here must be sought within the subdominant contributions.

\item {\it Quantum volatility of the AS$^2$ cosmology from the erratic $N$-dependence}

We emphasized earlier that the baby universe owes its existence to the erratic nature of the truncated state~\eqref{trucP}. This same erratic behavior also leads to the quantum volatile nature of the AS$^2$ spacetime. More explicitly, we can write 
\be 
\Fil{Z_\OO} = \sZ_1 + \sZ_2 \ ,
\ee
where $\sZ_1$ and $\sZ_2$ denote the contributions from the configurations in Fig.~\ref{fig:newsol}(a) and Fig.~\ref{fig:PETSL}(a), respectively. 
Following~\cite{VanLop26}, we take the configuration of Fig.~\ref{fig:newsol}(a) to be the leading one, i.e., 
\be 
{\sZ_2 \ov \sZ_1} \sim e^{-N^2 g_0} \ , \quad \text{with} \quad g_0 > 0  \ .
\ee
The ``variance'' of $Z_\OO$ is described on the gravity side by the wormhole configuration in Fig.~\ref{fig:PETSH}(a), and from~\eqref{ooco} we find
\be 
{\Fil{Z_\OO^2}_c \ov \sZ_1^2} \sim e^{- 2 N^2 g_0} \ .
\ee
That is, the magnitude of the ``fluctuations'' of $Z_\OO$ is of the exact same order as the contribution $\sZ_2$ from the AS$^2$ cosmology (in terms of the large-$N$ scaling). 
Note that this conclusion is in fact independent of whether $\sZ_1$ (i.e., the configuration of Fig.~\ref{fig:newsol}(a)) is dominant or not. Regardless of the presence or dominance of $\sZ_1$, the magnitude of $\sqrt{\Fil{Z_\OO^2}_c}$ is always of the same order as $\sZ_2$.

We stress that this filtered ``variance'' is distinct from the ordinary quantum variance of an observable in a fixed microscopic state. To give it an operational interpretation, we adopt a working notion of a low-energy gravitational observer whose accessible information is characterized by the large-$N$ filter. Such an observer is restricted to the semiclassical bulk algebra and to measurement protocols that remain within the semiclassical regime. In particular, the observer cannot resolve or control the exact finite-$G_N$ microscopic data encoded in the erratic parts of the dual CFT quantities.

The relation $\sqrt{\Fil{Z_\OO^2}_c} \sim \sZ_2$ means that the magnitude of the filtered fluctuations is of the same large-\(N\) order as the AS\(^{2}\) saddle contribution itself. We interpret this as saying that, for low-energy observers, bulk observables in the AS\(^{2}\) cosmology exhibit \(O(1)\) filtered fluctuations relative to their respective saddle-point values, both in the global AdS regions and in the baby universe, even as \(G_N\to0\).

In~\cite{EngLiu23}, the term quantum volatility was introduced to characterize spacetimes exhibiting \(O(G_N^0)\) fluctuations in geometric quantities. We use the term here in the broader operational sense that the semiclassical data accessible to a low-energy observer do not determine sharp predictions for observables associated with the AS\(^{2}\) sector.

If this failure of semiclassical predictability can itself be probed by measurements confined to the AdS region, it could provide an indirect signal of the closed-universe sector.

\een

\section{Black hole interiors} \label{sec:BHI}

In this section, we argue that the semiclassical geometry of a generic black-hole interior emerges by filtering boundary quantities with intrinsically erratic $N$-dependence, and that the resulting interior can be quantum volatile.
We first illustrate this idea using the BH phase of the PETS, and then comment on more general black holes.

Above the Hawking-Page temperature, the state~\eqref{0PETS} can be approximated by
 \be\label{2PETS}
 \ket{\Psi_\OO^{(\b_R, \b_L)}}
 \approx  {1 \ov \sqrt{Z_\OO Z_\b}}  \sum_{i,j} e^{- \ha \b_L  E_j} e^{- \ha \b_R E_i} \OO_{ij} \ket{i}_R  \ket{\tilde j}_L \ ,
 \ee
 where we have truncated the sums to heavy states $\ket{i}, \ket{j}$ with energies of order $O(N^2)$.
The corresponding partition function~\eqref{parti} can be expressed in terms of the matrix elements $\OO_{ij}$ as in~\eqref{Ecor1}. 
More explicitly, 
\be\label{PPae}
Z_\OO = \sum_{i} X_{ii}, \quad X_{ij} \equiv  {1 \ov Z_\b} \sum_k  B_{ik} e^{-\b E_k}  B^*_{jk} , \quad B_{ij} \equiv e^{- \ha \b_R (E_i - E_j)} \OO_{ij}  
\ .
\ee
From~\eqref{hdm}, this partition function may contain a disconnected contribution in addition to the connected contribution~\eqref{t000}. For our subsequent discussion, which of these configurations dominates (or whether the disconnected part vanishes entirely) is immaterial; what matters is only that the dominant configuration describes a black hole.

As in the standard bulk reconstruction story, bulk operators outside the horizons can be obtained from single-trace operators on the boundary, whereas those inside the horizons---belonging to the entanglement wedges of the $R$ and $L$ boundaries---can be reconstructed from modular-flowed single-trace operators of the form
\be \label{egb}
\hat \sO_R (s) = \rho_R^{-is} \sO_R \rho_R^{is} , \qquad \hat \sO_L (s) = \rho_L^{-is} \sO_L \rho_L^{is} ,
\ee
where  $\sO_R, \sO_L$ are single-trace operators, and $\rho_R$ and $\rho_L$ are the reduced density operators of $\mathrm{CFT}_R$ and $\mathrm{CFT}_L$ in the state~\eqref{0PETS}.  Focusing on the right boundary, the matrix elements of $\rho_R$ 
can be written as
\be \label{rhoR}
(\rho_R)_{ij} =  {X_{ij} \ov Z_\OO}   \ . 
\ee 
Below we would like to argue that while generically the two-point functions of single-trace operators $\sO_R$ in $\ket{\Psi_\OO^{(\b_R, \b_L)}}$ are given by the thermal answer with suppressed variance, the two-point functions of the modular-flowed operators~\eqref{egb} can, in principle, exhibit large fluctuations.

First consider the two-point function of a single-trace Hermitian operator $\sO_R$ in $\mathrm{CFT}_R$, evaluated in the state $\ket{\Psi_\OO^{(\b_R, \b_L)}}$, 
 \bega \label{grr}
 F_{RR} = \vev{\Psi_\OO^{(\b_R, \b_L)}\Big|\sO_R (t_1) \sO_R( t_2)\Big|\Psi_\OO^{(\b_R, \b_L)}} = \Tr \le(\rho_R \sO (t_1) \sO( t_2) \ri)
= {\hat F_{RR} \ov Z_\OO} \ ,
 \\
\hat F_{RR} = \Tr  \le(X \sO (t_1) \sO( t_2) \ri) =
 \sum_{i,j,k} X_{ij} e^{i E_j t_1} \sO_{jk} e^{- i E_k (t_1 - t_2)} \sO_{k i} e^{- i E_i t_2} \ ,
 \label{hfrr}
 \end{gather}
 where $\sO_{jk} = \vev{j\vert{}\sO\vert{}k} = \sO_{kj}^*$. We assume that the erratic parts of $\sO_{ij}$ and $\OO_{ij}$ are uncorrelated~(or that their correlations are not large enough to matter), i.e.,
 \be \label{oup}
 \Fil{\sO^m \OO^n} = \Fil{\sO^m} \Fil{\OO^n} \ .
 \ee
 We further assume ETH for $\sO_{ij}$ and, for simplicity, that its diagonal elements vanish, i.e., $\Fil{\sO_{ij}}=0$. Its two-point correlation over the erratic part is therefore
 \be \label{ETHvar}
 \Fil{\sO_{ij} \sO_{kl}^*} = \Fil{\sO_{ij} \sO^*_{kl}}_c
 = e^{- S (E_{ij})} f (\om_{ij};  E_{ij}) \delta_{ik} \delta_{jl} \ ,
 \ee
 where $E_{ij} = (E_i + E_j)/2$, $\om_{ij} = E_i - E_j$. $f(\om; E)$ is an order $O(N^0)$ function that depends weakly on $E$ and is supported for $\om \sim O(N^0)$.

When considering higher moments of $\hat F_{RR}$ under $\FF$, we can ignore contributions from the connected part of $(\sO_{jk} \sO_{ik}^*)^n$, which are exponentially suppressed.\footnote{There are two types of contributions. The first arise from inter-copy contractions, which reduce the number of independent energy sums and are therefore exponentially suppressed in the entropy. The second arise from genuine higher-order correlations of the erratic matrix elements. In generalized ETH, such irreducible higher-point correlations carry additional entropy suppression; see, e.g.,~\cite{FoiKur18,PapFoi22}. \label{ft:eth}}
  With this in mind, to understand the general behavior of the higher moments of $\hat F_{RR}$, we can directly replace 
$\sO_{jk} \sO_{ik}^*$ in~\eqref{grr} by $\Fil{\sO_{jk} \sO_{ik}^*}$. This effectively collapses the indices, leading to
 \bega
 \hat F_{RR} = \sum_{i,j} X_{ii}  e^{- S (E_{ij})} f (\om_{ij}; E_{ij})   e^{i (E_i - E_j) (t_1-t_2)}  \ .
 \end{gather}
 Since $f$ is supported only for energy differences $\om_{ij} \sim O(N^0)$, we can replace the sum over $j$ with an integral over $\om = E_i - E_j$, weighted by the density of states $e^{S(E_j)}$. Expanding the exponent $S(E_j) - S(E_{ij}) \approx -\ha \beta_i \om$ since $\om \ll E_i$, we obtain
 \bega \label{frrD}
\hat F_{RR} =\sum_i X_{ii} F_i \quad \implies \quad F_{RR} = {\hat F_{RR} \ov Z_\OO} = {\sum_i X_{ii} F_i \ov \sum_i X_{ii}} \ , \\
 F_i \equiv \int d\om \, e^{-\ha \om \b_i } f (\om ;  E_i-\om/2) e^{i \om (t_1 - t_2)}  \ , \quad \b_i = {\p S \ov \p E_i} \ .
 \end{gather}
 Given that $F_i$ depends weakly on $E_i$ and is independent of $\OO$, and that the partition sum $\sum_i X_{ii}$ is expected to be sharply peaked at a dominant saddle $E_*$, $F_{RR}$ evaluates to the standard thermal answer $F (E_*)$ with inverse temperature $\b_* = {\p S \ov \p E_*}$. Provided that the erratic nature of $\OO_{ij}$ does not lead to an $O(N^0)$ shift in $\b_*$, $F_{RR}$ has a suppressed variance and is insensitive to all other aspects of the erratic behavior. For example, $Z_\OO$ can, in principle, have large fluctuations itself. In other words, because the ratio is governed by the saddle point, as long as the saddle point does not significantly shift, the erratic fluctuations of $X_{ii}$ cancel between the numerator and the denominator, leaving only a smooth, thermal two-point function.

Now consider correlation functions of bulk operators in the black hole interior, which can be reduced to those of the modular-flowed boundary operators~\eqref{egb}. Consider
 \be \label{inet}
 \sG (s) = \vev{\Psi_\OO^{(\b_R, \b_L)}\Big|\hat \sO_R (s_1) \hat \sO_R( s_2)\Big|\Psi_\OO^{(\b_R, \b_L)}}
 = \Tr \le(\rho_R \hat \sO (s)\sO  \ri) \ ,
 \ee
where $\rho_R$ is given by~\eqref{rhoR} and $s=s_1 - s_2$. The matrix elements
 \be
 \hat \sO_{ij} (s) = \vev{i| \rho_R^{-is} \sO \rho_R^{is}|j}
 \ee
 involve $\OO$ through $\rho_R$. Thus, unlike~\eqref{oup}, powers of $\hat \sO_{ij}$ and $\OO$ no longer factorize under the filter $\FF$.
 
 To proceed, it is convenient to transition to the basis $\ket{\al}$ in which $\rho_R$ is diagonal, i.e.,
 \be
 \rho_R| \al \rangle=p_\al |\al\rangle \ , \quad p_\al = e^{-\sE_\al} \ , \quad \sO_{\al \b} = \vev{\al|\sO|\b} \ ,
 \ee
 which allows us to write
 \be\label{sgco}
 \sG(s) =
 \sum_{\al,\b}
 e^{-\sE_\al}
 \big|\sO_{\al \b}\big|^2
 e^{is(\sE_\al-\sE_\b)}  \ .
 \end{equation}
 Note that both $\sE_\al$ and $\sO_{\al \b}$ depend on $\OO$ in a highly nonlinear way.

We can simplify the problem by making the physically reasonable assumption that the matrix elements of $\sO$ in the $\ket{\al}$ basis satisfy ETH, i.e.,
\be\label{Meth}
\sO_{\al\b}
=
a(\sE_\al)\de_{\al\b}
+
e^{-\frac12\sS(\sE_{\al\b})}
F(\om_{\al\b},\sE_{\al\b})\sR_{\al\b}\,,
\quad
\om_{\al\b}=\sE_\al-\sE_\b\,,
\quad
\sE_{\al\b}=\frac12(\sE_\al+\sE_\b)\,.
\ee
Here $\sS(\sE)$ is the microcanonical entropy associated with the $\ket{\al}$ basis, so that the corresponding density of states scales as $e^{\sS(\sE)}$. Such a modular ETH ansatz can be expected to hold if the modular Hamiltonian $K\equiv-\log\rho_R$ is itself chaotic, so that the usual rationale underlying ETH can be applied to its eigenstates. The ansatz separates the dependence of $\sO_{\al\b}$ on $\OO_{ij}$ into two distinct sets of quantities: (i) the smooth, ``macroscopic'' quantities $a(\sE)$, $F(\om,\sE)$, and $\sS(\sE)$, and (ii) the erratic, ``microscopic'' variables $\sR_{\al\b}$.

Both sets of quantities can fluctuate as $\OO_{ij}$ is varied. The smooth quantities, however, characterize the collective behavior of an exponentially large number of levels and are therefore expected to be self-averaging. We will consequently assume that correlations between the microscopic variables $\sR_{\al\b}$ and the macroscopic quantities $a(\sE)$, $F(\om,\sE)$, and $\sS(\sE)$ are parametrically suppressed and can be neglected at the order of interest. As in standard ETH treatments, we take
\be\label{Rva}
\Fil{\sR_{\al\b}}=0\,,
\qquad
\Fil{\sR_{\al\b}\sR_{\de\ga}^*}
=
\de_{\al\de}\de_{\b\ga}\,,
\ee
with $\sR_{\al\b}=\sR_{\b\al}^*$ as required by Hermiticity.

Plugging~\eqref{Meth} into~\eqref{sgco} and replacing $\vert{}\sR_{\al \b}\vert{}^2$ with $\Fil{\vert{}\sR_{\al \b}\vert{}^2} = 1$ (for the same reasons discussed in footnote~\ref{ft:eth}), we obtain an expression for $\sG(s)$ solely in terms of macroscopic quantities. 
We can then replace the sum with an integral,
 \bega
 \sG (s) =  \int d \sE \, e^{\sS (\sE) - \sE}  A (\sE) +  \int d \sE d \om \, e^{\sS (\sE) - \sE} e^{i \om s - {\om \ov 2}} \hat F (\om, \sE)  \ , \\
 A (\sE) = a^2  (\sE) \ , \quad \hat F (\om,\sE) =  |F (\om, \sE) |^2  \ .
 \end{gather}
Note that the normalization of the density matrix implies
\be \label{ser}
\sum_\al p_\al=1
\quad\Longrightarrow\quad
\int d\sE\,e^{\sS(\sE)-\sE}=1\,.
\ee
For an extensive modular entropy, this integral can be evaluated by saddle point. The dominant saddle $\sE_0$ satisfies
\be
\sS'(\sE_0)=1\,.
\ee
The normalization condition further implies, to leading extensive order,
\be
\sS(\sE_0)=\sE_0+\text{subextensive corrections}\,,
\ee
where the corrections include the Gaussian fluctuation determinant around the saddle.

The entanglement entropy of $\rho_R$ is
\be
S_R=-\Tr(\rho_R\log\rho_R)
=\Tr(\rho_R K)
=\int d\sE\,\sE\,e^{\sS(\sE)-\sE}
\approx \sE_0\,.
\ee
That is, $\sE_0$ coincides, to leading order, with the entropy of the state. Since the entropy of the black-hole state is extensive, $\sE_0$ is itself extensive,  placing the saddle in the regime where the assumed thermodynamic scaling of $\sS(\sE)$ applies and the saddle-point approximation is justified self-consistently.

By definition, the macroscopic functions $A (\sE)$ and $\hat F (\om, \sE)$ vary slowly with $\sE$. Evaluating the integral at the saddle point, we can thus write
 \be \label{enn}
 \sG(s) \approx A (\sE_0) +  \int d \om \, e^{i \om s - {\om \ov 2}}  \hat F (\om, \sE_0)   \ .
 \ee
 We stress that the macroscopic functions $A (\sE)$ and $\hat F (\omega, \mathcal{E})$ inherently depend on $\OO_{ij}$. In general, there is no fundamental reason why these functions should have small variances under the erratic fluctuations of $\OO_{ij}$. Because $\sG(s)$ is determined by evaluating these functions at the saddle point $\mathcal{E}_0$, any $\mathcal{O}(1)$ variance in $A$ or $\hat F$ translates directly into an $\mathcal{O}(1)$ variance in $\sG (s)$ under the filter $\FF$.
 It is instructive to contrast this with the exterior correlation functions~\eqref{frrD}, whose small variance relies only on the stability of the thermodynamic saddle point itself. For the interior correlator~\eqref{enn}, a small variance would require much stronger conditions: not only must the saddle point $\sE_0$ be stable against erratic fluctuations, but the functions $A (\sE)$ and $\hat F (\omega, \mathcal{E})$ must independently have suppressed variances as well.

To gain some intuition, let us consider the case where the operator $B$ defined in~\eqref{PPae} is unitary, so that
\be
X=B\rho_\b B^\da\,,\qquad
\rho_R=X\,,\qquad
Z_\OO=1\,,
\ee
where $\rho_\b$ is the thermal density matrix. This corresponds physically to acting with a unitary operator on one side of the thermofield double (TFD) state. The eigenvectors of $\rho_R$ are simply
\be
\ket{i}_B=B\ket{i}\,,
\ee
while its eigenvalues, and hence its modular spectrum, are unchanged. We therefore have
\be
a(\sE_i)
={}_B\vev{i|\sO|i}_B
=\vev{i|B^\da\sO B|i}\,.
\ee
Thus, $a(\sE)$ is the expectation value of the ``dressed'' operator $B^\dagger\mathcal O B$ in a standard energy eigenstate. Since $B$ is itself constructed from the underlying erratic matrix elements of $\OO$, the dressed operator, and hence the smooth function $a(\sE)$, inherits a dependence on the same microscopic randomness. This example makes explicit how the macroscopic quantity $A(\sE)=a^2(\sE)$ can fluctuate under variations of the microscopic data, even though it is smooth as a function of $\sE$ for any given realization.

To summarize:
\ben

\item The very definition of the interior operators~\eqref{egb} through modular flow inherently depends on the erratic microscopic $N$-dependence of the underlying state. The smooth semiclassical geometry of the interior emerges only after filtering the resulting boundary quantities.

\item As discussed earlier (near the end of Sec.~\ref{sec:AS22}), a filtered variance is not an ordinary quantum variance in a fixed microscopic state. We have given it an operational interpretation as the uncertainty faced by a low-energy observer owing to the observer's inability to access the microscopic information discarded by the filter. For exterior observables, this uncertainty is suppressed by self-averaging. For interior observables, however, the filtered variance can remain unsuppressed. When this occurs, the interior geometry is ``quantum volatile'' in this operational sense.

\een

It is natural to conjecture that these statements hold more generally for black holes. For a generic two-sided black hole, bulk operators in the interior are again expected to be constructed from modular-flowed single-trace operators. While for such general states the modular operator can no longer be expressed explicitly in terms of heavy operators as in~\eqref{rhoR}, it is reasonable to expect that it continues to inherit the erratic $N$-dependence of the underlying microscopic state.

A notable exception is the eternal Schwarzschild black hole dual to the thermofield double (TFD) state above the Hawking--Page temperature. In this case, bulk operators in the future interior can be represented as linear superpositions of single-trace operators from the $R$ and $L$ CFTs without explicitly invoking modular flow. The TFD state, and the corresponding eternal Schwarzschild geometry, nevertheless constitute a highly special configuration. Under generic perturbations, the state is expected to evolve into a ``long'' black hole, for which the generic  behavior should be expected.

We expect similar considerations to apply to single-sided black holes. Consider, for instance, the AdS--Vaidya geometry, in which a single-sided black hole forms from the gravitational collapse of a thin shell of matter sourced from the boundary of empty AdS. In the CFT, such a shell can be modeled by the insertion of a heavy operator, extended over the boundary spatial directions, acting on the vacuum. After the black hole forms, bulk operators outside the horizon are, as usual, described by standard single-trace operators. For operators behind the horizon, although no explicit modular-flow representation is available, we expect their boundary representations to involve single-trace operators dressed by the heavy operator responsible for forming the black hole. Such dressed operators can naturally inherit sensitivity to the erratic $N$-dependence of the background microstate.

The quantum volatility of the black hole interior may have direct implications for the experience of an infalling observer. In particular, if the propagator of a bulk field exhibits $O(N^0)$ filtered fluctuations in the sense described above, then the effective predictions available to a low-energy observer for measurements of interior observables need not be sharply defined. We emphasize that this is not a firewall or a violent high-energy breakdown of semiclassical physics at the horizon. Rather, it reflects an intrinsic statistical volatility of interior correlators after microscopic information has been discarded by the filter.

Furthermore, the proper time available to an infalling observer before reaching the singularity is finite. Even if the observer could in principle probe quantum volatility through fluctuations in expectation values or response functions of local observables, there may be insufficient proper time to accumulate the statistics required to diagnose it reliably. Thus, even if the interior is quantum volatile from the standpoint of the underlying theory, this volatility may be only partially operationally accessible, with its observable consequences limited by the finite proper time before the singularity. This limitation is conceptually analogous to that faced by an observer in a closed universe approaching a Big Crunch.

\section{Conclusions and discussions} \label{sec:conc}



\noindent{\bf Summary of main results} 

\medskip 

In this paper, we postulated a large-$N$ split~\eqref{deCom} for the CFT partition function on a general compact Euclidean manifold $M$, and~\eqref{deiL} for general Lorentzian observables. Building on these ingredients, we proposed a filtered large-$N$ holographic dictionary~\eqref{mlarN} and its Lorentzian counterpart~\eqref{mlarN1}. The new dictionary proposes an intrinsic boundary definition of gravitational ``averages'' and can be used to explain the origin of wormhole contributions. This allows for an (in-principle) intrinsic boundary computation of their amplitudes, suggesting a resolution to the factorization puzzle.
Its Lorentzian formulation further leads to a generalized framework---``filtered emergence''---from which richer Lorentzian spacetime structures can arise.

The splits~\eqref{deCom} and~\eqref{deiL} are our main assumptions. Their structure generalizes the usual asymptotic form in terms of a transseries. Recall that a transseries is defined as a formal sum of several asymptotic expansions multiplied by exponentially small scales, e.g.
\be\label{transe}
F(N)=e^{-N^2 W_0} \sum_{n=0}^\infty \frac{f_n}{N^{2n}}
+ e^{-N^2 W_1} \sum_{n=0}^\infty \frac{g_n}{N^{2n}},
\qquad W_1>W_0 \ .
\ee
Mathematically, even though the leading asymptotic series is divergent and the subdominant contributions are exponentially smaller than any of its terms, it is still possible to make sense of these subdominant sectors both conceptually and operationally---the primary motivation for introducing transseries. In particular, the theory of resurgence governs the direct analytic links that can exist between the leading and subdominant series.

Here we are proposing that the large-$N$ asymptotics can be generalized further to
\be\label{newtran}
F(N)=e^{-N^2 W_0} \sum_{n=0}^\infty \frac{f_n}{N^{2n}}
+ e^{-N^2 W_1} \sum_{n=0}^\infty \frac{g_n}{N^{2n}}
+ e^{-N^2 W_2} h(N),
\ee
where, instead of an asymptotic expansion, the function $h(N)$ depends on $N$ in an erratic fashion. The first two terms are grouped into $F^{\rm smooth}$. It would be interesting to understand whether such erratic dependence can also be understood in terms of resurgence.

There may be mathematical ambiguities in making the decomposition into the smooth and erratic parts (and consequently in~\eqref{deCom} and~\eqref{deiL}), just as there are ambiguities in~\eqref{transe} when separating $F(N)$ into its leading and subdominant sectors. It is reasonable to expect that such ambiguities do not affect the qualitative structure of the two parts, in close analogy with the situation for transseries. Moreover, these ambiguities may themselves encode interesting physics.\footnote{For example, as in the choice of renormalization scheme, or as in transseries where the ambiguities are organized and resolved by resurgence.}


It would be highly desirable to identify these splits in explicit examples. 
Possible candidates include CFT$_2$ partition functions on manifolds of genus $g \ge 2$ and partition functions involving heavy operator insertions. A detailed analysis of the large-$N$ limits of superconformal and topologically twisted indices may also prove fruitful. 

We now summarize our main results concerning Euclidean wormholes:
\bi
\item The relations~\eqref{imRe}--\eqref{0nide} between the erratic parts of boundary partition functions and bulk wormhole amplitudes.

\item An infinite tower of inequalities constraining wormhole amplitudes, with the first few given in~\eqref{ineq0}--\eqref{ineq3}.

\item ``Correlations'' of OPE coefficients for heavy operators in general dimension $d$ and in $d=2$. In particular, in $d=2$ these correlations may be expressed in terms of the Liouville CFT structure constants and density of states~\eqref{Liou}.

\item Internal wormholes do not induce random couplings in the low-energy effective theory.

\item New contributions to holographic R\'enyi and entanglement entropies.

\item New boundary perspectives on the violation of global symmetries in quantum gravity.
\ei
In Lorentzian signature, the filtered dictionary leads to richer emergent spacetime structures. We examined three classes of examples:
(i) isolated closed universes;
(ii) baby universes entangled with AdS;
(iii) the interior of a long black hole. Their key features include:

\bi

\item The semiclassical Hilbert space of an isolated closed universe can be constructed from correlations of the erratic parts of boundary partition functions. Moreover, one may introduce a semiclassical Hilbert space that incorporates transitions among all isolated closed universes described by the CFT.

\item For a baby universe entangled with AdS, both the baby universe and the AdS region are quantum volatile. This provides a possible mechanism for AdS observers to infer the presence of the baby universe. 

This class of examples may also shed light on the physics of a fully evaporated black hole.

\item The interior of a black hole can be quantum volatile, whereas the exterior is not. 


\ei
These features refine how classical geometry should be understood in a theory of gravity. They suggest that the boundary between a genuinely classical spacetime and one that is secretly quantum volatile may be subtler than usually assumed.

It is instructive to contrast these three classes. The cases (ii) and (iii) admit a description in terms of a finite-$N$ state (such as PETS), whereas (i) appears not to have such a finite-$N$ Lorentzian description. This absence may reflect not a limitation of current understanding but rather a genuine feature:  the Lorentzian bulk description of a state in the boundary Hilbert space should involve a boundary. 
Isolated closed universes may only be realizable through an operator algebra (see, e.g.,~\cite{Liu25}) rather than through a state.

\medskip

\noindent {\bf Future perspectives}

\medskip

We conclude this section by highlighting several conceptual issues. 

\ben
\item {\it An explicit mathematical device for filtering on the CFT side}

The passage from~\eqref{cenT} to~\eqref{mlarN} can be viewed as implementing a filtering operation on both sides of the duality. As discussed in the introduction, on the gravity side this filtering is naturally realized by the gravitational path integral, which furnishes a semiclassical expansion that is smooth in $G_N$.

On the CFT side, we implemented the filter by postulating the intrinsic split~\eqref{deCom} and then projecting onto the smooth term. It is natural to wonder whether there exists an operational procedure that performs this projection directly at the level of $Z_{\rm CFT}[M]$, without first invoking the split. Identifying such an operational procedure---an analogue of the gravitational path integral on the boundary---would certainly be of great value.

\item {\it Spacelike singularities}

Closed universes and black hole interiors both contain spacelike singularities. The filtered large-$N$ perspective suggests a possible new angle on their interpretation. One might speculate that these singularities emerge precisely in regimes where the filtering procedure becomes ineffective: erratic large-$N$ fluctuations of boundary observables can no longer be smoothed into coherent $G_N$ dependence, causing semiclassical geometry to break down.
In this view, a spacelike singularity signals a limit in which the dictionary fails to produce a stable semiclassical spacetime. 

\item {\it Intrinsic gravity description of erratic $G_N$-dependence}

While the gravitational path integral provides a natural filtering mechanism, the question remains: what does erratic $G_N$-dependence look like from the gravity side? What is their gravitational origin? 

\een

\vspace{0.2in}   \centerline{\bf{Acknowledgements}} \vspace{0.2in}
We would like to thank 
Shoaib Akhtar, Thomas Banks, Gauri Batra, Alex Belin, Justin Berman, Nikolay Bobev, Xi Dong, Netta Engelhardt, Elliott Gesteau, Daniel Harlow,  Thomas Hartman, Luca Iliesiu, Kristan Jensen, Marc Klinger, Zohar Komargodski, Jonah Kudler-Flam, Adam Levine, Juan Maldacena, Joe Minahan,  Gregory Moore, Sameer Murthy, Xiaoliang Qi, Arvin Shahbazi-Moghaddam, Stephen Shenker, Douglas Stanford, Edward Witten, Cynthia Yan, Ying Zhao, and in particular Scott Collier for discussions. This work is supported by the Office of High Energy Physics of U.S. Department of Energy under grant Contract Number  DE-SC0012567 and DE-SC0020360 (MIT contract \# 578218), and was made possible through the support of grant \#63670 from the John Templeton Foundation.

\appendix


\section{
Mathematical aspects of the erratic decomposition} \label{sec:filter}



In our discussion, we have assumed that a boundary CFT observable $F$, such as the Euclidean partition function on a compact manifold, have the decomposition of the form 
\be \label{traS}
F = F^{(\rm smooth)}  + \de_e F 
\ee
where $F^{(\rm smooth)}$ denotes the part that can be expressed as a transseries (we suppress possible logarithms),  
\be 
F^{(\rm smooth)} =  \sum_{i} e^{-N^2 W_0^{(i)} }  Z_1^{(i)} \le(1 +  {1 \ov N^2} Z_2^{(i)} + \cdots  \ri) ,
\ee
and $\de_e F$ denotes the erratic part. Furthermore, products of $\de_e F$ may also have a decomposition 
into the smooth and erratic parts. 



To have some intuition how the structure~\eqref{traS} arises, consider an infinite sum of the form 
\be \label{fn}
F(N) = \sum_a f_a  (N) e^{i N^2 g_a}, 
\ee
where $f_a (N)$ depends smoothly on $N$, i.e., having an asymptotic expansion in $1/N^2$ and $\{g_a\}$  is a set of ($N$-independent)  random numbers (e.g. taken from a random distribution). 
We can then identify  
\be
F^{(\rm smooth)}  =  \sum_{a \in I} f_a,
\quad
\de_e F =  \sum_{a \not \in I} f_a e^{i N^2 g_a},  \quad I = \{ a | g_a = 0 \},
\ee 
where $F^{(\rm smooth)}$ has an asymptotic expansion in $1/N^2$ from those of $f_a (N)$.\footnote{Note that $I$ may be empty, in which case $F^{(\rm smooth)}$ vanishes identically and is trivially ``smooth''.}
The quantity \(\delta_eF\) involves an infinite sum of rapidly oscillating random phases and therefore generically does not admit a proper asymptotic expansion in \(1/N^2\). 

Now consider 
\bega 
G \equiv \de_e F (\de_e F)^* =  \sum_{a , b\not \in I} f_a f_b^*  e^{i N^2 (g_a - g_b)} = 
G^{(\rm smooth)}  + \de_e G , \\
G^{(\rm smooth)} = \sum_{\stackrel{a , b\not \in I}{g_a = g_b}} f_a f_b^* \ .
\end{gather} 
$G^{(\rm smooth)}$ has an asymptotic expansion in $1/N^2$ again following from those of $f_a (N)$.
We emphasize that the use of the phases $e^{i N^2 g_a}$ is purely illustrative: the true erratic dependence need not be phases nor of this particular structure.

\linespread{1.5}

\bibliographystyle{jhep}
\bibliography{all}

\end{document}